\documentclass[a4paper,fleqn]{cas-sc}

\usepackage[numbers]{natbib}
\usepackage{subcaption}
\usepackage[ruled,noend]{algorithm2e}
\usepackage{enumitem}
\usepackage{tabularx}

\newtheorem{proposition}{Proposition}
\newproof{proof}{Proof}
\newtheorem{remark}{Remark}

\providecommand{\Hat}[1]{\widehat{#1}}
\providecommand{\Bar}[1]{\overline{#1}}
\providecommand{\Tilde}[1]{\widetilde{#1}}

\begin{document}
\let\WriteBookmarks\relax
\def\floatpagepagefraction{1}
\def\textpagefraction{.001}

\shorttitle{Online adaptive Grassmann-based model reduction}
\shortauthors{A. Tiba et al.}

\title[mode=title]{Online adaptive non-intrusive model reduction via manifold interpolation and subspace updates: application to FSI convergence acceleration}

\author[1]{Azzeddine Tiba}[orcid=0000-0003-3578-141X]
\cormark[1]
\ead{azzeddine.tiba@gmail.com}
\ead[URL]{https://azzeddinetiba.github.io/}
\credit{Conceptualization, Formal Analysis, Investigation, Methodology, Software, Validation, Visualization, Writing – original draft}

\author[2]{Florian De~Vuyst}[orcid=0000-0003-0854-4670]
\credit{Formal analysis, Supervision, Writing – review \& editing}

\author[1]{Iraj Mortazavi}[orcid=0000-0003-3584-533X]
\credit{Formal analysis, Resources, Supervision, Writing – review \& editing}

\affiliation[1]{organization={MACS, CNAM},
                        addressline={2, Rue Conté},
            city={Paris},
            postcode={75003}, 
            country={France}}

\affiliation[2]{organization={Laboratory of Biomechanics and Bioengineering, Université de Technologie de Compiègne, CNRS},
            country={France}}

\cortext[cor1]{Corresponding author}

\begin{abstract}
We introduce a novel online adaptive non-intrusive reduced-order modeling strategy for parameterized dynamical systems involving parameter and time-dependent reduced bases. The proposed framework is based on a unified Grassmann manifold formulation combining three key components: interpolation of local reduced subspaces for unseen parameters, geodesic online subspace updates driven by incoming high-fidelity snapshots, and a latent-space regression strategy relying on Grassmann-distance weighting and Procrustes alignment to consistently aggregate predictions from multiple local models. The adaptive reduced-order model is embedded in a partitioned fluid-structure interaction framework, where it predicts fluid interface forces to provide accurate initial guesses for the nonlinear coupling iterations, thus achieving computational speedups with no loss of accuracy. The reduced basis and the regression operators are adapted independently during the simulation and without requiring the storage of high-dimensional streaming data, preserving computational efficiency while substantially improving predictive capabilities. Numerical results on reference FSI test cases demonstrate superior accuracy with respect to static and global reduced-order models, leading to a significant reduction in the number of fixed-point iterations required for convergence. The proposed framework offers a flexible and fully non-intrusive approach for the efficient simulation of nonlinear parameter-dependent multiphysics problems.
\end{abstract}


\begin{highlights}
\item Online adaptation is applied to non-intrusive reduced order models.
\item Local and online adaptive subspaces are found on geodesics on the Grassmann manifold.
\item Subspace and latent dynamics adapt independently.
\item No storage of streaming high-dimensional data is needed.
\item Speedups with no accuracy loss via better initialization of fixed-point iterations.
\end{highlights}


\maketitle

\section{Introduction}

Reduced Order Models (ROM)s have emerged as an increasingly vital methodology for reducing the computational costs associated with large-scale numerical simulations. In many engineering applications, high-fidelity (HF) full order models have prohibitive costs, especially when repeated evaluations are required, as is the case in parametric studies, design optimization, inverse problems, and real-time control. Reduced order modeling addresses this challenge by constructing significantly lower dimensional models that nonetheless capture the essential physics and the dominant structures in the problem of interest. In fact, for a broad class of problems, the solution usually evolves on a low dimension manifold embedded in the high-dimensional space~\citep{NOACK_AFANASIEV_MORZYŃSKI_TADMOR_THIELE_2003}.
ROMs commonly approximate these solution manifolds through linear subspaces, enabling the computation of solutions constrained to those spaces when solving the reduced system.
This approximation framework has proven remarkably effective in a wide range of problems, yielding highly accurate approximations at a fraction of the computational cost of the full-order model~\citep{bennerBook}.

In many situations, the restriction of the solution on a linear subspace turns out to be insufficient, unless using subspaces with large dimensions, at the expense of the computational gain of the ROM. This is the case for example when solving transport-dominated problems~\citep{GREIF2019216, chapter3BookTransport}, or nonlinear problems with complex structures in the solution manifold~\citep{gu2011model, haller2016nonlinear, JAIN201780}. To address this issue, several approaches have been proposed to better capture the nonlinear behavior of the manifold. These approaches can be broadly classified into two main categories according to how the nonlinearities are incorporated. This classification is independent of the reduced-order modeling strategy, as both projection-based and Galerkin-free methods can be found within each category.
\begin{itemize}
    \item The first class of methods attempt to approximate the manifold itself via its nonlinear representation in the high-dimensional space, \textit{e.g} using piece-wise linear projection functions~\citep{chenjieGuManiROMs},  polynomial-approximated manifolds~\citep{JAIN201780, BARNETT2022111348, haller} or autoencoders~\citep{kashimeAutoenc, gonzalez2018deepconvolutionalrecurrentautoencoders, LEE2020108973}.
    \item The second class of methods use local subspaces (as opposed to global subspaces), in the sense that different linear reduced bases are used for different regions of the parameter domain~\citep{amsallem2008interpolation, OULGHELOU2018416, Daniel2020}, the state-space domain~\citep{amsallemLocalOriginal, COLANERA2025118393} or the time domain~\citep{dihlmann2011model}. This essentially means that the manifold is approximated using piece-wise linear subspaces.
\end{itemize}

In that second class, we can also find methods that use online adaptation in order to have a time-varying basis.
The existing approaches that used online adaptive bases have done so for various purposes, but they have all demonstrated remarkable advantages over static bases ROMs.
In~\citep{amsallem2015fast}, subspace update was used when a transition occurs between different subregions of the state-space associated to different local bases. The online data used to update the subspace associated to the "current" subregion is obtained from the ROM solution computed on the previous subregion, which enhances the basis quality otherwise computed using offline data only.
In~\citep{onlineGalerking} and~\citep{zimmerCox}, the update is used to adapt the Discrete Empirical Interpolation Method (DEIM) basis and its interpolation points using an augmented set of points during the online simulation.
A similar update method was adopted in~\citep{onlineGalerkinTransport} for transport-dominated problems. In that work, the update method was applied on both for DEIM and the ROM solution as they shared the same space.~\citet{HUANG2023112356} then adapted the approach of~\citep{onlineGalerkinTransport} to better account for the non-local effects in problems with distributed multi-scale
features. Similarly, online hybrid approaches have been proposed to switch between ROMs and Full Order Models (FOMs), while using the results of the latter to update the reduced basis~\citep{BAI2022100, RIFFAUD2025113677, deformableUpdate}.

More recently, online adaptive subspaces are now being adopted in Galerkin-free methods as well.~\citet{taleb} used a recursive approach to update both the reduced basis and the linear model in a Dynamic Mode Decomposition (DMD) framework. Similarly,~\citep{Tang2025, Hedayat2026} proposed updating the reduced basis in a purely data-driven model when new observations are obtained in a digital twin context, or when the HF operator is available for online queries. We note that subspace updating is not the only strategy for online adaptation in data-driven ROMs. In recent years, emerging approaches have used adaptation mechanisms to enhance ROM accuracy through alternative means. For example,~\citet{SCHERDING2025109404} employed online adaptive \emph{radial basis function} (RBF) networks to improve the extrapolation capabilities of data-driven ROMs. In~\citep{BermanCoLora}, Neural Networks (NNs) are fine-tuned using efficient low-rank weight updates, enabling improved accuracy of data-driven ROMs with limited offline training data. We also note recent adjacent works that developed ROMs with dynamic low-rank representations, either by solving the governing equations directly ~\cite{RAMEZANIAN2021113882, AITZHAN2025113549}, or by constructing the ROMs in a fully data-driven manner~\cite{dechant2026dynamicsubspaceapproachlowrank}.

In this work, we propose an online adaptive non-intrusive ROM that combines three components into a unified framework: a local basis constructed by interpolation on the Grassmann manifold to initialize the subspace for a new parameter configuration, an online basis update via geodesic gradient descent on that same manifold using incoming HF snapshots, and a latent-space regression strategy that aggregates locally trained operators through Procrustes-aligned, Grassmann-distance-weighted combinations to ensure consistency across heterogeneous reduced coordinate frames.
While the recent developments in \citep{Tang2025, Hedayat2026} proposed online adaptive non-intrusive ROMs, they did not treat parametric locality. More notably, those strategies are restricted on the simultaneous update of both the subspace and the regression components, and need storage of high-dimensional snapshots. The current approach circumvents both those difficulties, through a strategy where the manifold structure governs both the initialization, the update, and the coordination of the regression outputs. And that composition constitutes, to the authors' knowledge, a novel methodological contribution.

The application context further motivates this construction: in partitioned Fluid-Structure Interaction (FSI) simulations, HF snapshots are naturally generated at runtime by the fluid and solid solvers.
The current strategy is used in a non-intrusive fashion to predict fluid forces and is coupled to a solid-state ROM to provide an accelerated initial guess for the nonlinear FSI coupling scheme, building on the data-driven framework introduced in~\citep{TIBA2025109522}, where adaptation on the regression operators alone was shown to improve accuracy. The present work extends that contribution by incorporating adaptive bases, addressing the additional challenge of strong parameter dependence through local subspace representations that outperform the global bases used in~\citep{TIBA2025109522}. The proposed framework is evaluated in terms of solution accuracy against static and global ROM baselines, and in terms of computational gain through accelerated convergence of the FSI iterations.

The paper is organized as follows: In section \ref{section1}, we present the proposed online adaptation approach in the context of data-driven ROMs. In section \ref{section2}, we explain the practical computational steps of the ROM construction  and prediction. The next section \ref{section3} contains the explanation of how the proposed ROM is applied on the prediction of initial guesses in a partitioned FSI simulation. Then, we show the evaluation of the ROM accuracy and computational gains on three FSI-related test cases in section \ref{section4}. Finally, we present our conclusions about the proposed approach and ideas for further discussions in section \ref{section5}.

\section{Parametric data-driven reduced order models with online time update}\label{section1}
\subsection{General formulation}
We consider a parameter-dependent, controlled dynamical system representing the FOM of the form
\begin{equation}\label{objective_system}
 {\dot{\boldsymbol{x}}_{\boldsymbol{\theta}}} (t) = \mathcal{F}( {\boldsymbol{x}}_{\boldsymbol{\theta}}(t), {\boldsymbol{u}}(t), \boldsymbol{\theta}, t )
\end{equation}
where $\boldsymbol{x}_{\boldsymbol{\theta}}(t) \in \mathbb{R}^{N_x}$, $\boldsymbol{u}(t) \in \mathbb{R}^{N_u}$, 
and $\boldsymbol{\theta} \in \mathbb{R}^{N_p}$ are the state, control input, and 
parameter vector respectively. Upon time discretisation with step $\Delta t$, we denote 
$\boldsymbol{x}_{\boldsymbol{\theta}}^n = \boldsymbol{x}_{\boldsymbol{\theta}}(n \Delta t)$ and $\boldsymbol{u}^n = \boldsymbol{u}(n \Delta t)$.
We have assumed here that the control input is independent of the state-space and the parameter $\boldsymbol \theta$.

We define the input space as the Cartesian product:
\begin{equation}
    \mathcal{Z} = \mathbb{R}^{N_x} \times \mathbb{R}^{N_u}
\end{equation}
and we aim to rapidly approximate the dynamics by a discrete-time, step-dependent operator defined as
\begin{equation}
    \Hat{F}^n_{\boldsymbol{\theta}} : \mathcal{Z} \longrightarrow \mathbb{R}^{N_x} \quad 
    \left(\boldsymbol{x}^{n-1},\, \boldsymbol{u}^{n}\right) 
    \longmapsto \boldsymbol{x}^{n}
\end{equation}
such that the discrete dynamical system reads:
\begin{equation}\label{eq:discrete_operator}
    \boldsymbol{x}^{n}_{\boldsymbol{\theta}} = \widehat{F}^n_{\boldsymbol{\theta}}\!\left(\boldsymbol{x}_{\boldsymbol{\theta}}^{n-1},\, 
    \boldsymbol{u}^{n}\right), \qquad \forall\, n \in \{1, \dots, N_t\}
\end{equation}
where $N_t$ is the total number of time steps. The superscript $n$ on $F^n$ 
indicates that the internal structure of the operator may vary across time steps. In this work, as motivated by our application of accelerating coupling iterations online, we focus our attention on a ROM that operates in a non-autonomous setting, where it is continuously fed high-fidelity state and control data $(\boldsymbol{x}^{n-1}, \boldsymbol{u}^n$) as input.

As the state-space lives in a very high-dimensional space (\textit{i.e} $N_x$ is very large), we apply dimensionality reduction via encoders and decoders that we consider to be local in the parametric space

We define the encoder $\mathcal{E}_{\boldsymbol{\theta}}$ and the decoder $\mathcal{D}_{\boldsymbol{\theta}}$ as:
\begin{equation}
    \mathcal{E}_{\boldsymbol{\theta}} : \mathbb{R}^{N_x} \longrightarrow \mathbb{R}^{r} 
    \quad \boldsymbol{x} \longmapsto  
     \boldsymbol{x}^{r}
\end{equation}
\begin{equation}
    \mathcal{D}_{\boldsymbol{\theta}} : \mathbb{R}^{r} \longrightarrow \mathbb{R}^{N_x} \quad 
     \boldsymbol{x}^{r}
    \longmapsto  
     \boldsymbol{x}
\end{equation}
where $r$ is chosen to be very small compared to its associated high-dimensional spaces: $r \ll N_x$.

We then define a regression operator that, given the reduced coordinates of the previous state and the current control input in their respective latent spaces, returns the current reduced coordinates of the state space.
\begin{equation}
    \mathcal{I}_{\boldsymbol{\theta}} : \mathbb{R}^{r} \times \mathbb{R}^{r_u} \longrightarrow \mathbb{R}^{r} \quad 
     \left(\boldsymbol{x}^{r, n-1}, \boldsymbol{u}^{r, n}\right)
    \longmapsto  
     \boldsymbol{x}^{r, n}.
\end{equation}

In~\citep{TIBA2025109522}, we only updated the (non-parametric) ROM by using a time-dependent regressor operator, while keeping the other components fixed. In this work, we will apply time-adaptation on both the regression and the encoder-decoder operators. Therefore, the approximated model can be written in the following form:
\begin{equation}\label{eq:learnt_system}
    \boldsymbol{x}^{n}_{\boldsymbol{\theta}} = \widehat{F}^n_{\boldsymbol{\theta}}\!\left(\boldsymbol{x}_{\boldsymbol{\theta}}^{n-1},\, 
    \boldsymbol{u}^{n}\right) = \mathcal{D}^n_{\boldsymbol{\theta}}\left(\, \mathcal{I}^n_{\boldsymbol{\theta}}\left( \mathcal{E}_{ \boldsymbol{\theta}}^n\left(\boldsymbol{x}^{n-1}\right), \,\,\,  \boldsymbol{u}^{n} \right)\right), \qquad \forall\, n \in \{1, \dots, N_t\}.
\end{equation}

\begin{figure}
    \centering
    \includegraphics[width=.35\textwidth]{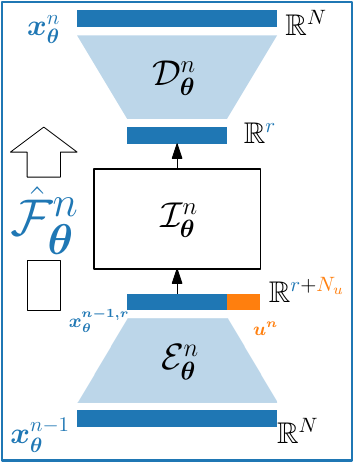}
    \caption{An illustration of the different components of the ROM as a parametric discrete dynamical system.}
    \label{fig:rom-structure}
\end{figure}

In Fig. \ref{fig:rom-structure}, we present an illustration of the components of $\widehat{F}^n_{\boldsymbol{\theta}}$.

\subsection{Proper Orthogonal Decomposition}\label{sec:pod}
We use the Proper Orthogonal Decomposition (POD) method~\citep{lumley1967structure} to compute the low-dimensional representation of the state-space and the control input.
We store $m$ snapshots of data $\boldsymbol{x}^n$ of the state-space in the snapshot matrix as $\boldsymbol{X} = \left[\boldsymbol{x}^1 - \Bar{\boldsymbol{x}}, \cdots ,  \boldsymbol{x}^m  - \Bar{\boldsymbol{x}} \right] \in \mathbb{R}^{N_x \times m}$ where $\Bar{\boldsymbol{x}}$ is the mean of the snapshots. POD constructs an orthonormal basis representing the optimal linear subspace $\mathcal{V}$ on which $\boldsymbol{X}$ can be projected.
\begin{equation}
    \boldsymbol{\Phi} = arg\, \min_{\boldsymbol{\Phi}_x \in \mathbb{R}^{N_x\times r}}||\boldsymbol{X} - \boldsymbol{\Phi}_x \boldsymbol{\Phi}^T\boldsymbol{X}||_F^2
\end{equation}
\begin{equation}
    \text{subject to }\; \boldsymbol{\Phi}^T \boldsymbol{\Phi} = \boldsymbol{I}_{r}.
\end{equation}
where $\boldsymbol{I}_{r}$ is the $r\times r$ identity matrix and $||\cdot||_F$ is the Frobenius norm. Following the method of snapshots~\citep{sirovich_turbulence_1987}, the POD basis is obtained through a truncated Singular Value Decomposition (SVD) of the snapshot matrix.
\begin{equation}
    \boldsymbol{X} \approx  \boldsymbol{\Phi} \boldsymbol{\Lambda} \boldsymbol{\Psi}^{T},
\end{equation}
where $T$ denotes the transpose.
The truncated POD basis is of rank $r$, and the rank is chosen following an energy criteria based on the singular values:
\begin{equation}\label{SVDdecay}
\begin{aligned}
r = arg \min_{r\in[1, min(n, m)]} \quad & S = \frac{\sum_{i=1}^{r} \lambda_i^2}{\sum_{i=1}^{min(n, m)} \lambda_i^2}\\
\textrm{s.t.} \quad &  S \leq \varepsilon\\
\end{aligned}
\end{equation}
where $\lambda_i$ are the elements of $\boldsymbol{\Lambda}$. The encoders and decoders are then expressed as:
\begin{equation}\label{encoder_temp}
    \mathcal{E}(\boldsymbol{x}) =  {\boldsymbol{\Phi}}^T (\boldsymbol{x} - \Bar{\boldsymbol{x}}),
\end{equation}
\begin{equation}\label{decoder_temp}
    \mathcal{D}(\boldsymbol{x}^r) =  {\boldsymbol{\Phi}} \boldsymbol{x}^r + \Bar{\boldsymbol{x}}.
\end{equation}
We note that we omitted here the parameter subscript $\boldsymbol{\theta}$ and the time superscript $n$ , since parameter and time adaptation are not introduced until the next section.

\subsection{Locality in the parameter space: Grassmann interpolation}\label{grassmannInterpolationSection}

In this work, we use parametrically-local ROMs to obtain accurate predictions for previously unseen parameter values and to account for potential nonlinearities in the parameter space. To achieve this, we interpolate the subspaces corresponding to different parameters. Thus, the interpolated subspaces must remain on the manifold of all subspaces of a given dimension, also known as the Grassmann manifold.
In this section, we review key concepts of the Grassmann manifold that are relevant to our subspace interpolation approach. For further details and theoretical results, we refer the reader to~\citep{EdelmanA, Absil2004, AbsilMahonySepulchre, Lee2012, zimmermann2022manifoldinterpolationmodelreduction}.

The set of all $r$-dimensional subspaces of $\mathbb{R}^N$ forms the Grassmann manifold:
\[
  Gr(N, r) = \{ \mathcal{V} \in \mathbb{R}^N: \mathcal{V} \textnormal{ is a subspace with } dim(\mathcal{V}) = r \}.
\]
To associate an element $\mathcal{V}$ of $Gr(N, r)$ with a basis matrix $ \boldsymbol{\Phi}$ representing it, we first recall the definition of Stiefel manifold, formed by the set of orthonormal matrices:
\[
  St(N, r) = \{ \boldsymbol{\Phi} \in \mathbb{R}^{N \times r}  \textnormal{ s.t } \boldsymbol{\Phi}^T \boldsymbol{\Phi} = \boldsymbol{I}_r \}.
\]
Obviously, an orthonormal basis $\boldsymbol{\Phi}$ is not a unique representative of $\mathcal{V} = span(\boldsymbol{\Phi})$. We can, however, use the following equivalence class
\begin{equation}\label{equiv_class}
    \left[ \boldsymbol{\Phi} \right] = \{ \boldsymbol{U} \in \mathbb{R}^{N \times r} : \boldsymbol{U} = \boldsymbol{\Phi} \boldsymbol{Q},\,\boldsymbol{\Phi}\in St(N, r) , \, \boldsymbol{Q} \in  \mathbb{R}^{r \times r},\, \boldsymbol{Q}^T \boldsymbol{Q} = \boldsymbol{I}_r  \}
\end{equation}
to identify $\mathcal{V}$. We can now associate an element $\mathcal{V}$ of $Gr(N, r)$ with a basis representative if we use the following expression of the Grassmann manifold
\begin{equation}
    Gr(N, r) = \{ \left[ \boldsymbol{\Phi} \right]: \boldsymbol{\Phi} \in St(N, r) \}.
\end{equation}
At a point $\mathcal{V}$ of ${Gr}(N,r)$, the tangent space can be expressed using a representative of $\mathcal{V}$:
\begin{equation}\label{tangent_space_grassmann}
    T_{\mathcal{V}} {Gr}(N,r) := \left\{ \boldsymbol{\Delta} \in \mathbb{R}^{N \times r} \;\middle|\; \boldsymbol{\Phi}^{\top}\boldsymbol{\Delta} = \boldsymbol{0} \right\} \subset \mathbb{R}^{N \times r}.
\end{equation}

The curve on the Grassmann manifold, achieving the minimum path between two points on ${Gr}(N,r)$ is called the geodesic. In fact, it satisfies the solution of the differential equation minimizing that path.
Out of this geodesic, three important concepts emerge, the Grassmannian distance, the Grassmannian logarithmic maps and the exponential maps. \medskip

We first consider two points on the ends of the geodesic $\mathcal{V}_0, \mathcal{V}_1 \in {Gr}(N,r)$ represented respectively by  $\boldsymbol{\Phi}_0, \boldsymbol{\Phi}_1 \in {St}(N,r)$. We then take  $\boldsymbol{\Delta}_0 \in T_{\mathcal{V}_0} {Gr}(N,r)$ as the initial velocity at $\mathcal{V}_0$.
Let  $\boldsymbol{U} \boldsymbol{\Sigma} \boldsymbol{V}^T = \operatorname{SVD}( {\boldsymbol{\Phi}_1}^T \boldsymbol{\Phi}_0 )$,
$\boldsymbol{P} \boldsymbol{S} \boldsymbol{R}^T = \operatorname{SVD}( \boldsymbol{\Delta}_0 )$, 
$\boldsymbol{L} = (\boldsymbol{I}_r - {\boldsymbol{\Phi}_0}^T \boldsymbol{\Phi}_0) \boldsymbol{\Phi}_1 \boldsymbol{U} \boldsymbol{V}^T$
and $\widehat{\boldsymbol{U}} \widehat{\boldsymbol{\Sigma}} \widehat{\boldsymbol{V}}^T = \operatorname{SVD}( \boldsymbol{L} )$
\begin{itemize}
    \item The principal angles between $\mathcal{V}_0$ and  $\mathcal{V}_1$ are given by \citep{BSMF_1875__3__103_2, GOLUB19943}:
    \begin{equation}\label{subsp_angle_expression}
        \gamma_i = \arccos(\sigma_i), \qquad \forall\, i \in \{1, \dots, r\}
    \end{equation}
    where $\sigma_i$ are the singular values in $\boldsymbol{\Sigma}$.
    \item The distance between the two subspaces on the geodesic can be computed as \citep{EdelmanA}: 
    \begin{equation}
        d(\mathcal{V}_0, \mathcal{V}_1) = \sqrt{\sum_i^{r} \gamma_i^2}.
    \end{equation}
    \item Knowing the initial point $\mathcal{V}_0$ and the initial velocity $\boldsymbol{\Delta}_0$ for the geodesic, the solution to the geodesic equation allows to find the final point $\mathcal{V}_1$ and thus the exponential map as \citep{EdelmanA}:
    \begin{equation}\label{exp_map}
        \operatorname{Exp}_{\mathcal{V}_0} : T_{\mathcal{V}_0}\mathrm{Gr}(N,r) \to \mathrm{Gr}(N,r), 
        \quad \boldsymbol{\Delta}_0 \mapsto \mathcal{V}_1 =  \operatorname{span} \left(  \boldsymbol{\Phi}_0 \boldsymbol{R} \cos \left(\boldsymbol{S} \right) \boldsymbol{R}^T + \boldsymbol{P} \sin \left(\boldsymbol{S} \right)\boldsymbol{R}^T  \right).
    \end{equation}
    where $\cos(\cdot)$ and $\sin(\cdot)$ are applied element-wise on the diagonal entries of $\boldsymbol{S}$.
\item Alternatively, the geodesic equation can also be solved as a boundary value problem. Given the two end points of the geodesic $\mathcal{V}_0$ and $\mathcal{V}_1$, that solution allows to find the initial velocity $\boldsymbol{\Delta}_0$, which defines the logarithmic map \citep{zimmermann2022manifoldinterpolationmodelreduction}:
    \begin{equation}
        \operatorname{Log}_{\mathcal{V}_0} :  \mathrm{Gr}(N,r) \to T_{\mathcal{V}_0}\mathrm{Gr}(N,r), 
        \quad \mathcal{V}_1  \mapsto \boldsymbol{\Delta}_0 =  \widehat{\boldsymbol{U}} \arcsin\!\!\left( \widehat{\boldsymbol{\Sigma}} \right) \widehat{\boldsymbol{V}}^T.
    \end{equation}
    where $\arcsin(\cdot)$ is again applied element-wise.
\end{itemize}

Data interpolation on manifolds by "flattening" the data into tangent spaces can be traced back to~\citet{JuppKent}. It was extended to Grassmann manifolds in the works of~\citet{RahmanGrassmann} and \citet{Hüper2007}, and then introduced to the model order reduction community by~\citet{amsallem2008interpolation}. The subspace interpolation approach performs standard interpolation algorithms on the tangent vectors (velocities) associated with different input parameters. The resulting interpolated velocity is then used to construct a new geodesic, whose end point corresponds to the interpolated subspace. This method approximates the subspace itself while accounting for parameter variations, rather than simply applying element-wise interpolation on the basis vectors. Specifically, a reference point $\mathcal{V}_{ref} = [ \boldsymbol{\Phi}_{ref} ]$ is chosen among the training subspaces, then the different tangent vectors are computed using the logarithmic map. An interpolation operator is then constructed on the Grassmann manifold tangent space:
\begin{equation}\label{interp_tangent_grass}
    \varphi : \mathbb{R}^{N_p} \to T_{\mathcal{V}_{ref}}\mathrm{Gr}(N,r), 
    \quad \boldsymbol{\theta} \mapsto \boldsymbol{\Delta}.
\end{equation}
Then the predicted subspace (as well as its representative basis) for an unseen parameter $\boldsymbol{\theta}$ are computed as:
\begin{equation}\label{grassmann_interp_predict}
    \mathcal{V}_{\boldsymbol{\theta}} = \operatorname{Exp}_{\mathcal{V}_{ref}} \left( \varphi \left( \boldsymbol{\theta} \right) \right) = [ \boldsymbol{\Phi}_{\boldsymbol{\theta}} ].
\end{equation}
We note here that in this work, $\varphi \left(\cdot\right)$ can be evaluated, within or beyond the convex hull of the training samples. Nonetheless, we continue to use the term interpolation, in line with the existing literature \citep{amsallem2008interpolation, zimmermann2022manifoldinterpolationmodelreduction}.
Thereafter, we use the following definitions instead of (\ref{encoder_temp}) and (\ref{decoder_temp}):
\begin{equation}\label{encoder_temp_2}
    \mathcal{E}_{\boldsymbol{\theta}}(\boldsymbol{x}) =  {\boldsymbol{\Phi}_{\boldsymbol{\theta}}}^T (\boldsymbol{x} - \Bar{\boldsymbol{x}})
\end{equation}
\begin{equation}\label{decoder_temp_2}
    \mathcal{D}_{\boldsymbol{\theta}}(\boldsymbol{x}^r) =  {\boldsymbol{\Phi}_{\boldsymbol{\theta}}} \boldsymbol{x}^r + \Bar{\boldsymbol{x}}.
\end{equation}

\subsection{Regression in the latent space}

\subsubsection{Regression methods}\label{regr_methods_part}

The regressor approximates the relationship between the low-dimensional representations 
of the previous state-space and the current control input to the current state-space
\begin{equation}\label{operator_definition_temp}
    \mathcal{I}: \mathbb{R}^{r} \times \mathbb{R}^{N_u} \rightarrow \mathbb{R}^{r}\;;\; 
    \left( {\boldsymbol{x}}^{n-1, r},  {\boldsymbol{u}}^{n, r} \right) \rightarrow  
    {\boldsymbol{x}}^{n, r}.
\end{equation}
Different existing methods can accomplish this task. In our experiments, the regression 
methods that provided the best accuracy are RBF regression 
\citep{wahba_spline_1990, audouzeRBF} and $q$-th order polynomial regression. For the 
former, letting $\boldsymbol{z}^n = \left({\boldsymbol{u}}^{n,r} ,\, {\boldsymbol{x}}^{n-1,r}\right) \in \mathbb{R}^{N_u+r}$ denote the combined input 
vector, the function is modelled as
\begin{equation}\label{regr_rbf_temp}
    \boldsymbol{x}^{n, r} = \mathcal{I}({\boldsymbol{z}}^{n}) = \sum_i^m \boldsymbol{\alpha_i}\, 
    \phi\!\left(\|{\boldsymbol{z}}^{n} - {\boldsymbol{z}}_{i}\|\right) + 
    \boldsymbol{P}({\boldsymbol{z}}^{n})
\end{equation}
where $\phi(\cdot)$ is a kernel function, $\boldsymbol{P}$ is a first-order polynomial, 
and ${\boldsymbol{z}}_{i}$ are the RBF centers, chosen as the training samples of the 
combined reduced inputs, resulting in a linear system to be solved for the RBF weights 
$\alpha_i$.

Alternatively, a polynomial regression of order $q$ is used. The relationship between 
the state-space and the control input is thus modelled as a $q$-th order polynomial:
\begin{equation}\label{regr_poly_temp}
    \boldsymbol{x}^{n, r} = \mathcal{I}({\boldsymbol{z}^n}) = \boldsymbol{W}\,\boldsymbol{\Gamma}_q({\boldsymbol{z}^n})
\end{equation}
where $\boldsymbol{\Gamma}_q(\boldsymbol{z}^n)$ denotes the vector of all monomials of 
$\boldsymbol{z}^n$ up to degree $q$. The polynomial coefficients are arranged in 
$\boldsymbol{W} \in \mathbb{R}^{r \times \hat{r}}$, where 
$\hat{r} = \binom{r + N_u + q}{q}$ is the total number of monomial terms.
The number of polynomial coefficients grows rapidly with the input dimension 
$r + N_u$ and the polynomial order $q$, making overfitting a concern when training 
data are limited. We therefore employ Ridge (L2) regularisation~\citep{hoerlRidge} to 
obtain a well-conditioned and generalisable model. The minimisation reads:
\begin{equation}\label{ridge}
    \boldsymbol{W}_i = \underset{W_i}{\arg\min} \left\|{\boldsymbol{x}}^{n,r}_i - 
    \sum_{j=1}^{\hat{r}} {W}_{ij}\,[\boldsymbol{\Gamma}_q({\boldsymbol{z}^n})]_j 
    \right\|_2^2 + \lambda \sum_{j=1}^{\hat{r}} {W}_{ij}^2 
    \qquad \forall\; i \in \{1, \ldots, r\}.
\end{equation}
The regularisation parameter $\lambda$ controls the trade-off between data fidelity 
and coefficient magnitude, and is selected by cross-validation. We note that NNs represent another viable option for the considered regression problem, given their successful use in scientific machine learning. In principle, NN-based regressors could be integrated within the proposed framework without difficulty. Nevertheless, the relatively small dataset sizes considered in this work, together with the need for efficient retraining discussed in section \ref{modelUpdateSection}, motivated the choice of computationally lighter regression methods. Detailed comparative study between different regression models is beyond the scope of the present work.

\subsubsection{Parameter dependence and latent space consistency}
Now, we should adapt this regression to incorporate parameter dependence. Furthermore, as presented in the previous section, multiple reduced bases are considered as the localized models, this means that, in order to offload the computational cost to the offline phase, we should use multiple regressions $\mathcal{I}_{\boldsymbol{\theta_k}}$ associated to each parameter $\boldsymbol{\theta_k}\,; \forall k \in \{1, \cdots p \}$. We propose to construct each regression operator as a linear combination of the output of the operators associated to all the observed parameters.
\begin{equation}\label{regr_express_param_temp}
    \boldsymbol{x}^{n, r}_{\boldsymbol{\theta}} = \mathcal{I}_{\boldsymbol{\theta}}({\boldsymbol{z}}^{n}) = \sum_k^p {w_k}\, \boldsymbol{Q}_k \mathcal{I}_{\boldsymbol{\theta_k}}({\boldsymbol{z}}^{n}).
\end{equation}
where $\boldsymbol{Q}_k$ are orthogonal matrices $\boldsymbol{Q}_k^T\boldsymbol{Q}_k = \boldsymbol{I}_r$, here called the Procrustes alignment matrices. In fact, as the output of the regression belongs to the latent space associated to the new parameter $\boldsymbol{\theta}_{*}$, and this latter space is unknown at the offline phase, the trained operators $\mathcal{I}_k$ do not inherently give output quantities that are consistent with the target latent space. Therefore, we use the Procrustes alignment matrices $\boldsymbol Q_k$ as orthogonal rotation operators that map the different latent embeddings onto a common coordinate frame.
The alignment matrices are computed as the solution of the Procrustes optimization problem
\begin{equation}\label{procrustes_problem}
    \min_{\boldsymbol{Q}_k^T\boldsymbol{Q}_k = \boldsymbol{I}_r} || \boldsymbol{\Phi}_{ \boldsymbol{\theta}} \boldsymbol{Q}_k - \boldsymbol{\Phi}_{\boldsymbol{\theta_k}} ||_F^2.
\end{equation}
The solution can be written as
\begin{equation}\label{procrustes_solution}
    \boldsymbol{Q}_{k} = \boldsymbol{U}_k \boldsymbol{V}_k^T
\end{equation}
with
\begin{equation}
    \boldsymbol{U}_k \boldsymbol{\Sigma}_k \boldsymbol{V}_k^T = \textnormal{SVD}(\boldsymbol{\Phi}_{ \boldsymbol{\theta}}^T \boldsymbol{\Phi}_{ \boldsymbol{\theta_k}}).
\end{equation}

The coefficients $w_k$ of linear combination (\ref{regr_express_param}) are computed as the weights that are inversely proportional to the Grassmann distance from the current subspace $\{\mathcal{V}_{\theta}\}$ to the training subspaces $\mathcal{V}_{\theta_k}$
\begin{equation}\label{weights_locality}
    w_k =  \frac{d(\mathcal{V}_{\theta}, \mathcal{V}_{\theta_k}) ^{-z}}{\sum_i^{p} d(\mathcal{V}_{\theta}, \mathcal{V}_{\theta_i})^{-z}}
\end{equation}
with $z \in \mathbb{N}^{+}$ a hyperparameter controlling the locality of the interpolation.
In other words, the contribution of each regression associated to $\theta_k$ will be correlated to the geometric proximity ( in a Grassmann sense ) of the training subspace $\mathcal{V}_{\theta_k}$  to the target subspace $\mathcal{V}_{\theta}$.
In the aggregate prediction (\ref{regr_express_param_temp}), the manifold structure governs both the weighting and the alignment of the regression outputs, ensuring geometric consistency across the heterogeneous reduced coordinate frames.
Our choice for expressing the weights $w_k$ is inspired by the approach proposed in~\citep{OULGHELOU2021109924}. In that work, however, the same weighting scheme was used as part of a global optimization problem over the all the considered subspaces $\mathcal{V}_k$. For simplicity, $\mathcal{I}_{\boldsymbol{\theta}_k}(\cdot)$ and $d(\mathcal{V}_{\theta}, \mathcal{V}_{\theta_k})$ will hereafter be denoted by $\mathcal{I}_{k}(\cdot)$ and $d_k$ respectively.

\begin{remark}\label{remarkParamNumber}
The adopted approach uses contributions associated to all training parameter samples, which can cause large overhead if $p$ is large. In that case, a nearest-neighbor or clustering technique can be used to prune the contributions of distant subspaces. This was not pursued in the current work as $p$ is moderate in the tested cases.
\end{remark}

\subsection{Model online adaptation:}\label{modelUpdateSection}

In this work, we propose to update the ROM when new observations of the state-space are available: $\{\boldsymbol{x}^{n-1},\boldsymbol{u}^{n}, \boldsymbol{x}^{n}\}$. As the chosen ROM here is composed of the regression and the encoder-decoder operators separately, their update will be treated separately as well.

\subsubsection{Regression retraining}\label{regressionRetrainingSect}
Similar to what was done in~\citep{TIBA2025109522}, the update of the regression component is performed through the retraining of $\mathcal{I}_k$ when new data become available. This retraining is carried out at a prescribed update frequency $\tau$, treated as a hyperparameter of the method. The procedure remains computationally efficient since the regression is learned in the reduced spaces $\mathbb{R}^{r} \times \mathbb{R}^{N_u}$ and $\mathbb{R}^{r}$, whose dimensions are significantly smaller than those of the original system. Moreover, the considered regression techniques discussed in (\ref{regr_methods_part}) only require the solution of moderately sized least-squares problems or linear systems within these reduced spaces, resulting in a low computational overhead for each update.

The regression operator will now depend on the time step $n$, denoted as $\mathcal{I}^n(\cdot)$. Let $\tau_n$ denote the (variable) number of observations collected at time step $n \in \{1, \ldots, N_t\}$, and let $\bar{\kappa}_n = \sum_{k=1}^{n} \kappa_k$ be the cumulative observation count. The operator $\mathcal{I}_n$ is piece-wise constant and is updated whenever $\bar{\kappa}_n$ crosses a multiple of $\tau$. That is, $\mathcal{I}_n$ remains constant over intervals where the index $\hat{\kappa} = \lfloor \bar{\kappa}_n / \tau \rfloor$ is constant, and is re-trained when $\hat{\kappa}$ increments. We thus replace the operator definition in (\ref{operator_definition_temp}) by
\begin{equation}
    \mathcal{I}^n: \mathbb{R}^{r} \times \mathbb{R}^{N_u} \rightarrow \mathbb{R}^{r}\;;\; 
    \left( {\boldsymbol{x}}^{n-1, r},  {\boldsymbol{u}}^{n, r} \right) \rightarrow 
    {\boldsymbol{x}}^{n, r}.
\end{equation}
Similarly, we replace (\ref{regr_rbf_temp}), (\ref{regr_poly_temp}) and (\ref{regr_express_param_temp}) by 
\begin{equation}
    \boldsymbol{x}^{n, r} = \mathcal{I}^n({\boldsymbol{z}}^{n}) = \sum_i^m \boldsymbol{\alpha_i}^n\, 
    \phi\!\left(\|{\boldsymbol{z}}^{n} - {\boldsymbol{z}}_{i}\|\right) + 
    \boldsymbol{P}^n({\boldsymbol{z}}^{n}),
\end{equation}
\begin{equation}\label{regr_poly}
    \boldsymbol{x}^{n, r} = \mathcal{I}^n({\boldsymbol{z}^n}) = \boldsymbol{W}^n\,\boldsymbol{\Gamma}_q({\boldsymbol{z}^n})
\end{equation}
and
\begin{equation}\label{regr_express_param}
    \boldsymbol{x}^{n, r}_{\boldsymbol{\theta}} = \mathcal{I}^n_{\boldsymbol{\theta}}({\boldsymbol{z}}^{n}) = \sum_k^p {w_k}\, \boldsymbol{Q}_k \mathcal{I}^n_{\boldsymbol{\theta_k}}({\boldsymbol{z}}^{n}).
\end{equation}
respectively.

\subsubsection{Subspace update}\label{subspace_update_subsection}

To obtain a time-varying basis, we use a basis update approach that keeps the basis rank fixed and only rotates the basis in space, instead of adapting the dimension itself. This allows the use of pre-trained regressions on latent spaces of dimensions known at the offline phase. Specifically, at each time step $n$, with a new snapshot $\boldsymbol{x}^n$, the ROM basis $\boldsymbol{\Phi}^{n-1}$ is updated via a rank-one update to obtain $\Bar{\boldsymbol{\Phi}}^{n}$. This new intermediate basis will not be actually used in the next time step, \textit{i.e} by taking $\boldsymbol{\Phi}^{n} = \Bar{\boldsymbol{\Phi}}^{n}$ but will be kept in memory and only "\textit{activated}" after a user-defined frequency of update $K$. This is because each time $\boldsymbol{\Phi}^n$ is updated, the output of regression $\mathcal{I}(\cdot)$ should be transformed towards the new coordinate system, and we want these operations to be done at an optimised rate. 
The intermediary basis $\Bar{\boldsymbol{\Phi}}^{n}$ allows the subspace update to proceed snapshot-by-snapshot, circumventing the storage of a high-dimensional batch of snapshots in $\mathbb{R}^{N_x}$.

Several directions have been explored in the literature to update the basis in adaptive ROMs.
The works in~\citep{amsallem2015fast, BAI2022100, Tang2025} for example used the incremental SVD approach of~\citet{BRAND200620} while~\citep{onlineGalerking, onlineGalerkinTransport, HUANG2023112356} employed an additive low-rank update that minimizes the residual of the DEIM approximation on an augmented set ensemble of interpolation points.
The projection approximation subspace tracking (PAST) algorithm~\citep{pastAlg}, which accommodates non-stationary data through a forgetting factor, has been adopted in~\citep{taleb}.
More closely related to the present work,~\citep{zimmerCox} performs subspace updates directly on the Grassmann manifold, applying the Grassmannian Rank-One Update Subspace Estimation (GROUSE) method~\citep{balzano2010online} to update the DEIM subspace and its interpolation points. We adopt a similar geometric approach and employ GROUSE as our basis update strategy, which we briefly recall in what follows.

Based on a new observation $\boldsymbol{x}^n$, GROUSE aims to find an updated subspace that minimizes the projection error on the Grassmann manifold:
\begin{equation}
    \mathcal{J}(\mathcal{V}, n) = ||\boldsymbol{\Phi} \boldsymbol{x}^{n, r} - \boldsymbol{x}^n||^2.
\end{equation}
with $\boldsymbol{\Phi}$ the basis representative of $\mathcal{V}$. Denoting the projected value $\boldsymbol{x}^{n}_{||\mathcal{V}} = \boldsymbol{\Phi} \boldsymbol{x}^{n, r}$ and the residual $\boldsymbol{x}^n_{\perp \mathcal{V}} = \boldsymbol{x}^{n} - \boldsymbol{x}^{n}_{||\mathcal{V}}$, the gradient descent step on the Grassmann manifold (see~\citep{EdelmanA}) for $\mathcal{J}(\mathcal{V}, n)$ with a step size $\eta^n$ is:
\begin{equation}
    \boldsymbol{\Phi} \leftarrow \boldsymbol{\Phi} + \left( \sin\left( \eta^n  \sigma \right) \frac{\boldsymbol{x}^n_{\perp \mathcal{V}}}{||\boldsymbol{x}^n_{\perp \mathcal{V}}||} + \left( \cos \left( \eta^n \sigma \right) -1 \right) \frac{\boldsymbol{x}^{n}_{||\mathcal{V}}}{||\boldsymbol{x}^{n}_{||\mathcal{V}}||} \right) \frac{ \left( \boldsymbol{x}^{n, r} \right)^T}{||\boldsymbol{x}^{n, r}||}
\end{equation}
where $\sigma = ||\boldsymbol{x}^n_{\perp \mathcal{V}}||\, ||\boldsymbol{x}^n_{|| \mathcal{V}}||$. In~\citep{balzano2015local}, it has been demonstrated that an optimal step length satisfies 
\begin{equation}
    \sin{\eta^n \sigma} = {\frac{||\boldsymbol{x}^n_{\perp \mathcal{V}}||}{||\boldsymbol{x}^n||}}.
\end{equation}
We note that looking at the "global" cost function
\begin{equation}
    \overline{\mathcal{J}}(\mathcal{V}) = \sum_{n=1}^{N_t} \min_{\boldsymbol{x}^{n, r}}||\boldsymbol{\Phi} \boldsymbol{x}^r - \boldsymbol{x}^n||^2
\end{equation}
shows that the GROUSE update is precisely a stochastic gradient step on $\Bar{\mathcal{J}}(\mathcal{V})$ over the Grassmann manifold.

It is worth noting that GROUSE was originally proposed in the context of highly incomplete observations, where each snapshot consists of only a small random subset of the $N_x$ components~\citep{balzano2010online}, and its convergence theory is developed primarily for that setting. In the present framework, however, full-state snapshots are available at every step, which means the geodesic update is computed from the complete residual $(\boldsymbol{I}_N - \boldsymbol{\Phi} \boldsymbol{\Phi}^T) \boldsymbol{x}^n$ rather than a subsampled approximation thereof. This represents a favorable simplification as the update direction is exact, and the method retains all of its geometric properties.

We chose GROUSE as the subspace update algorithm in this work for different reasons: First, the initialization of the subspace tracking procedure is handled naturally by the present framework. In fact, the initial basis for each new parameter $\boldsymbol{\theta}$ is obtained by interpolation on the Grassmann manifold, (see Section \ref{grassmannInterpolationSection}) a procedure that places the starting point at a geometrically meaningful location on $Gr(N, r)$ before any simulation snapshot is observed. Since GROUSE is itself a gradient descent on this same manifold, the two operations are homogeneous: the interpolation delivers a well-positioned point on $Gr(N, r)$, and GROUSE then traces a path from that point guided by the incoming dynamics. The practical consequence is that the initial residuals  $(\boldsymbol{I}_N - \boldsymbol{\Phi} \boldsymbol{\Phi}^T) \boldsymbol{x}^n$ are small, and the tracking procedure benefits from a warm start.

Second, in this work, a new snapshot $\boldsymbol{x}^n \in \mathbb{R}^N$ is used to update the working basis $\boldsymbol{\Phi}^n$ and immediately discarded at each time step, so that no high-dimensional data is retained beyond the basis itself. The rank $r$ is held fixed throughout the simulation, as the reduced coordinates need to belong to the same space, on which the regression operator is trained, see section \ref{regressionRetrainingSect}. These two constraints, fixed rank and single-pass snapshot processing, rule out exact incremental SVD methods such as that of~\citet{BRAND200620}, which achieve an exact subspace membership guarantee $\boldsymbol{\Phi} \boldsymbol{\Phi}^T \boldsymbol{x}^n = \boldsymbol{x}^n$ only when the rank is allowed to grow; once a truncation back to $r$ is enforced at every step, this guarantee is forfeited and the method reduces to an 
$\mathcal{O}(Nr + r^3)$ approximation that compounds truncation error across updates. The PAST algorithm~\citep{pastAlg} avoids the SVD cost but does not maintain orthonormality of the basis, requiring periodic re-orthogonalization whose cost and timing introduce additional design choices. GROUSE by contrast, operates natively at fixed rank by performing a geodesic step on $Gr(N, r)$
and orthonormality is preserved exactly by construction, while the per-step cost remains 
$\mathcal{O}(Nr)$ with no auxiliary factorization. Moreover, the $K$ sequential geodesic steps performed between two activations of the basis act as $K$ successive refinements of the subspace toward the observed dynamics, so that the promoted basis reflects an integrated view of the last $K$ snapshots, and avoids a single aggressive rotation.

Now, similar to what we wrote in section \ref{regressionRetrainingSect}, we modify the definitions (\ref{encoder_temp_2}) and (\ref{decoder_temp_2}) to include the time-dependence of the reduced bases. The operators $\mathcal{E}_{\boldsymbol{\theta}}^n$ and $\mathcal{D}_{\boldsymbol{\theta}}^n$ are now piece-wise constant and are updated whenever $\bar{\tau}_n$ crosses a multiple of $K$. We write
\begin{equation}\label{encoder}
    \mathcal{E}_{\boldsymbol{\theta}}^n(\boldsymbol{x}) =  {\boldsymbol{\Phi}^n_{x, \boldsymbol{\theta}}}^{T} (\boldsymbol{x} - \Bar{\boldsymbol{x}})
\end{equation}
\begin{equation}\label{decoder}
    \mathcal{D}_{\boldsymbol{\theta}}^n(\boldsymbol{x}^r) =  {\boldsymbol{\Phi}^n_{x, \boldsymbol{\theta}}} \boldsymbol{x}^r + \Bar{\boldsymbol{x}}.
\end{equation}

\section{ROM construction}\label{section2}
\SetKw{KwBy}{by}

\subsection{Offline phase}
The inputs to the offline phase (Algorithm \ref{alg:offline}) are the snapshot data collected from $p$ high-fidelity simulations, each associated with a training parameter $\boldsymbol{\theta}_k$. For each parameter $\boldsymbol{\theta}_k$, we have two snapshot matrices $\boldsymbol{X}_k^{\text{pre}}, \boldsymbol{X}_k^{\text{post}} \in \mathbb{R}^{N_x \times m_k}$, where $m_k$ denotes the number of snapshots collected for the $k$-th simulation. Each pair of columns in the two matrices represent the input (previous) and the output (current) of the discrete dynamical system. In general, the number of snapshots $m_k$ may differ across simulations. We also have control input snapshots $\boldsymbol{U}_{\text{in}, k} \in \mathbb{R}^{N_u \times m_k}$ associated with the time instants of the snapshots $\boldsymbol{X}_k^{\text{post}}. $\footnote{In the current context of partitioned coupling, the matrices  $\boldsymbol{X}_k^{pre}$ contain only solver outputs at the coupling convergence, and $\boldsymbol{X}_k^{post}$ and $\boldsymbol{U}_{\text{in}, k}$ contain every solver output, including the intermediate coupling iteration solutions.}

First, we normalize the snapshots based on a global mean vector $\bar{\boldsymbol{X}}$ and a global maximum vector $\boldsymbol{x}_{\max}$ from all available snapshots. For notational simplicity, we continue to use $\boldsymbol{X}_k^{\text{post}}$ to denote the normalized snapshots in the remainder of this section and in the algorithm.
For each parameter $\boldsymbol{\theta}_k$, the POD is applied to the snapshot matrix $\boldsymbol{X}_k^{\text{post}}$ (any combinations of the two matrices $\boldsymbol{X}_k^{\text{post}}, \boldsymbol{X}_k^{\text{pre}}$ can be used, depending on which is "\textit{richer}" of information), yielding an orthonormal basis $\boldsymbol{\Phi}_k \in \mathbb{R}^{N_x \times r}$ representative of the local subspace $\mathcal{V}_k \in Gr(N_x, r)$. The rank $r$ is chosen as the maximum of the ranks determined by the energy criterion in \eqref{SVDdecay} for each parameter.

Once all local bases are computed, the Grassmann interpolation framework (see Section~\ref{grassmannInterpolationSection}) is set up. A reference subspace $\mathcal{V}_{ref}$ is chosen among the training subspaces. 
The logarithmic map $\operatorname{Log}_{\mathcal{V}_{ref}}(\mathcal{V}_k) = \boldsymbol{\Delta}_k$ is computed, mapping each training subspace to a tangent vector in $T_{\mathcal{V}_{ref}} Gr(N_x, r)$. These tangent vectors live in a vector space, and standard fitting techniques can be applied to them. The RBF regression (\ref{interp_tangent_grass}) is then trained on these pairs, so that the subspace associated to any new parameter $\boldsymbol{\theta}_*$ can be recovered at prediction time via the exponential map.

Finally, for each parameter $\boldsymbol{\theta}_k$, the regression operator $\mathcal{I}_k(\cdot)$ is trained on input-output pairs formed by projecting the snapshots onto the local reduced basis $\boldsymbol{\Phi}_k$. The input data in the latent space is $\boldsymbol{X}^r_k = \left[\boldsymbol{U}_{\text{in},k}^T,\, (\boldsymbol{X}_k^{\text{pre}})^T \boldsymbol{\Phi}_k \right]^T$ and the output data is $\boldsymbol{Y}^r_k = \boldsymbol{\Phi}_k^T \boldsymbol{X}_k^{\text{post}}$.

The outputs of the offline phase: the set of local bases $\{\boldsymbol{\Phi}_k\}$, the set of trained regression operators $\{\mathcal{I}_k\}$, and the Grassmann regressor $\varphi(\cdot)$ constitute the complete dictionary that the next phases draw upon. A summary is illustrated in the upper left of Fig. \ref{fig:illustrativeSummary}.

\definecolor{mynicegreen}{RGB}{48,168,48}
\begin{algorithm}
\SetKwInput{KwData}{Input}
\caption{Offline phase of $\widehat{F}^n_{\boldsymbol{\theta}}$}\label{alg:offline}
\KwData{
\begin{itemize}[noitemsep, topsep=0pt, parsep=0pt, partopsep=0pt]
    \item Parameters $\{\boldsymbol{\theta}_1, \boldsymbol{\theta}_2, \cdots \boldsymbol{\theta}_p\}$
    \item State space snapshots pairs $\left\{ \{ \boldsymbol{X}_1^{\text{pre}}, \boldsymbol{X}_1^{\text{post}} \}, \{ \boldsymbol{X}_2^{\text{pre}}, \boldsymbol{X}_2^{\text{post}} \},  \cdots  \{ \boldsymbol{X}_p^{\text{pre}}, \boldsymbol{X}_p^{\text{post}} \} \right\}$,
    \item Control input snapshots $\left\{ \boldsymbol{U}_{\text{in}, 1}, \boldsymbol{U}_{\text{in}, 2},  \cdots  \boldsymbol{U}_{\text{in}, p} \right\}$,
    \item State-space energy criterion $\epsilon$,
    \item Reference point index $ref$.
\end{itemize}}

\KwResult{
\begin{itemize}[noitemsep, topsep=0pt, parsep=0pt, partopsep=0pt]
\item $\left\{ \boldsymbol{\Phi}_{1},  \boldsymbol{\Phi}_{2}, \cdots  \boldsymbol{\Phi}_{p} \right\}$

\item $\left\{ \mathcal{I}_{1},  \mathcal{I}_{2}, \cdots  \mathcal{I}_{p} \right\}$

\item $\varphi(\cdot)$
\end{itemize}}
\hrulefill

\nl Compute the mean vector of all the snapshots
$ \Bar{\boldsymbol{X}} = \frac{1}{p} \Sigma_k^p \Sigma_i^{m_k} \boldsymbol{x}_k^{\text{post}, i} $

\nl Compute the maximum vector of all the snapshots
$\boldsymbol{x}_{\max} = \begin{pmatrix}
\max_{k=1,2,\ldots,p} \left( \max_{i=1,2,\ldots,m_k} |x_{1,k}^{\text{post}, i}| \right)  \\
\max_{k=1,2,\ldots,p} \left( \max_{i=1,2,\ldots,m_k} |x_{2,k}^{\text{post}, i}| \right)  \\
\vdots \\
\max_{k=1,2,\ldots,p} \left( \max_{i=1,2,\ldots,m_k} |x_{N_x,k}^{\text{post}, i}| \right)  \\
\end{pmatrix}$

\For{$k\gets1$ \KwTo $p$}{
\nl Use  $\Bar{\boldsymbol{X}}$ and $\boldsymbol{x}_{\max}$ to normalize $\boldsymbol{X}_k^{\text{post}}$.
 
\nl Compute the POD modes $\boldsymbol{\Phi}_{k}$, corresponding to the subspace $\mathcal{V}_k$:
\begin{itemize}
\item Truncated SVD : $\boldsymbol{X}_k^{\text{post}} = \boldsymbol{\Phi}_k \boldsymbol{\Lambda} \boldsymbol{\Psi}^{T}$, $r_k$ is determined based on $\epsilon$. See (\ref{SVDdecay}).
\end{itemize}}

\nl Truncate all bases to $r = \text{max}_{k=1,2,\ldots,p}(r_k)$

\For{$k\gets1$ \KwTo $p$}{
\uIf{$k\, != \, ref$}{
\nl Compute the logarithmic map associated with each subspace 
: \, $\operatorname{Log}_{\mathcal{V}_{ref}}(\mathcal{V}_k)$

\begin{itemize}[noitemsep, topsep=0pt, parsep=0pt, partopsep=0pt]
    \item SVD : $\boldsymbol{\Phi}_{k} ^T \boldsymbol{\Phi}_{\text{ref}} = \boldsymbol{U}_k \boldsymbol{\Sigma}_k \boldsymbol{V}_k^T $
    \item $\boldsymbol{L}_k = \left(\boldsymbol{I}_r - \boldsymbol{\Phi}_{\text{ref}} \boldsymbol{\Phi}_{\text{ref}}^T \right) ( \boldsymbol{\Phi}_{k} \boldsymbol{U}_k  \boldsymbol{V}_k^T ) $
    \item Thin SVD : $\boldsymbol{L}_k = \widehat{\boldsymbol{U}}_k \widehat{\boldsymbol{\Sigma}}_k \widehat{\boldsymbol{V}}_k^T $
    \item $\boldsymbol{\Delta}_{k} = \operatorname{Log}_{\mathcal{V}_{ref}}(\mathcal{V}_k) =\widehat{\boldsymbol{U}}_k \arcsin \left( \widehat{\boldsymbol{\Sigma}}_k \right) \widehat{\boldsymbol{V}}_k^T$
\end{itemize}
}
\nl Form regression input data in the reduced space :  \,  $\boldsymbol{X}^r_k = \left[\boldsymbol{U}_{\text{in}}^T,\,\, (\boldsymbol{X}_k^{\text{pre}}) ^{T} \boldsymbol{\Phi}_{k}\right]^T$ \label{input_reduced_data_line}

\nl Form regression output data in the reduced space : \,  $ \boldsymbol{Y}^r_k = \boldsymbol{\Phi}_{k} ^T \boldsymbol{X}_k^{\text{post}}$ \label{output_reduced_data_line}

\nl Train the offline (initial) regression operator \, $\mathcal{I}_{k}(\cdot) :  \boldsymbol{X}^r_k \rightarrow \boldsymbol{Y}^r_k$
}

\nl Train an RBF regression from parameters to tangent vectors
$\varphi(\cdot) : \boldsymbol{\theta}_k \rightarrow \boldsymbol{\Delta}_{k}$
\end{algorithm}

\subsection{Pre-prediction phase}

The pre-prediction phase (Algorithm~\ref{alg:pre-pred}) groups all computations that depend solely on the new parameter $\boldsymbol{\theta}_*$ and are therefore independent of the time evolution of the simulation. For transient simulations, these quantities are computed and stored once, at the beginning of the simulation, as soon as $\boldsymbol{\theta}_*$ is known, and then reused at every time step without further cost.

The first task of this phase is to determine the initial reduced basis $\boldsymbol{\Phi}_{*}^0$ associated with the unseen parameter $\boldsymbol{\theta}_*$. This is achieved through the evaluation of the Grassmann interpolation (or extrapolation) (\ref{grassmann_interp_predict}). This basis then serves as the initial basis for the online subspace tracking procedure.

Once $\boldsymbol{\Phi}_{*}^0$ is available, the Grassmann distances $d_k$ from each training subspace $\mathcal{V}_k$ to the new subspace $\mathcal{V}_*^0$ are computed. These distances are then used to form the geometric weights following (\ref{weights_locality}). 

The last quantity computed in this phase is the set of Procrustes alignment matrices $\{\boldsymbol{Q}_k\}_{k=1}^p$. As noted in Section~\ref{grassmannInterpolationSection}, the reduced coordinates produced by each locally trained regression $\mathcal{I}_k$ belong to the latent space associated with $\boldsymbol{\Phi}_k$, which in general differs from the coordinate frame of the new subspace $\boldsymbol{\Phi}_{*}^0$. The matrices $\boldsymbol{Q}_k$ are orthogonal rotations, solutions to the Procrustes alignment problem between $\boldsymbol{\Phi}_{*}^0$ and each $\boldsymbol{\Phi}_k$, that map all local latent embeddings onto the common coordinate frame of the new subspace. This alignment is necessary to make the outputs of the different regressors mutually consistent before combining them.

We note here a special case of the Procrustes alignment between the reference point of the Grassmann interpolation $\mathcal{V}_{\text{ref}}$ and the predicted point $\mathcal{V}_{*}$. We show in the following proposition that the Procrustes matrix $\boldsymbol{Q}_{\text{ref}}$ is the identity matrix.

\begin{proposition}\label{proposition1}
Let $\mathcal{V}_{\text{ref}}$ be the reference point of the Grassmann geodesic interpolation, represented by a basis $\boldsymbol{\Phi}_{\text{ref}}$, and $\mathcal{V}_{*}$ the subspace obtained from the exponential map (\ref{exp_map}) and represented by the basis $\boldsymbol{\Phi}_{*}$ computed using line \ref{computeNewSub}, Algorithm \ref{alg:pre-pred}. Then the Procrustes alignment matrix $\boldsymbol{Q}_{\text{ref}}$ satisfying (\ref{procrustes_problem}) is equal to the identity matrix $\boldsymbol{I}_r$.
\end{proposition}
\begin{proof}
Recall that we denoted the tangent vector at $\mathcal{V}_{ref}$ as $\boldsymbol \Delta$. We also recall that $\boldsymbol \Delta = \boldsymbol{P} \boldsymbol{S} \boldsymbol{R}^T$ using a thin SVD. By the property (\ref{tangent_space_grassmann}) of the tangent space of the Grassmann manifold,
\[
\boldsymbol{0} = \boldsymbol{\Phi}_{ref}^T \boldsymbol{\Delta} = \boldsymbol{\Phi}_{ref}^T \boldsymbol{P} \boldsymbol{S} \boldsymbol{R}^T = \boldsymbol{\Phi}_{ref}^T \boldsymbol{P}.
\]
As a consequence, 
\[
\boldsymbol{\Phi}_{*}^T \boldsymbol{\Phi}_{ref} = \boldsymbol{R} \sin(\boldsymbol{S})\boldsymbol{P}^T \boldsymbol{\Phi}_{ref} + \boldsymbol{R} \cos(\boldsymbol{S})\boldsymbol{R}^T = \boldsymbol{R} \cos(\boldsymbol{S})\boldsymbol{R}^T =  \cos(\boldsymbol{S})\boldsymbol{I}_r .
\]
Then, from (\ref{procrustes_solution}), we conclude that
\[
\boldsymbol{Q}_{ref} = \boldsymbol{I}_r .
\]
\qed
\end{proof}

This means that no mapping is needed to transform the the output of the regression associated to $\boldsymbol{\theta}_{\text{ref}}$. This effect will be visualised in the results section \ref{section4}.

The three outputs of the pre-prediction phase: the initial basis $\boldsymbol{\Phi}_{*}^0$, the weights $\{w_k\}$, and the alignment matrices $\{\boldsymbol{Q}_k\}$ are fully determined before the first simulation time step is taken, and are passed directly to the online phase. The pre-prediction phase is illustrated in the upper right of Fig.~\ref{fig:illustrativeSummary}.

\begin{algorithm}
\SetKwInput{KwData}{Input}
\caption{Pre-Prediction step of  $\widehat{F}^n_{\boldsymbol{\theta}}$}\label{alg:pre-pred}
\KwData{
\begin{itemize}[noitemsep, topsep=0pt, parsep=0pt, partopsep=0pt]
    \item Local bases dictionary $\left\{ \boldsymbol{\Phi}_{1},  \boldsymbol{\Phi}_{2}, \cdots  \boldsymbol{\Phi}_{p} \right\}$
    
    \item Regressor in the tangent space of the Grassmann manifold $\varphi(\cdot)$,
    
    \item Weight power hyperparameter $z$.
\end{itemize}
    }

\KwResult{
\begin{itemize}[noitemsep, topsep=0pt, parsep=0pt, partopsep=0pt]

\item New subspace $\boldsymbol{\Phi}_{*}^0$

\item Weights $\left\{ w_{1},  w_{2}, \cdots  w_{p} \right\}$

\item Procrustes alignment matrices $\left\{ \boldsymbol{Q}_{1}, \boldsymbol{Q}_{2}, \cdots  \boldsymbol{Q}_{p} \right\}$
\end{itemize}
}

\hrulefill

\nl \label{computeNewSub} Find the new subspace for the unseen parameter $\boldsymbol{\Phi}_{*}^0$ (associated to $\mathcal{V}_*^0)$
\begin{itemize}[noitemsep, topsep=0pt, parsep=0pt, partopsep=0pt]
    \item  $\boldsymbol{\Delta}_{*} = \operatorname{Log}_{\mathcal{V}_{ref}}(\mathcal{V}_*) = \varphi(\boldsymbol{\theta}_*)$
    \item Thin SVD : $\boldsymbol{\Delta}_* = \boldsymbol{P} \boldsymbol{S} \boldsymbol{R}^T$
    \item   $\boldsymbol{\Phi}_{*}^0 = \operatorname{Exp}_{\mathcal{V}_{ref}} \left( \boldsymbol{\Delta}_{*} \right) = \boldsymbol{\Phi}_{{ref}} \boldsymbol{R} \cos(\boldsymbol{S}) \boldsymbol{R}^T + \boldsymbol{P} \sin(\boldsymbol{S})\boldsymbol{R}^T$
\end{itemize}

\For{$k\gets1$ \KwTo $p$}{

\nl  Compute the subspace angles $\boldsymbol{\gamma}_k$ between $\mathcal{V}_k$ and the new subspace $\mathcal{V}_*^0$:
\begin{itemize}[noitemsep, topsep=0pt, parsep=0pt, partopsep=0pt]
        \item SVD : $ \boldsymbol{\Phi}_{*}^{0\,T} \boldsymbol{\Phi}_{k}  = \boldsymbol{U}_k \boldsymbol{\Sigma}_k \boldsymbol{V}_k^T$ 
        \item $\boldsymbol{\gamma}_k = \arccos{\boldsymbol{\Sigma_k}}$
\end{itemize}\label{lin:computeAngles}

\nl Compute the Grassmann distances $d_k$ from  $\mathcal{V}_k$ to $\mathcal{V}_*^0$
    \begin{itemize}[noitemsep, topsep=0pt, parsep=0pt, partopsep=0pt]
        \item $d_k = \sqrt{\sum_i^{r} \gamma_{k, i}^2}$
    \end{itemize}

\nl Compute the Procrustes alignment matrices $\boldsymbol{Q}_{k}$ \label{computeQ} 
\begin{itemize}[noitemsep, topsep=0pt, parsep=0pt, partopsep=0pt]
        \item $\boldsymbol{Q}_{k} = \boldsymbol{U}_k \boldsymbol{V}_k^T$
\end{itemize}
}

\For{$k\gets1$ \KwTo $p$}{
\nl Find the weights $w_k$ associated to each training basis:
\begin{itemize}[noitemsep, topsep=0pt, parsep=0pt, partopsep=0pt]
    \item $w_k =  d_k ^{-z} / \sum_i^{p} d_i^{-z}$
\end{itemize}\label{computeW}

}
\end{algorithm}

\subsection{Online phase}
The online phase is the phase that runs concurrently with the transient simulation. At each time step $n$, the ROM receives a new full-order snapshot $\boldsymbol{x}^n$ from the solver, and uses it to both make a prediction and progressively refine its internal representation of the dynamics. Three interleaved processes take place: prediction (Algorithm~\ref{alg:rom-pred}), regression retraining (Algorithm~\ref{alg:regress-update}), and subspace activation (Algorithm~\ref{alg:basis-update}), governed by two user-defined frequencies $\tau$ and $K$.

\subsubsection{Prediction} At each time step, the ROM prediction proceeds as described in Algorithm~\ref{alg:rom-pred}. For each training parameter $\boldsymbol{\theta}_k$, the previous state $\boldsymbol{x}^{n-1}$ is encoded into the latent space of $\boldsymbol{\Phi}_k$, the locally trained regression $\mathcal{I}_k$ is evaluated, and the result is rotated into the coordinate frame of the current subspace $\boldsymbol{\Phi}_*^n$ via the Procrustes alignment matrix $\boldsymbol{Q}_k$. The Grassmann-weighted combination of all these outputs then yields the aggregate prediction (\ref{regr_express_param_temp}). In addition to this offline-based combination, a current regression $\mathcal{I}_*^n$, trained on the online HF data collected so far, is also evaluated. The final reduced prediction is a convex combination of the two:
\begin{equation}
    \overline{\boldsymbol{x}^{n,r}} = \xi\, \mathcal{I}_*^n\!\left(\mathcal{E}_{\boldsymbol{\theta}_*}^n(\boldsymbol{x}^{n-1}),\, \boldsymbol{u}^r\right) + (1-\xi)\, \boldsymbol{x}^{n,r},
\end{equation}
where $\xi \in (0, 1)$ is a hyperparameter. The rationale for this blending is that, at the start of the simulation, the online data buffer is still small and $\mathcal{I}_*^n$ is poorly trained; in that regime, a small value of $\xi$ ensures that the prediction relies predominantly on the offline-trained combination. As the buffer grows and the online regression becomes more reliable, $\xi$ can either be kept fixed, or be allowed to depend on the time step $n$ through a schedule that increases monotonically with the cumulative number of online observations $\bar\kappa_n$ (defined in Section~\ref{regressionRetrainingSect}). We consider the following logistic ramp, expressed here in its equivalent hyperbolic-tangent form:
\begin{equation}\label{eq:xi_schedule}
    \xi^n = \tanh\!\left( \frac{\bar\kappa_n / m_0}{ \epsilon_\xi} \right),
\end{equation}
where $m_0$ is a reference online-sample count and $\epsilon_\xi$ is a  hyperparameter controlling the steepness of the ramp; $\xi^n$ increases smoothly from $0$, before any online data is collected, towards $1$ once $\bar\kappa_n$ approaches $m_0$. The final full-state prediction is then recovered by applying the decoder $\mathcal{D}_{\boldsymbol{\theta}_*}^n$.

\subsubsection{Subspace adaptation} Each incoming HF snapshot $\boldsymbol{x}^n$ is immediately used to perform a GROUSE geodesic update step on the Grassmann manifold (Algorithm~\ref{alg:rank1-update}), yielding an intermediate updated basis $\overline{\boldsymbol{\Phi}}_{*}^n$. Crucially, this intermediate basis is stored in memory but \emph{not yet activated}: the working basis $\boldsymbol{\Phi}_*^n$ used for encoding and decoding remains unchanged. This separation allows the subspace to accumulate $K$ successive geodesic refinements before any coordinate-frame change is imposed on the rest of the algorithm. In addition, each HF snapshot $\boldsymbol{x}^n$ (along with $\boldsymbol{x}^{n-1}$) is projected onto the current working basis and appended to the online data arrays (along with the input $\boldsymbol{u}^{n}$):
\begin{equation}
    \boldsymbol{X}_{*}^{n,r} \leftarrow \left[\boldsymbol{X}_{*}^{n,r},\;  
    \left[ \left( \boldsymbol{u}^n \right)^T,\,\, (\boldsymbol{x}^{n-1}) ^{T} \boldsymbol{\Phi}_{k}^n \right]^T      \right],       \qquad    
    \boldsymbol{Y}_{*}^{n,r} \leftarrow \left[\boldsymbol{Y}_{*}^{n,r},\; \boldsymbol{\Phi}_{\boldsymbol{\theta}_*}^{n\,T} \boldsymbol{x}^n\right]
\end{equation}
To control memory consumption, these arrays are capped at a maximum capacity $d$: once $d$ columns are stored, new snapshots are appended and the oldest ones are discarded in a first-in-first-out fashion. At maximum capacity, $\boldsymbol{X}_{*}^{n,r} \in \mathbb{R}^{r+r_u \times d}$ and  $\boldsymbol{Y}_{*}^{n,r} \in \mathbb{R}^{r \times d}$.

\subsubsection{Regression retraining} Whenever the cumulative number of collected HF snapshots reaches a new multiple of $\tau$, the online regression $\mathcal{I}_*^n$ is retrained from scratch on the current contents of the data arrays $\boldsymbol{X}_{*}^{n,r}$ and $\boldsymbol{U}_{*}^{n,r}$ (Algorithm~\ref{alg:regress-update}). Since the regression is performed in the low-dimensional latent space $\mathbb{R}^r$, this retraining incurs negligible cost compared to a HF solve (see section \ref{regressionRetrainingSect}). Note that we choose $\tau < K$, so that the regression is updated multiple times between two successive subspace activations.

\subsubsection{Subspace activation} Whenever the cumulative number of collected HF snapshots reaches a new multiple of $K$, the intermediate basis $\overline{\boldsymbol{\Phi}}_{*}^n$ is \emph{activated}: Rather than discarding the current basis and regression, they are appended to the offline dictionary:
\begin{equation}
    \left\{\boldsymbol{\Phi}_1, \ldots, \boldsymbol{\Phi}_p\right\} \leftarrow \left\{\boldsymbol{\Phi}_1, \ldots, \boldsymbol{\Phi}_p, {\boldsymbol{\Phi}}_{*}^{n}\right\}, \qquad \left\{\mathcal{I}_1, \ldots, \mathcal{I}_p\right\} \leftarrow \left\{\mathcal{I}_1, \ldots, \mathcal{I}_p, \mathcal{I}_*^{n}\right\}, \qquad p \leftarrow p+1.
\end{equation}
Then, $\overline{\boldsymbol{\Phi}}_{*}^n$ replaces the working basis, i.e., $\boldsymbol{\Phi}_*^n \leftarrow \overline{\boldsymbol{\Phi}}_{*}^n$, and the current online regression $\mathcal{I}_*^n$ is reinitialized (Algorithm~\ref{alg:basis-update}). 
The newly activated basis thus acts as a fictitious new training parameter, enriching the dictionary with a subspace that is already well adapted to the online dynamics. After this dictionary augmentation, the Procrustes alignment matrices $\{\boldsymbol{Q}_k\}$ and Grassmann weights $\{w_k\}$ are recomputed with respect to the new working basis $\boldsymbol{\Phi}_*^n$, following lines \ref{lin:computeAngles}--\ref{computeW} of Algorithm~\ref{alg:pre-pred}. Finally, the online data already stored in the buffer is rotated into the new coordinate frame:
\begin{equation}\label{rotateLatentData}
    \boldsymbol{X}_{*}^{n,r} \leftarrow   
    \left[ \boldsymbol{X}_{*}^{n,r}[:N_u,\, :] ^T,\,\,  \left(  \boldsymbol{Q}_p \boldsymbol{X}_{*}^{n,r}[N_u:,\,  :]   \right)^T   \right]^T,  \qquad \boldsymbol{Y}_{*}^{n,r} \leftarrow \boldsymbol{Q}_p \boldsymbol{Y}_{*}^{n,r},
\end{equation}
where $\boldsymbol{Q}_p$ is the Procrustes matrix aligning the previous working basis to the newly activated one. $[:N_u,\,  :]$ indicates the first $N_u$ lines of the array, and $[N_u:,\,  :]$ the remaining lines. This ensures that the online regression, retrained after each activation, always operates in a coordinate frame that is consistent with the current subspace. Perhaps more significantly, this allows the independent adaptation of the subspace and the latent regression. Moreover, we dropped the need for storing streaming data in the high-dimensional space, since it is rotated in the latent space in (\ref{rotateLatentData}), and can be then used consistently with the updated basis.

\begin{remark}
Since the activation of the new basis will only be done after a number $K$ of updates, we ensure that the new Grassmann distance (between $\boldsymbol{\Phi}_p$ and ${\boldsymbol{\Phi}}_{*}^{n}$) is non-zero, and thus the weights in (\ref{weights_locality}) are well defined.
\end{remark}
\begin{remark}
As in Remark \ref{remarkParamNumber}, we point out that with the incrementation of the size of the dictionary $p$, the prediction cost grows linearly with simulation length as more bases get activated online. That size can be pruned using a maximum horizon as well, similarly to our approach for the size of the online data. This was not pursued in the current work. As we will see in the results section, even without this additional pruning, the update overhead remains negligible.
\end{remark}

\begin{figure}
    \centering
    \makebox[\textwidth][c]{%
        \includegraphics[width=1.12\textwidth]{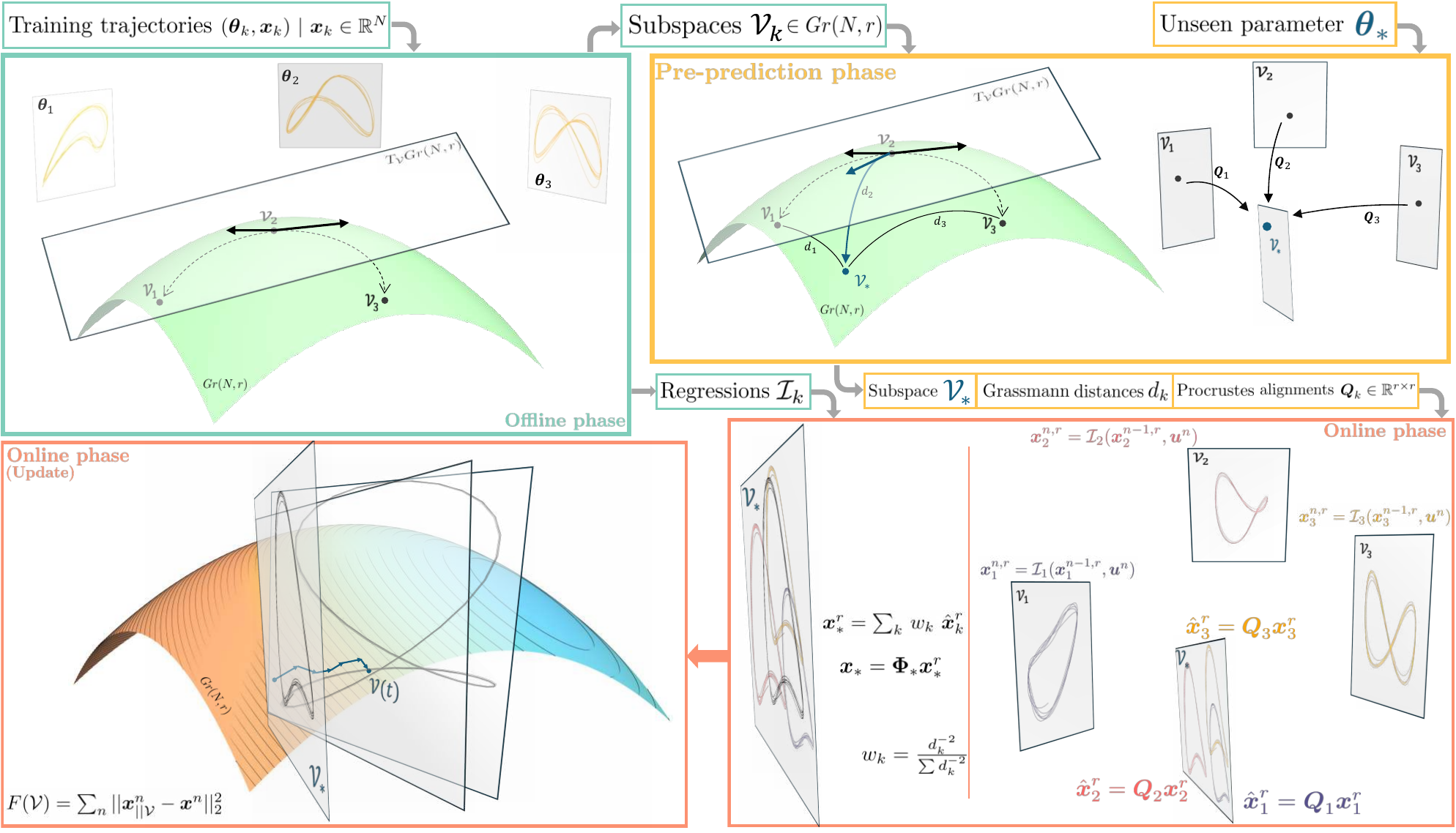}%
    }
    \caption{Illustrative summary of the proposed ROM with online adaptive bases.}
    \label{fig:illustrativeSummary}
\end{figure}

The interplay between the three frequencies, prediction at every step, regression retraining every $\tau$ steps, and basis activation every $K$ steps allows the online phase to continuously refine both the latent-space model and the subspace representation, while keeping all per-step operations at $\mathcal{O}(Nr)$ cost. A summary of the different phases of the proposed ROM is illustrated in Fig.~\ref{fig:illustrativeSummary}.

\begin{algorithm}
\SetKwInput{KwData}{Input}
\caption{Prediction step of  $\widehat{F}^n_{\boldsymbol{\theta}}$}\label{alg:rom-pred}
\KwData{
\begin{itemize}[noitemsep, topsep=0pt, parsep=0pt, partopsep=0pt]
    \item New subspace $\boldsymbol{\Phi}_{*}^n$ (New $\left\{  \mathcal{E}_{\boldsymbol{\theta_*}}^n , \, \mathcal{D}_{\boldsymbol{\theta_*}}^n \right\}$ ), 
    \item Dictionary of regressions $\left\{ \mathcal{I}_{1},  \mathcal{I}_{2}, \cdots  \mathcal{I}_{p} \right\}$,
    \item Current regression $\mathcal{I}_{*}^n$
    \item Solid state encoder $\mathcal{E}_S$,
    \item Dictionary of calibration operators $\left\{ \boldsymbol{Q}_{1}, \boldsymbol{Q}_{2}, \cdots  \boldsymbol{Q}_{p} \right\}$,
    \item Geometric proximity weights $\left\{ w_{1}, w_{2}, \cdots  w_{p} \right\}$,
    \item Regression weight hyperparameter $0 < \xi < 1$,
    \item Input  $\left(\boldsymbol{x}^{n-1},\, \boldsymbol{u}^{n}\right)$
\end{itemize}
}
\KwResult{
$\boldsymbol{x}^n$

}
\hrulefill

\For{$k\gets1$ \KwTo $p$}{
\nl $ \boldsymbol{x}^{n-1, r} = \mathcal{E}_{\boldsymbol{\theta_k}}^n \left( \boldsymbol{x}^{n-1} \right) $

\nl $ \boldsymbol{x}^{r}_k = \mathcal{I}_{k} \left( \boldsymbol{x}^{n-1, r}, \boldsymbol{u}^{r} \right) $

\nl $ \widehat{\boldsymbol{x}}^{r}_k = \boldsymbol{Q}_k  \boldsymbol{x}^{r}_k $
}

\nl ${\boldsymbol{x}}^{n, r} = \sum_i^p w_i \widehat{\boldsymbol{x}}^{r}_i $

\nl $\overline{\boldsymbol{x}^{n, r}} = \xi \, \mathcal{I}_{*}^n \left( \mathcal{E}_{\boldsymbol{\theta_*}}^n \left( \boldsymbol{x}^{n-1} \right)  , \boldsymbol{u}^{r} \right) + (1 - \xi){\boldsymbol{x}}^{n, r}$ \label{weightingCurrentI}

\nl ${\boldsymbol{x}}^{n} = \mathcal{D}_{\boldsymbol{\theta_*}}^n \left(  \overline{\boldsymbol{x}^{n, r}} \right) $

\end{algorithm}

\begin{algorithm}
\SetKwInput{KwData}{Input}
\caption{Regression update step of  $\widehat{F}^n_{\boldsymbol{\theta}}$}\label{alg:regress-update}
\KwData{
\begin{itemize}[noitemsep, topsep=0pt, parsep=0pt, partopsep=0pt]
    \item Online data $\boldsymbol{X}_{*}^{n, r}$ and $\boldsymbol{Y}_{*}^{n, r}$,
    \item Online regression $\mathcal{I}_{*}$.
\end{itemize}
}
\hrulefill

\uIf{$\mathcal{I}_{*}^n\, != \, \{ \emptyset \}$}{
\nl Retrain $\mathcal{I}_{*}^n ( \cdot )$ on $\boldsymbol{X}_{*}^{n, r}$ and $\boldsymbol{Y}_{*}^{n, r}$
}\Else{
\nl Train $\mathcal{I}_{*}^n ( \cdot )$ on $\boldsymbol{X}_{*}^{n, r}$ and $\boldsymbol{Y}_{*}^{n, r}$
  }

\end{algorithm}

\begin{algorithm}
\SetKwInput{KwData}{Input}
\caption{Subspace update step of  $\widehat{F}^n_{\boldsymbol{\theta}}$}\label{alg:basis-update}
\KwData{
\begin{itemize}[noitemsep, topsep=0pt, parsep=0pt, partopsep=0pt]
    \item Online data $\boldsymbol{X}_{*}^{n, r}$ and $\boldsymbol{Y}_{*}^{n, r}$,
    \item Local bases dictionary $\left\{ \boldsymbol{\Phi}_{1},  \boldsymbol{\Phi}_{2}, \cdots  \boldsymbol{\Phi}_{p} \right\}$,
    \item Regression dictionary $\left\{ \mathcal{I}_{1},  \mathcal{I}_{2}, \cdots  \mathcal{I}_{p} \right\}$,
    \item Current basis $\boldsymbol{\Phi}_{*}^n$,
    \item Updated basis $\overline{\boldsymbol{\Phi}}_{*}^n$,
    \item Procrustes matrices $\left\{ \boldsymbol{Q}_{1}, \boldsymbol{Q}_{2}, \cdots  \boldsymbol{Q}_{p} \right\}$,
\end{itemize}
}
\hrulefill

\nl Add the latest update subspace to the "dictionary" of modes $\left\{ \boldsymbol{\Phi}_{1},  \boldsymbol{\Phi}_{2}, \cdots  \boldsymbol{\Phi}_{p} \right\} \leftarrow \left\{ \boldsymbol{\Phi}_{1},  \boldsymbol{\Phi}_{2}, \cdots  \boldsymbol{\Phi}_{p},  \boldsymbol{\Phi}_{*}^n \right\}$

\nl Add the new regression to the "dictionary" of regressions $\left\{\mathcal{I}_{1}, \cdots  \mathcal{I}_{p} \right\} \leftarrow  \left\{ \mathcal{I}_{1}, \cdots  \mathcal{I}_{p} ,\mathcal{I}_{*}^n \right\}$

\nl Activation of the updated subspace $\boldsymbol{\Phi}_{*}^n \leftarrow \overline{\boldsymbol{\Phi}}_{*}^n $

\nl $p \leftarrow {p+1}$

\nl Repeat lines \ref{lin:computeAngles} to \ref{computeW}, Algorithm \ref{alg:pre-pred} to find the new Procrustes matrices $\left\{ \boldsymbol{Q}_{1}, \cdots  \boldsymbol{Q}_{p} \right\}$ and weights $\left\{ {w}_{1}, \cdots {w}_{p} \right\}$

\nl Rotate the collected data
\begin{itemize}
    \item[] $\boldsymbol{X}_{*}^{n,r} \leftarrow
    \left[ \boldsymbol{X}_{*}^{n,r}[:N_u,\, :]^T,\,\,  \left(  \boldsymbol{Q}_p \boldsymbol{X}_{*}^{n,r}[N_u:,\,  :]   \right)^T   \right]^T$
    \item[] $\boldsymbol{Y}_{*}^{n, r} \leftarrow \boldsymbol{Q}_{p} \boldsymbol{Y}_{*}^{n, r}$
\end{itemize}

\nl $\mathcal{I}^n_{*}(\cdot) \leftarrow \emptyset$

\end{algorithm}

\section{Application on convergence acceleration of partitioned FSI simulations}\label{section3}
In FSI problems, the kinematic and dynamic coupling conditions on the interface of the solid and fluid computational domains are ${\Gamma_{fsi}}$: 
\begin{equation}\label{fsiCondition1}
    \begin{cases}
	  \sigma_f \cdot \boldsymbol{n}  =  - \sigma_s \cdot \boldsymbol{n} \, \, \, \, at \, \, \, \, {\Gamma_{fsi}} \\
	  \boldsymbol{v} = \dot{\boldsymbol{d}} \, \, \, \, at \, \, \, \, {\Gamma_{fsi}}\\
   \end{cases}
\end{equation}
where $\sigma_f$ and $\sigma_s$ are the Cauchy stresses applied by the fluid and the solid respectively, $\boldsymbol{v}$ is the fluid velocity and $\dot{\boldsymbol{d}}$ is the solid velocity. An additional coupling condition is imposed in the case mesh-based methods for the fluid problem:
\begin{equation}\label{fsiCondition2}
      \boldsymbol{\omega} = {\boldsymbol{d}} \, \, \, \, at \, \, \, \,{\Gamma_{fsi}}\\
\end{equation}
where $\boldsymbol{\omega}$ is the fluid mesh displacement field and $\boldsymbol{d}$ is the solid displacement.

The \textit{Dirichlet-Neumann} coupling formulation represents the two solvers as distinct operators that communicate the solid displacement, solid velocity and fluid forces on the interface. We represent the fluid solver operator as $\mathcal{F}$: 
\begin{equation}
  \mathcal{F}: \mathbb{R}^{2N_S} \rightarrow \mathbb{R}^{N_F}\;;\; (\boldsymbol{d}_{|\Gamma_{fsi}}, \dot{\boldsymbol{d}}_{|\Gamma_{fsi}}) \rightarrow \boldsymbol{f}_{|\Gamma_{fsi}} 
\end{equation}
where $\boldsymbol{d}_{|\Gamma_{fsi}}$ is the displacement field $N_S$ is the number of interface degrees of freedom and $\boldsymbol{f}_{|\Gamma_{fsi}}$ represents the fluid viscous and pressure forces at $\Gamma_{fsi}$:
\begin{equation}
  \boldsymbol{f}_{|\Gamma_{fsi}} = \sigma_f \cdot \boldsymbol{n}_{f|\Gamma_{fsi}}.
\end{equation}
%
Similarly, the solid operator $\mathcal{S}$ is defined as:
\begin{equation}\label{solid_operator}
  \mathcal{S}: \mathbb{R}^{N_F} \rightarrow \mathbb{R}^{2N_S}\;;\; \boldsymbol{f}_{|\Gamma_{fsi}} \rightarrow (\boldsymbol{d}_{|\Gamma_{fsi}}, \dot{\boldsymbol{d}}_{|\Gamma_{fsi}})  .
\end{equation}
Clearly, the operator are written using a simplified definition that hides the time dependence of the solid and fluid solvers.

In the case of big added-mass effect, \textit{e.g} when considering incompressible flow and light structures, a \textit{strong coupling} is needed to overcome numerical instabilities \citep{causin_added-mass_2005, BrummelenAdded}, in the sense that the coupling conditions should be enforced exactly using iterative schemes. This nonlinear problem can be written as the fixed-point:
\begin{equation}\label{dirich_neum}
    \left( \mathcal{F} \circ \mathcal{S} \right) \left(\boldsymbol{f}_{|\Gamma_{fsi}} \right) = \boldsymbol{f}_{|\Gamma_{fsi}}.
\end{equation}
The subscript $|\Gamma_{fsi}$ will be dropped hereafter for simplicity.

\subsection{Enhanced predictor using adaptive ROMs}
The main idea of \citep{TIBA2025109522} is to use ROMs to predict an initial guess of the iterative scheme for solving (\ref{dirich_neum}). In fact, in transient simulations, a finite-differences extrapolation (commonly linear or quadratic) is usually applied on the interface field to start the nonlinear iterations in the current time step. A "reduced FSI coupling" is launched at the start of each time step, where each subsystem $\mathcal{F}$ and $\mathcal{S}$ in (\ref{dirich_neum}) is replaced by a reduced order model $\hat{\mathcal{F}}$ and $\hat{\mathcal{S}}$, and the solution
\begin{equation}
    \widehat{\mathcal{F}}(\widehat{\mathcal{S}}(\boldsymbol{f}^{0, n})) = \boldsymbol{f}^{0, n}, \qquad \forall\, n \in \{1, \dots, N_t\}
\end{equation} 
is sought via fixed-point iterations at a fraction of the computational time needed for the HF solver calls. The new ROM-assisted predictor will achieve speedups with no accuracy loss as long as the fixed-point FSI coupling is tightly converged. The first superscript in $\boldsymbol{f}^{0, n}$ indicates the index of the fixed-point iteration, and the second indicates the time step. A similar approach was used in \citep{DELAISSE2022106720} using physics-based surrogates.

In this work, we extend the approach in \citep{TIBA2025109522} to make an online adaptive ROM with subspace update for the fluid. The control input $\boldsymbol{u}$ for $\widehat{F}^n_{\boldsymbol{\theta}}(\cdot)$ is taken as the POD coefficient vector of the solid state space $\boldsymbol{d}^r = \mathcal{E}_S \left( \left[ \boldsymbol{d}, \dot{\boldsymbol{d}} \right] \right) \in \mathbb{R}^{r_S}$ where $\mathcal{E}_S(\cdot)$ is the encoder of the solid ROM.
\begin{equation}\label{eq:equivStateFSI}
    \boldsymbol{x} \equiv \boldsymbol{f}, \qquad \boldsymbol{u} \equiv \boldsymbol{d}^r, \qquad N_x \equiv N_F, \qquad N_u \equiv r_S
\end{equation}

For the solid ROM, we choose to use a global, static subspace, although with an online adaptive regression. In addition to simplifying the predictor, this choice is motivated by the fact that the solid displacement is typically smoother and exhibit less complex dynamics than the fluid forces. Furthermore, the solid ROM can be simplified to only predict the reduced coordinates of the solid state-space $( \boldsymbol{d}, \dot{\boldsymbol{d}} )$ since the displacement and velocities are not needed in their high-dimensional space, only their representation as an input to $\widehat{F}^n_{\boldsymbol{\theta}}(\cdot)$. The input to $\widehat{S}^n$ are the reduced coordinates of the fluid forces $\boldsymbol{f}^r$.

\begin{equation}\label{eq:learnt_system_fsi}
    \widehat{S}^n\!\left( \left[ \boldsymbol{d}^{n-1}, \dot{\boldsymbol{d}}^{n-1} \right] ,\, 
    \boldsymbol{f}^{n, r}\right) =  \mathcal{I}^n_S \left( \mathcal{E}_{S}^n\left( \left[ \boldsymbol{d}^{n-1}, \dot{\boldsymbol{d}}^{n-1} \right] \right), \,\,\,  \boldsymbol{f}^{n, r} \right), \qquad \forall\, n \in \{1, \dots, N_t\}.
\end{equation}

We write the discrete form of the reduced coupling using the discrete equivalent of the approximated fluid and solid operators:
\begin{equation}
    \widehat{F}^n_{\boldsymbol{\theta}} \left( \boldsymbol{f}^{n-1}, \widehat{S}^n \left( \boldsymbol{d}^{n-1}, \dot{\boldsymbol{d}}^{n-1}, \boldsymbol{f}^{0, n} \right)  \right) = \boldsymbol{f}^{0, n}, \qquad \forall\, n \in \{1, \dots, N_t\}.
\end{equation}
The ROM operators' notation now include the time dependence by taking the previous time step state as an input.

In Algorithm \ref{globalAlg}, the iterative FSI scheme with the ROM-predictor is detailed and in Algorithm \ref{local}, we show the reduced coupling's procedure of the reduced models. An illustration of this procedure is given in Fig. \ref{fig:illust-pred-ROM}.

\section{Results}\label{section4}

In this section, we evaluate the proposed framework on three test cases. The first is a flow-induced vibrations benchmark used to assess ROM accuracy in isolation, without any FSI coupling. The second and third test cases involve full partitioned FSI simulations, where the ROM is used as a predictor for the iterative coupling scheme, and the evaluation additionally includes the computational gain in terms of FSI convergence acceleration. The simulation in both cases is computed using the simulation framework \texttt{KratosMultiphysics}  \citep{dadvand_object-oriented_2010}.

Across all three test cases, we report the quantities defined below.

\subsubsection{Subspace evolution} To characterise the temporal adaptation of the online basis, we report three complementary angle-based metrics. The \emph{online recursive distance} measures the geodesic distance on the Grassmann manifold between the subspace at two consecutive GROUSE update steps 
\begin{equation}
    d\left( \left[ \overline{\boldsymbol{\Phi}}_{*}^{n-1}\right],\, \left[\overline{\boldsymbol{\Phi}}_{*}^{n}\right] \right),
\end{equation}
and reflects how much the subspace is being corrected at each step. The \emph{accumulated subspace distance} measures the total geodesic distance travelled from the initial approximated subspace $\boldsymbol{\Phi}_{*}^0$:
\begin{equation}
    d\left( \left[ \boldsymbol{\Phi}_{*}^{0}\right],\, \left[\overline{\boldsymbol{\Phi}}_{*}^{n}\right] \right),
\end{equation}
and quantifies the overall departure of the adapted subspace from its initialisation. To give a quantitative picture of the magnitude of the subspace rotation, we also report the \emph{maximum accumulated subspace angle} (see (\ref{subsp_angle_expression})), measured on the direction that achieved the biggest rotation from its counterpart in the initial subspace:
\begin{equation}
    \max_{i \in \{1, \dots, r\}}{\gamma_i}.
\end{equation}

\subsubsection{Subspace quality: relative projection error} The representational quality of the current basis is assessed independently of the regression, by measuring how much of an incoming HF snapshot lies outside the current subspace. The \emph{relative orthogonal projection error} is defined as:
\begin{equation}
    e_{\perp}^n = \frac{\left\|\boldsymbol{x}^{n}_{\perp\mathcal{V}} \right\|}{\left\| \boldsymbol{x}^n \right\|}.
\end{equation}
This quantity is compared across the following subspace strategies:
\begin{itemize}
    \item \textbf{Global static}: a single POD basis computed from all available training snapshots, fixed throughout the simulation.
    \item \textbf{Local static}: a Grassmann-predicted basis $\boldsymbol{\Phi}_{*}^0$ specific to the new parameter $\boldsymbol{\theta}_*$, fixed throughout the simulation.
    \item \textbf{Proposed (local + GROUSE)}: the Grassmann-predicted basis continuously updated online via GROUSE geodesic gradient descent steps.
\end{itemize}

\subsubsection{Full-field prediction accuracy} The end-to-end prediction quality of the ROM is evaluated using the \emph{relative prediction error} at each time step:
\begin{equation}
    e^n = \frac{\left\| \hat{\boldsymbol{x}}^n - \boldsymbol{x}^n \right\|}{\left\| \boldsymbol{x}^n \right\|},
\end{equation}
where $\hat{\boldsymbol{x}}^n$ is the full-state ROM prediction and $\boldsymbol{x}^n$ the reference HF snapshot. The following methods are compared:
\begin{itemize}
    \item \textbf{Proposed (linear / nonlinear (Nln))}: the proposed framework with a linear ($q=1$) latent-space regression, or a nonlinear regression using a second-order polynomial ($q=2$) fit or RBF interpolation (see \ref{regr_methods_part})
    \item \textbf{Global static (linear / nonlinear)}: the global static basis with linear and nonlinear regressions respectively.
    \item \textbf{Local static (linear)}: the local predicted static basis with a linear regression.
    \item \textbf{Global static + parameter input (linear)}: the global static basis with a linear regression taking $\boldsymbol{\theta}$ as an additional input, providing a parametric regression baseline without subspace adaptation.
    \item \textbf{Recursive DMD with control (rDMDc)}~\citep{taleb}: an online DMD-based adaptive ROM that updates both the basis (using PAST) and a linear operator recursively at each step. Furthermore, we initialize the rDMDc basis with a local basis from the  Grassmann interpolation. The forgetting factors for PAST and the linear regression are both set to $0.95$.
\end{itemize}
Since this work focuses on the basis update, all these compared methods use an adaptive regression with the same frequency. 

\subsection{Results reproduction}

We provide two repositories containing scripts for reproducing the present results. The repositories provide a detailed description for the software environment we used for each example. The first repository for the offline (no FSI coupling) results is \href{https://github.com/azzeddinetiba/Online-adaptive-GF-ROMs/tree/v0.3.2}{Online-adaptive-GF-ROMs}\cite{tibaOnlineAdaptROMRepo}  and the second is for reproducing the results inside the setting of a FSI coupling predictor:
\href{https://github.com/FsiROM/FSI-ROM-Predictor/tree/v0.3.2}{FSI-ROM-Predictor}\cite{tibaOnlinePredictorRepo}. The results reported in this paper correspond specifically to version v0.3.2 of both repositories. The ROM operations used in this work are implemented in the \href{https://github.com/azzeddinetiba/ROM_AM}{\texttt{rom\_am}}\cite{tiba2023romam} package, with the corresponding version specified in the software requirements of each repository.

\subsection{Oscillating cylinder in a fluid flow}
We begin by testing our approach on an FSI benchmark of Vortex Induced Vibrations (VIV) on a 2D cylinder for which simulation data is readily available. Thus, the ROM predictions in this case will only be used to evaluate the accuracy of the model, independently of any FSI coupling scheme.

\begin{figure}
    \centering
    \includegraphics[width=0.6\textwidth]{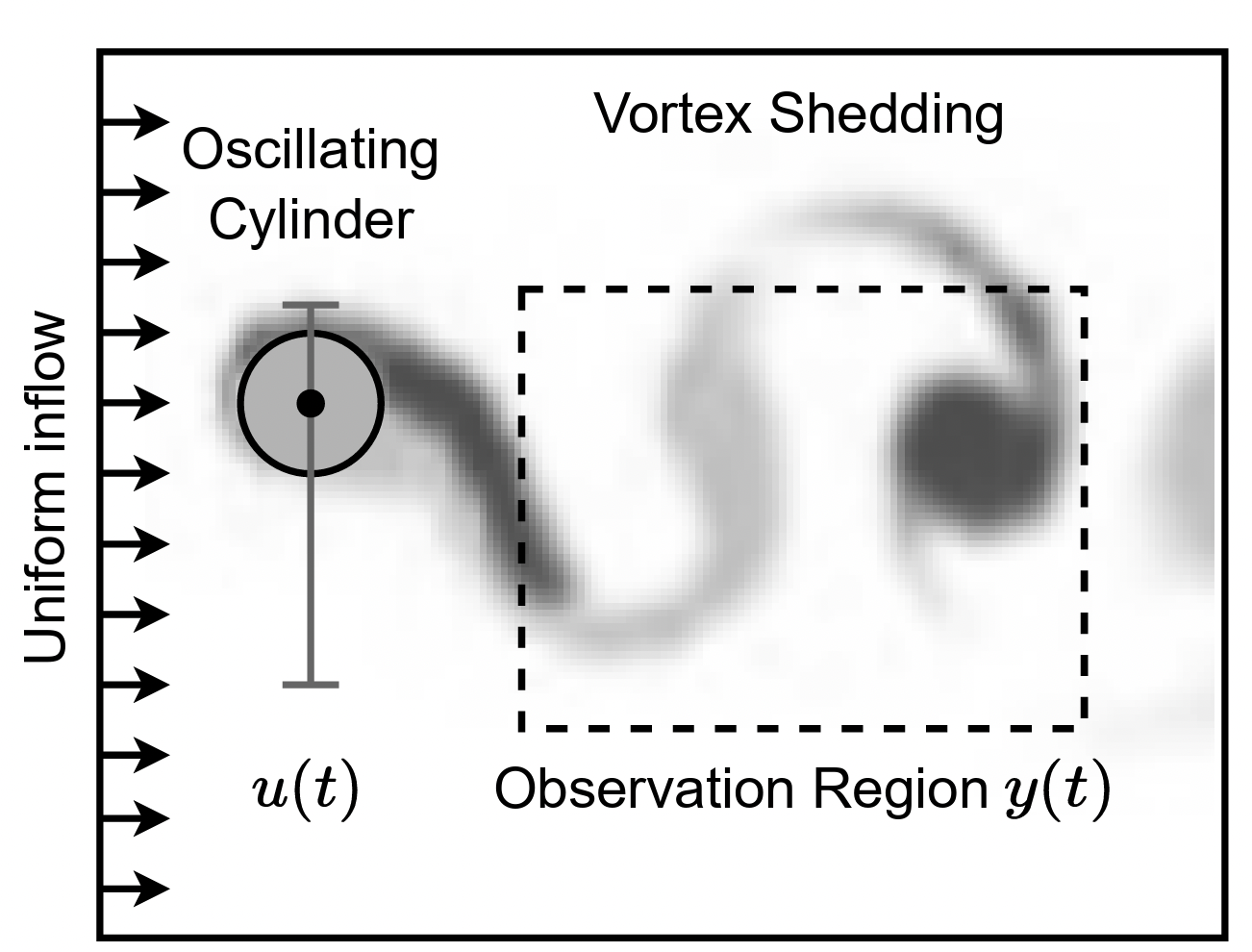}
    \caption{An unsteady fluid flow in the wake of a moving cylinder. From~\citep{beintema2024data}.}
    \label{fig:vivScheme}
\end{figure}

The benchmark data is provided by~\citep{nlnBenchmarks}. The data is obtained from the results of a finite volume Navier-Stokes simulation with a deformable computation domain. Details about the simulation are found in~\citep{decuyper2017nonlinear}. See Fig.~\ref{fig:vivScheme} for an illustration of the test case configuration.

The interesting thing about the VIV problem here is the lock-in phenomenon, where the shedding frequency becomes \textit{locked} to the excitation when the excitation frequency is sufficiently close to the natural shedding frequency $f_{St}$. We recall that for a stationary cylinder,
\begin{equation}
    f_{St} = St \frac{V_{\infty}}{D}
\end{equation}
where $St$ is the Strouhal number, $V_{\infty}$ is the upstream velocity and $D$ is the cylinder diameter. The amplitude of the cylinder oscillation is $A = 0.2 D$.

\begin{figure}
    \centering
    \includegraphics[width=0.65\textwidth]{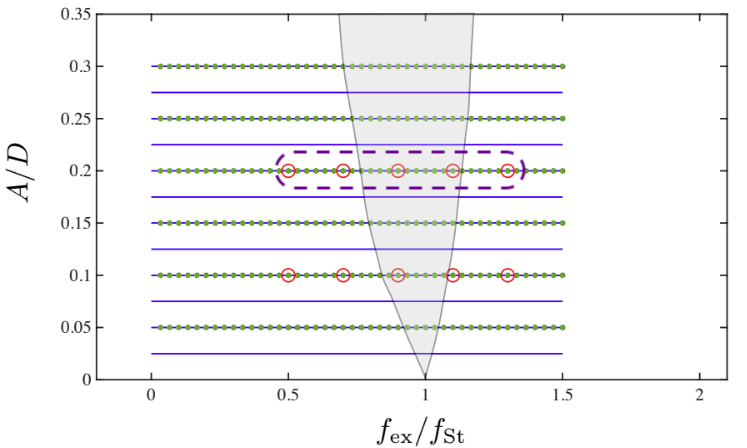}
    \caption{Graphical representation in the frequency-amplitude plane of imposed motion of considered simulations. In our work, we consider the operating points represented by the top red circles, corresponding to monosine oscillations (sweep and hold). The lock-in region, as observed by~\citep{KUMAR} is indicated by the grey shaded region. From~\citep{beintema2024data}.}
    \label{fig:vivFreq}
\end{figure}

We consider $p = 4$ scalar training parameters $\theta_k = f_{ex, k} \in \{0.5 f_{St},\, 0.7f_{St},\, 1.1f_{St},\, 1.3f_{St}\}$ and predict for an unseen parameter near the barycenter of the training set: $\theta_* = f_{ex,*} = 0.9\,f_{St}$ (see Fig.~\ref{fig:vivFreq}). The snapshots are spaced with a time step $\Delta t = 0.025~s$. $m_k=800$ snapshots for each parameter are used as training data, centered around the mid-time of each simulation, ensuring the same amount of data from the sweep and the hold phases are associated to each parameter. The prediction is computed for the time range $[0,~25s]$. Both the prediction frequency and the training frequency $0.7\,f_{St}$ fall within the lock-in region (shown in grey in Fig.~\ref{fig:vivFreq}), making this an especially challenging case for a ROM: the dynamics exhibit a qualitative regime change during the simulation that a static basis struggle to track.

\begin{figure}
    \centering
    \includegraphics[width=0.6\textwidth]{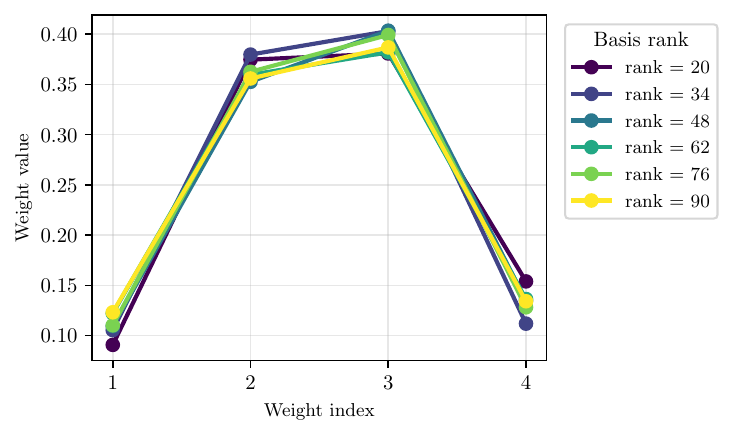}
    \caption{Grassmann-distance weights  using different POD basis ranks.}
    \label{fig:weightsPlot}
\end{figure}

The state-space is formed by both the velocity and pressure fields on the observation region (discretized on a $105\times61$ grid):
\begin{equation}
    \boldsymbol{x}
    =
    \begin{bmatrix}
        \mathbf{v}_h \\
        \mathbf{p}_h
    \end{bmatrix}
    \in \mathbb{R}^{19215},
\end{equation}
where $\mathbf{v}_h \in \mathbb{R}^{2 * 105 * 61}$ and
$\mathbf{p}_h \in \mathbb{R}^{105 * 61}$ denote the discrete velocity and pressure fields, respectively. The control input here is the $y$-position of the cylinder, so $N_u=1$.
The fluid basis rank is set to $r = 64$, corresponding to an energy threshold of $\epsilon = 99.9\%$. The reference point for the Grassmann interpolation is set to $\text{ref} = 1$ (i.e., $\mathcal{V}_{\text{ref}} = \mathcal{V}_1$, associated with $\theta_1 = 0.5\,f_{St}$). The Grassmann-distance weights computed in line~\ref{computeW} of Algorithm~\ref{alg:pre-pred} for different basis ranks are shown in Fig. \ref{fig:weightsPlot}.
They reflect the geometric proximity of each training subspace to the interpolated one, with the two nearest training frequencies ($0.7\,f_{St}$ and $1.1\,f_{St}$) receiving the largest contributions. Remarkably, the weight associated to the third frequency, located in the lock-in region, is slightly higher, which suggests that the Grassmann interpolation accurately reproduced the relative placement of the four frequencies with respect to the lock-in region, as shown in Fig. \ref{fig:vivFreq}. The remaining hyperparameters are set to $z = 2$, $d = 1000$, $\tau = 80$, and $K = 170$. For this test case, $\xi$ follows the adaptive schedule of Eq.~(\ref{eq:xi_schedule}), with $m_0 = \min_k m_k = 800$ and $\epsilon_\xi = 0.4$. The nonlinear regression is an RBF with a cubic kernel.

\begin{figure}
    \centering
    \includegraphics[width=0.85\textwidth]{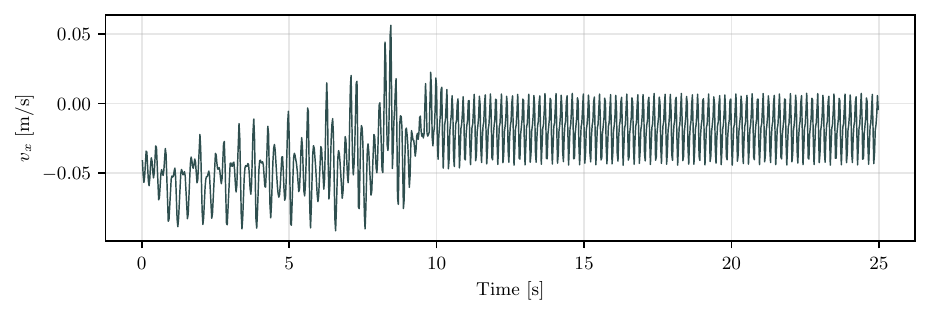}
    \caption{[VIV] -- X-velocity $v_x$ at a near-cylinder probe throughout the online simulation. The sharp transition around $t \approx 10~s$ marks the onset of the lock-in regime, where the shedding frequency locks onto the excitation frequency and the oscillation amplitude increases significantly.}
    \label{fig:vySignal}
\end{figure}

Before presenting the quantitative metrics, we briefly illustrate the nature of the dynamics encountered in this test case (test parameter). Fig.~\ref{fig:vySignal} shows the X-velocity $v_x$ recorded at a near-cylinder probe throughout the online simulation. Two distinct regimes are clearly visible: a first phase ($t \leq 10~s$) of moderate-amplitude quasi-periodic oscillations, followed by a sharp transition to a larger-amplitude, strongly periodic regime: the signature of lock-in.

\subsubsection{Latent space geometry and Procrustes alignment:} 
\begin{figure}
    \centering
    \includegraphics[width=.8\textwidth]{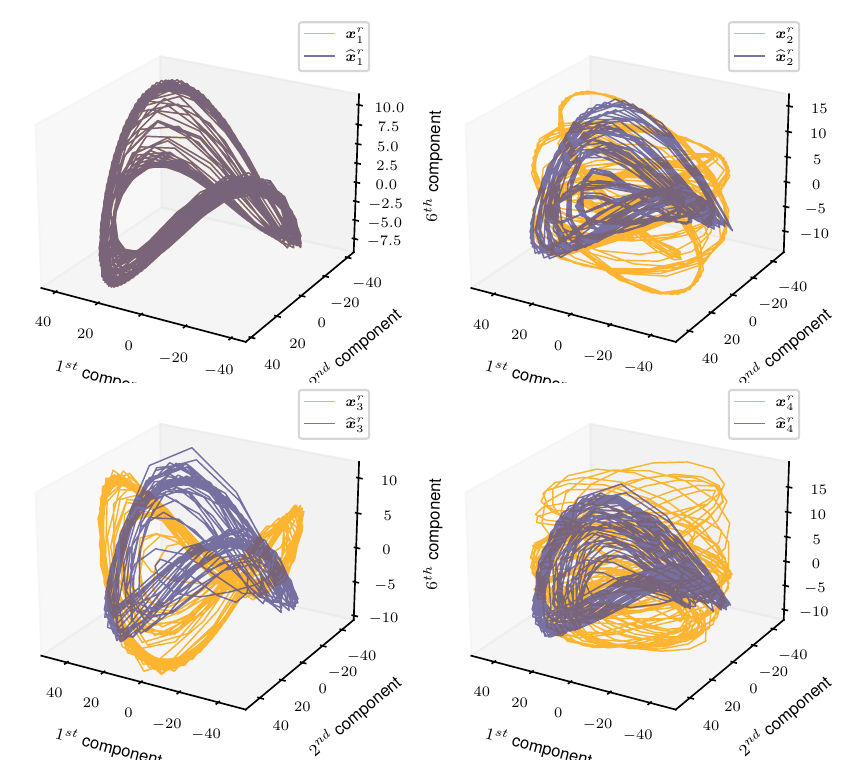}
    \caption{[VIV] -- Parametric trajectories in the reduced space before and after application of the Procrustes alignment operator $\widehat{\boldsymbol{x}}^r_k = \boldsymbol{Q}_k \boldsymbol{x}^r_k$. The trajectory associated with the reference parameter ($\text{ref} = 1$) is unchanged, as $\boldsymbol{Q}_{\text{ref}} = \boldsymbol{I}_r$.}
    \label{fig:compareReducedSnaps3D}
\end{figure}

Fig.~\ref{fig:compareReducedSnaps3D} visualises the trajectories of the training snapshots in the reduced coordinate spaces of their respective local bases. Before applying the Procrustes alignment matrices $\boldsymbol{Q}_k$, the four trajectories exhibit very different shapes and orientations, reflecting the fact that each local basis defines its own coordinate frame with no imposed consistency. After alignment, the trajectories all adopt markedly similar shapes and are brought into a common frame. The one exception, as expected from Proposition~\ref{proposition1}, is the trajectory associated with the reference parameter $\theta_1$ ($\text{ref} = 0$), for which the Procrustes matrix is the identity: its trajectory is unchanged by the alignment.

\subsubsection{Subspace evolution} 
\begin{figure}
    \centering
    \includegraphics[width=\linewidth]{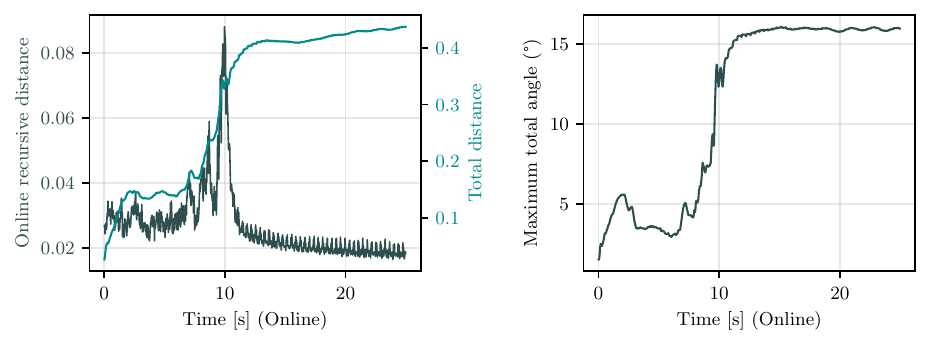}
    \caption{[VIV] -- Left: Online distance of the updated subspace from that used in the previous iteration, shown together with the the total distance of the updated subspace to the initial subspace. \, Right: The maximum subspace angle between the current $\mathcal{V}^n_{*}$ and the initial subspsace $\mathcal{V}^0_{*}$.}
    \label{fig:combinedAngleShiftOnlineNln}
\end{figure}

The temporal evolution of the updated subspace is illustrated in Fig.~\ref{fig:combinedAngleShiftOnlineNln}. The left figure reports the online recursive distance, as well as the accumulated subspace distance, while the right figure reports the maximum accumulated subspace angle.
During the first online steps before $t\approx8~s$, the recursive distance remains small and stable, indicating that the GROUSE updates are making only minor corrections — the initial predicted subspace is already well aligned with the incoming dynamics. Around online step $t=10~s$, a sharp spike occurs. This coincides with the onset of the lock-in regime, where the dynamics undergo a qualitative change: the shedding frequency locks onto the excitation frequency, leading to a structural shift in the solution manifold that the subspace must track. After this transient, the distance stabilises at a lower value for the remainder of the simulation, indicating that the subspace has converged to a representation that fits the locked-in dynamics.

The accumulated rotation tells a complementary story: a small bump is visible around step $t=2~s$ (reaching $\approx 5^{\circ}$), followed by the large jump associated with the lock-in event, after which the total rotation stabilises near $15^{\circ}$. This confirms that the majority of the subspace adaptation is triggered by the lock-in transition, rather than being uniformly distributed across the simulation.

\begin{figure}
    \centering
    \includegraphics[width=\textwidth]{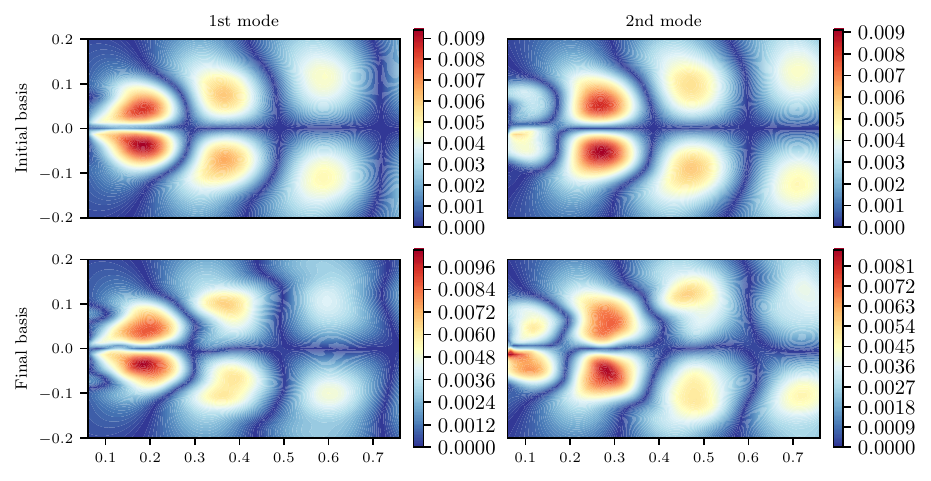}
    \caption{[VIV] -- Absolute value of the X-velocity of two selected POD modes (the two modes whose spatial structure changed the most) for the initial predicted basis $\boldsymbol{\Phi}_{*}^0$ (top row) and the final GROUSE-adapted basis $\overline{\boldsymbol{\Phi}}_{*}^{N_t}$ (bottom row). The structural changes are consistent with the transition from pre-lock-in to locked-in flow dynamics.}
    \label{fig:modesComparison}
\end{figure}

Fig.~\ref{fig:modesComparison} compares the absolute value of the X-velocity of two selected POD modes from the initial predicted basis $\boldsymbol{\Phi}_{*}^0$ (top row) and the final GROUSE-adapted basis $\overline{\boldsymbol{\Phi}}_{*}^{N_t}$ (bottom row). The two modes shown are those whose spatial structure changed the most between the two bases, as measured by the inner product between corresponding mode shapes. The two modes transition from a broad, nearly symmetric structure concentrated around the cylinder body to a more elongated, asymmetric pattern extending slightly further into the wake.
These changes confirm that the GROUSE updates are genuinely rotating the subspace toward a qualitatively different set of flow structures.

\subsubsection{Projection error} 
\begin{figure}
    \centering
    \includegraphics[width=\textwidth]{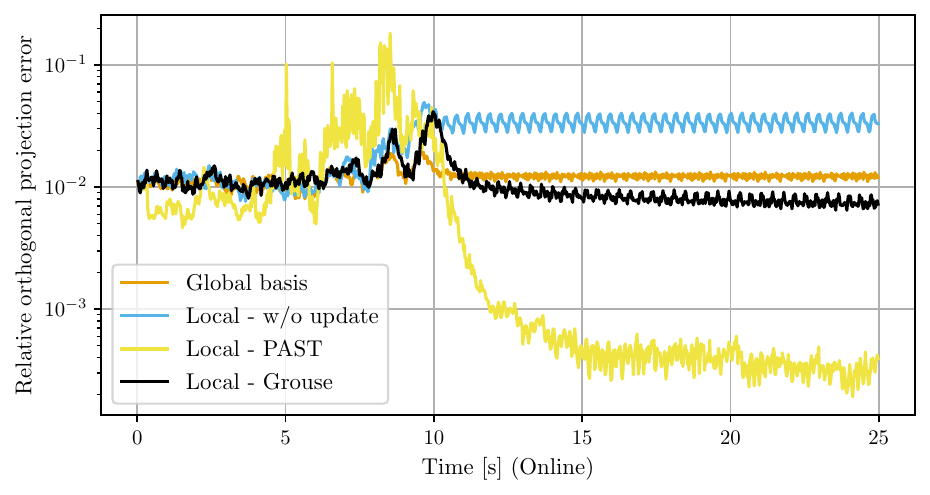}
    \caption{[VIV] -- Relative orthogonal projection error throughout the online simulation, comparing a global static basis, a local static basis, and the proposed method using GROUSE and PAST for subspace tracking.}
    \label{fig:fluidNlnOrthNorms}
\end{figure}

Fig.~\ref{fig:fluidNlnOrthNorms} reports the relative orthogonal projection norm. Prior to the lock-in event, all methods produce comparable projection errors. At the onset of lock-in, all methods exhibit a spike in projection error, as the solution manifold undergoes its qualitative shift. After this transient, a clear divergence in behaviour is observed: the global and static local bases stagnate, as they have no mechanism to track the new dynamics, while the proposed method continues to improve. With an increasing number of accumulated snapshots informing the GROUSE and PAST updates, the projection error using the dynamic subspace decreases steadily, and the error using PAST ultimately reaches the lowest value among all compared approaches. This demonstrates that the online subspace adaptation provides a tangible and sustained improvement in representational quality, precisely in the regime where it matters most.

\subsubsection{Latent space predictions} 
\begin{figure}
    \centering
    \includegraphics[width=\textwidth]{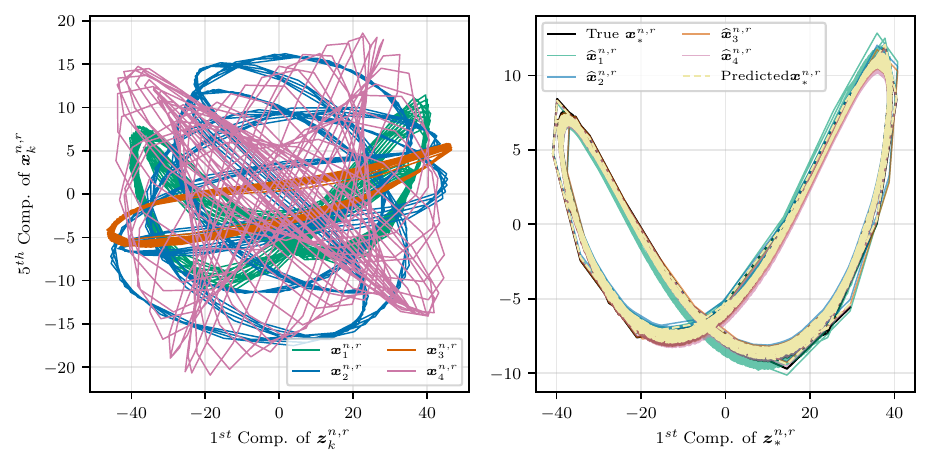}
    \caption{[VIV] -- Left: The different trajectories in the different reduced spaces associated to the parameters. \, Right: Individual predicted aligned states from each locally trained regression and their Grassmann-weighted combination, compared to the reference HF reduced trajectory.}
    \label{fig:compareReducedPrediction}
\end{figure}

Fig.~\ref{fig:compareReducedPrediction} compares the individual predictions $\boldsymbol{Q}_k \mathcal{I}_k(\boldsymbol{z}^n)$ of each locally trained regression with the combined prediction and the reference full-order reduced trajectory. The Procrustes alignment effect on the different regressions' is demonstrated through the comparison between the left and right figures, where it is shown that the different trajectories are brought to the common system. After this alignment, all individual predictions are nearly indistinguishable from one another. More importantly, the combined prediction closely matches the true reduced trajectory, confirming that the alignment and aggregation procedure correctly reconstructs the target latent dynamics.

\subsubsection{Full-field prediction accuracy} 
\begin{figure}
    \centering
    \includegraphics[width=\textwidth]{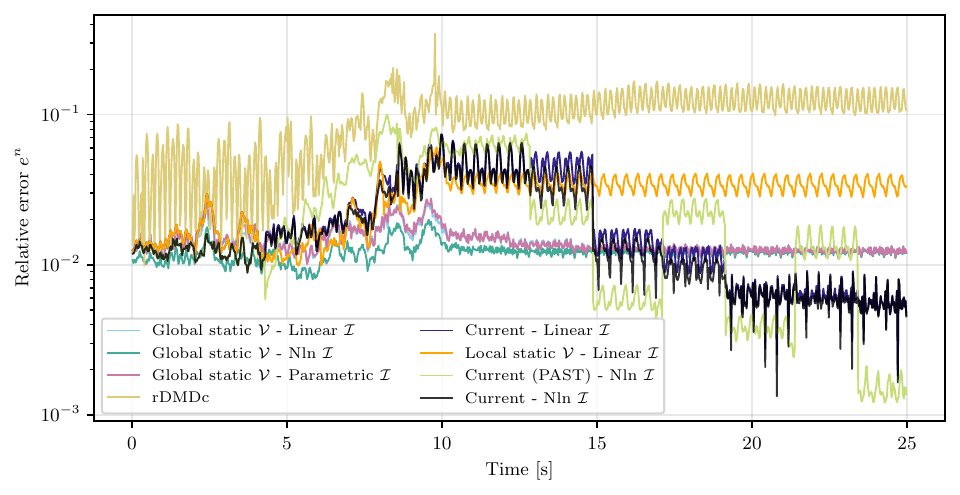}
    \caption{[VIV] -- Relative prediction error of the concatenated velocity-pressure field through the online simulation steps, comparing the proposed method against global static, local static, parametric-input, recursive DMD, and PAST-based baselines.}
    \label{fig:fluidNlnAccuracy}
\end{figure}

Fig.~\ref{fig:fluidNlnAccuracy} reports the relative prediction error.
The temporal pattern of the error curves closely mirrors that of the projection error in Fig.~\ref{fig:fluidNlnOrthNorms}: all methods perform comparably before the lock-in event, experience a sharp increase at the onset of lock-in, and then diverge. During the transition, the global basis ROMs provide the best accuracy. After the transient, the proposed method with nonlinear regression continues to improve and achieves the lowest and most consistently decreasing error, confirming that the combination of a geometrically adapted subspace and an expressive latent-space regressor is necessary to track the post-lock-in dynamics.The proposed approach outperforming rDMDc suggests that the strategy of accumulating $K$ snapshots before activating the update helps robustify accuracy against the more aggressive, single-snapshot recursive updates of rDMDc.

Fig.~\ref{fig:fluidNlnAccuracy} also includes a variant of the proposed method in which GROUSE is replaced by the PAST algorithm for the subspace tracking step. An apparently paradoxical result emerges: when considering the projection error alone (Fig.~\ref{fig:fluidNlnOrthNorms}), PAST ultimately achieves a lower residual than GROUSE by nearly an order of magnitude, suggesting a better subspace fit. Yet this advantage does not translate into prediction accuracy. In Fig.~\ref{fig:fluidNlnAccuracy}, the PAST-based variant exhibits large oscillations and occasionally gives a much higher error than the GROUSE-based method.

\begin{figure}
    \centering
    \includegraphics[width=\textwidth]{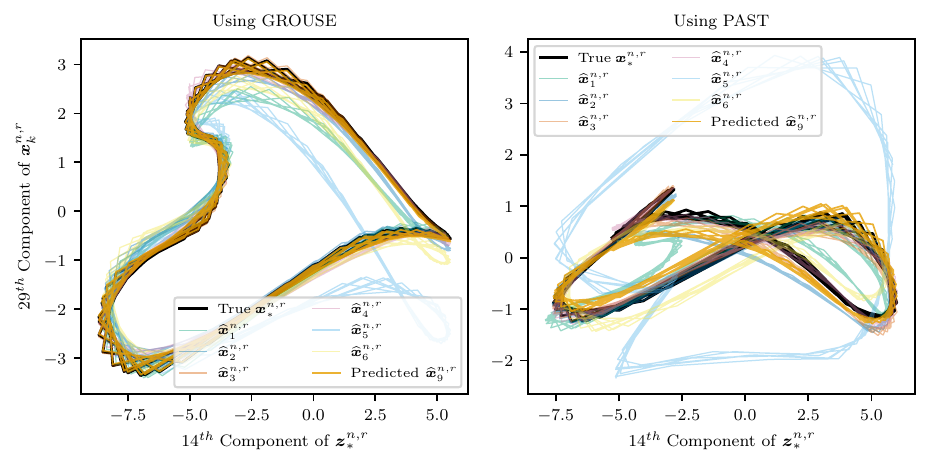}
    \caption{[VIV] -- Aligned reduced predictions $\widehat{\boldsymbol{x}}_k^{n,r} = \boldsymbol{Q}_k \mathcal{I}_k(\boldsymbol{z}^n)$ (last 100 steps, components 14 and 29) using GROUSE (left) and PAST (right). With GROUSE, the individual predictions cluster around the true trajectory and their weighted combination (orange) is accurate. With PAST, the loss of orthonormality ($\|\boldsymbol{\Phi}^T\boldsymbol{\Phi} - \boldsymbol{I}_r\| = 3.62$) corrupts the Procrustes alignment, scattering the predictions and degrading the combined output.}
    \label{fig:compareGrouseAndPast}
\end{figure}

The explanation lies in the loss of orthonormality. The PAST algorithm does not enforce $\boldsymbol{\Phi}^T \boldsymbol{\Phi} = \boldsymbol{I}_r$; in our experiment, $\| \overline{\boldsymbol{\Phi}}^T \overline{\boldsymbol{\Phi}} - \boldsymbol{I}_r \|_F = 3.2$ for PAST at the last online snapshot, compared to $4.07 \times 10^{-12}$ for GROUSE. This deviation corrupts the two operations that rely on exact orthonormality: (i) the Procrustes alignment matrices $\boldsymbol{Q}_k$, computed from the SVD of $\boldsymbol{\Phi}_*^T \boldsymbol{\Phi}_k$, and (ii) the Grassmann distance-based weights $w_k$. The effect is visualised in Fig.~\ref{fig:compareGrouseAndPast}: with GROUSE, the aligned individual predictions $\widehat{\boldsymbol{x}}_k^{n,r}$ ( especially for the trailing POD coefficients ) cluster tightly and the weighted combination nearly coincides with the true trajectory; with PAST, the individual predictions are scattered, reflecting erroneous alignments, and the resulting combination deviates significantly from the reference.

This comparison empirically validates the design choice discussed in Section~\ref{subspace_update_subsection}: within the proposed framework, exact orthonormality of the tracked basis is necessary for the downstream Procrustes alignment and Grassmann weighting to function correctly. In the remainder of this work, only the GROUSE-based variant is used.

\subsection{FSI Lid driven cavity}

The second test case is a lid-driven cavity with a flexible bottom wall, a classical FSI benchmark that combines recirculating flow with structural deformation. Unlike the VIV case, here the ROM is deployed as a predictor within the partitioned FSI coupling scheme described in Section~\ref{section3}, and we evaluate both ROM accuracy and the resulting computational speedup.

The simulations were executed sequentially (with NumPy and SciPy operations relying on OpenBLAS for low-level computations) on a laptop equipped with an Apple M1 chip (8 cores) and 16 GB of memory.

\subsubsection{Problem setup}

\begin{figure}
    \centering
    \includegraphics[width=.6\linewidth]{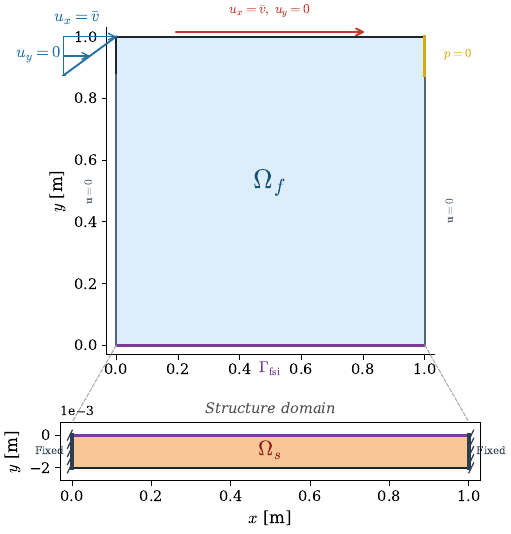}
    \caption{[Lid Driven Cavity] -- Geometry and boundary conditions of the wall-driven cavity problem with a flexible bottom.}
    \label{fig:LidGeometry}
\end{figure}

The test case, introduced in~\citep{Wall2000} and detailed in~\citep{dissertation-2117-94177}, is a modified lid-driven cavity with a flexible bottom wall; see Fig.~\ref{fig:LidGeometry} for the geometry and boundary conditions. The configuration departs from the classical rigid-walled benchmark in two respects: the originally constant top lid velocity is replaced by an oscillatory one, and the inlet region is extended from the top corners over a horizontal strip of height $0.125\,L$, where $L = 1$~m is the cavity side length. The imposed inlet velocity profile reads
\begin{equation}
    \bar{v} = 1 - \cos\!\left(\frac{2\pi t}{5}\right).
\end{equation}

The fluid domain is governed by the incompressible Navier--Stokes equations in an Arbitrary Lagrangian--Eulerian (ALE) frame, with a Newtonian constitutive law. The spatial discretization uses stabilized equal-order linear finite elements based on a Variational Multiscale (VMS) formulation, and time integration is performed with the second-order BDF2 scheme. Mesh deformation is handled by fictitious linear elastic solid elements, with the mesh velocity discretized using the $2^{nd}$ order Bossak scheme \citep{bossak}, matching the time integration scheme on the solid domain, which enforces velocity continuity at the fluid--structure interface. The reference fluid parameters are a density $\rho_\mathrm{f} = 1~\mathrm{kg/m^3}$ and a dynamic viscosity $\mu = 10^{-2}~\mathrm{Kg/ms}$.

The structural domain is modelled as a 2D plane-stress solid with a Saint-Venant--Kirchhoff constitutive law and large-displacement kinematics, discretized with bilinear quadrilateral (Q1) elements (two elements through the thickness).
The structural properties are: density $\rho_s = 500~\mathrm{kg/m^3}$, elastic modulus $E = 250~\mathrm{Pa}$, and Poisson ratio $\nu = 0$.

The two solvers are coupled in a partitioned Gauss--Seidel scheme, accelerated by the IQNILS quasi-Newton method \citep{degroote_performance_2009}. The FSI convergence criterion (line~\ref{alg:tolerance}, Algorithm~\ref{globalAlg}) is set to $e = 10^{-7}$. The reference time step is $\Delta t_\mathrm{ref} = 0.01~\mathrm{s}$; the influence of $\Delta t$ on the computational speedup is investigated in the FSI convergence acceleration study below.

As indicated in (\ref{eq:equivStateFSI}), the state-space considered here is the discretized fluid forces (with $X$ and $Y$ components) on the interface of dimension $N_x = 62$. The forces' reduced basis has a dimension of $r = 8$, covering $\epsilon=99.99\%$ of the energy. The solid encoder is used with a latent dimension of $N_u = r_S = 26$, corresponding to $\epsilon=99.9999\%$\footnote{Higher values are used for $\mathcal{E}_S$ for the sole reason of testing the performance of the proposed ROM with higher dimensions of the control input.}. The parameter space is two-dimensional, $\boldsymbol{\theta} = [\rho, \mu]$, where $\rho$ is the fluid density and $\mu$ the dynamic viscosity. We use $p = 15$ training configurations (detailed in Table \ref{tab:parameters},  Appendix~\ref{appendix:liddriven}) and predict for $110$ unseen parameters, some of which are outside the convex hull of the training samples (See Fig. \ref{fig:lidDrivenGrid}). For the ROM results, we also use the test parameter $\boldsymbol{\theta}_* = [1~\mathrm{kg/m^3},\; 10^{-2}~\mathrm{Kg/ms}]$ that corresponds to the value of the parameter used in the benchmark \citep{Wall2000, dissertation-2117-94177}. Each one of the $p$ training configurations is built from $200$ time steps of the reference simulation up to $t = 90~s$, together with the intermediate fixed-point iterations, generating a number of snapshots in the range $m_k \in [ 719, ~1168 ]\; \forall k \in \{1, \cdots p \}$.
Fig.~\ref{fig:LidDRivencompareReducedSnaps3D} visualises the trajectories of the training snapshots before and after the Procrustes alignment operations.
The remaining hyperparameters are set to $z = 2$, $d = 2100$, $\tau = 50$, and $K = 120$. Here, we use a fixed $\xi = 0.5$: the adaptive schedule of Eq.~(\ref{eq:xi_schedule}) was also tested for this case but yielded a marginally lower accuracy, so the simpler fixed value was retained. The nonlinear regression is a $2^{nd}$ order polynomial regression. Unless indicated otherwise, the reported results of this ROM are based on the test parameter $\boldsymbol{\theta}_*$ of the reference benchmark. The test phase is evaluated over the time interval immediately following the transient response and extending to $t=100~s$, with a slight extrapolation over the final $10 ~s$.

\subsubsection{Subspace evolution and projection error}

The temporal evolution of the adapted subspace is reported in Fig.~\ref{fig:totalAngleRotationLidDriven}. The accumulated subspace rotation (Fig.~\ref{fig:totalAngleRotationLidDriven}) shows a sharp increase when encountering the first snapshots of the simulation, before stabilising around $6^{\circ}$ as the flow generates similar snapshots during its periodic regime. The total rotation remains moderate, indicating that the Grassmann-predicted initial basis already provides a good starting point for this parameter configuration. In Fig.~\ref{fig:modesComparisonLidDriven}, we show a comparison of the forces modes before and after the basis updates.

The relative projection error (Fig.~\ref{fig:LidDrivenOrthNorms}) confirms the same trends observed in the VIV case: the proposed method with GROUSE updates consistently achieves a projection errors as small as $10^{-6}$, lower than both the global static and local static baselines.
Due to the sequential nature of the updates, the projection error exhibits oscillations inherent to single-sample gradient updates, which nonetheless remain strictly below the errors using the static bases at every step.

\subsubsection{Latent space predictions}

Fig.~\ref{fig:compareLidDrivenReducedPrediction} shows the individual aligned predictions $\boldsymbol{Q}_k \mathcal{I}_k(\boldsymbol{z}^n)$ from each locally trained regression alongside the combined prediction and the reference trajectory. The Procrustes alignment brings the individual predictions into a common coordinate frame, and in this case, the right figure  shows the importance of the choice of the geometric proximity weights. A training subspace that is close to $\mathcal{V}_*^0$ in the Grassmannian sense will receive a larger weight, and thus contribute more strongly to the final prediction. The resulting weighted combination closely follows the true reduced trajectory.

\subsubsection{Full-field prediction accuracy}

The relative prediction error is reported in Fig.~\ref{fig:lidDrivenAccuracy}. The proposed method with nonlinear regression achieves the lowest error throughout the simulation, with a consistent improvement over both global and local static baselines. The error $e^n$ reaches $10^{-4}$ at the end of the online phase. The basis update mechanism allows the ROM to progressively refine its predictions as more HF data becomes available, resulting in a decreasing error trend that is less present for the static bases methods.

\subsubsection{FSI convergence acceleration}

\begin{figure}
    \centering
    \includegraphics[width=\textwidth]{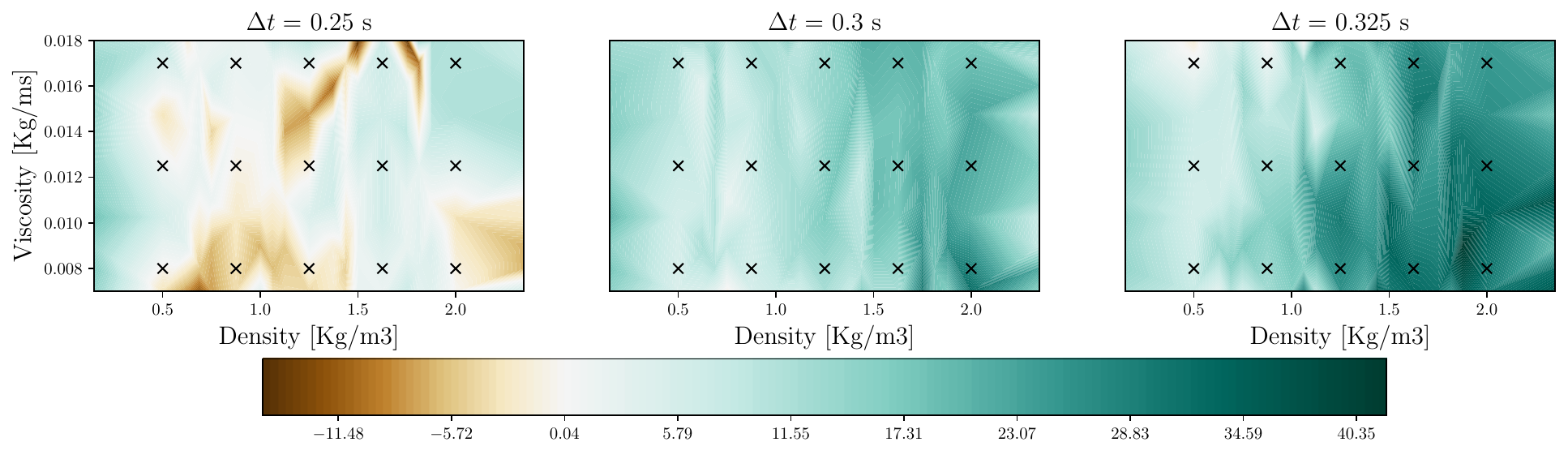}
    \caption{Percentage of iterations gained compared to the quadratic predictor when comparing in terms of 3 different time steps. The training $\boldsymbol{\theta}_k$ are represented by the cross marks.}
    \label{fig:comparisonLidDrivenTimeSteps}
\end{figure}

\begin{figure}
    \centering
    \includegraphics[width=\textwidth]{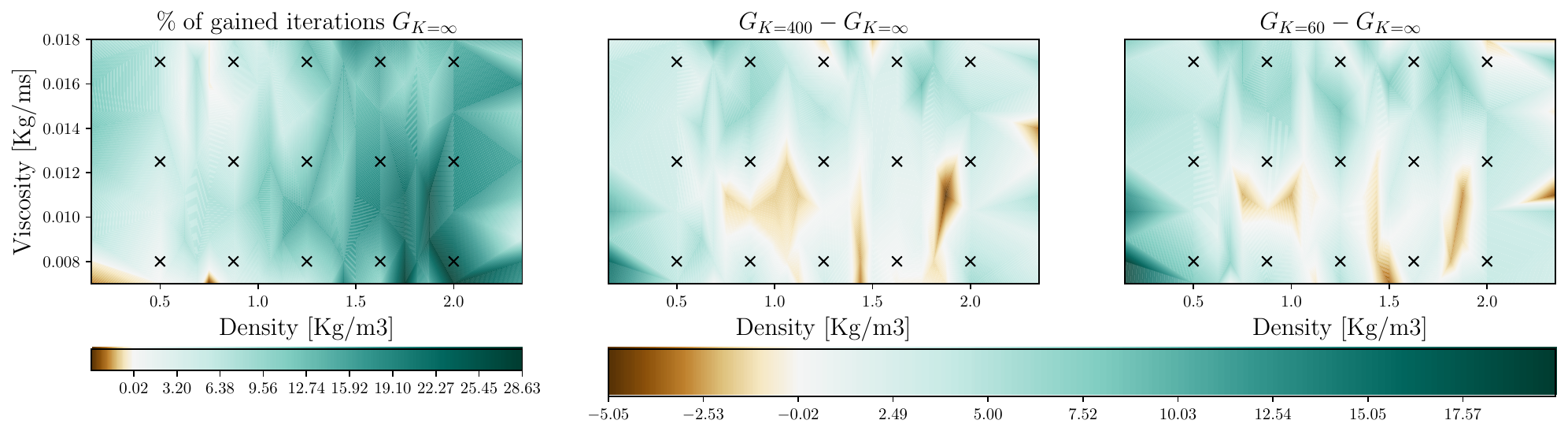}
    \caption{Percentage of iterations gained compared to the quadratic predictor when comparing in terms of 3 different values of the subspace update frequency $K$. The training $\boldsymbol{\theta}_k$ are represented by the cross marks. Case of $\Delta t = 0.3s$.}
    \label{fig:comparisonLidDriven}
\end{figure}

We now turn to the evaluation of the ROM-based predictor in terms of FSI iteration savings. The percentage of iterations gained compared to a standard quadratic extrapolation predictor ($\boldsymbol{f}^{0, n} = 3\boldsymbol{f}^{n-1} - 3 \boldsymbol{f}^{n-2} + \boldsymbol{f}^{n-3} $) is reported for different time step sizes in Fig.~\ref{fig:comparisonLidDrivenTimeSteps} and for different subspace update frequencies $K$ in Fig.~\ref{fig:comparisonLidDriven}.

Fig.~\ref{fig:comparisonLidDrivenTimeSteps} shows that the iteration gain depends on the time step size $\Delta t$: larger time steps lead to less accurate finite differences-based extrapolators, which in turn increases the relative benefit of a better initial guess. For the largest time step considered, the ROM predictor consistently saves up to 40\% of the coupling iterations across the parameter space. We can also observe that the gain is highest for higher fluid density values, as those cases are more strongly coupled due to the added mass effect and therefore require more fixed-point iterations, amplifying the relative benefit of a better initial guess.

Fig.~\ref{fig:comparisonLidDriven} illustrates the sensitivity to the subspace update frequency $K$ for $\Delta t = 0.3~s$. The metric is $G_K$, the percentage of gained iterations associated to a frequency $K$. Indeed, subspace updates allow for a slightly bigger gain. Both the two tested values of $K$ yield a substantial iteration gain over the quadratic predictor, with moderate differences between them, confirming that the method is not overly sensitive to this hyperparameter.

\begin{figure}
    \centering
    \includegraphics[width=.6\textwidth]{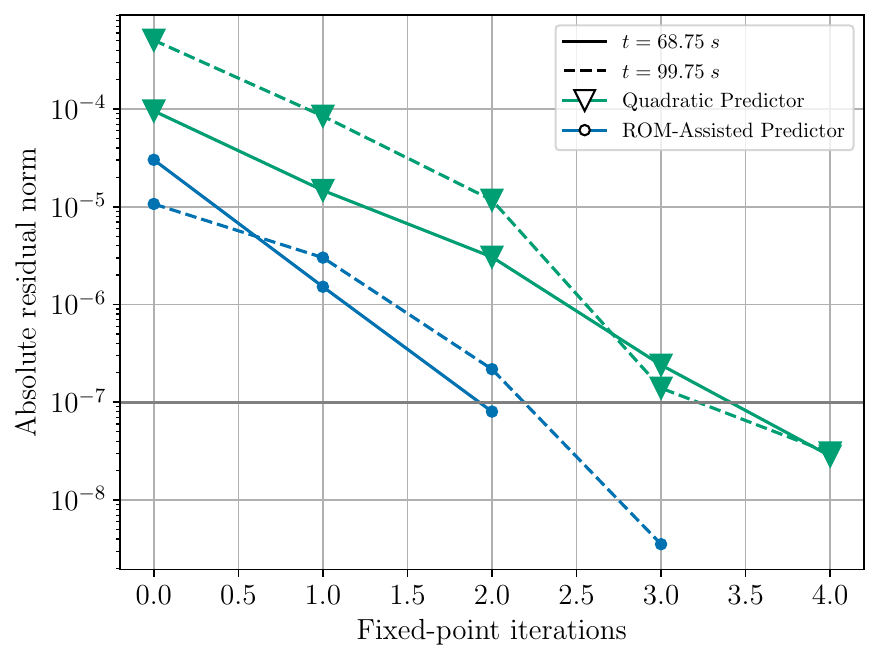}
    \caption{Convergence of the FSI fixed-point iterations at two time steps, $t = 68.75~s$ and  $t = 99.75~s$. The ROM-assisted predictor results in better initial guesses, as shown from the lower residual norms.}
    \label{fig:lidDrivenConvergenceResiduals}
\end{figure}

We show in Fig. \ref{fig:lidDrivenConvergenceResiduals} two examples on two time steps of the faster convergence obtained with the better initial guess from the ROM-assisted predictor.

\begin{figure}
    \centering
    \begin{subfigure}[t]{0.46\columnwidth}
        \centering
        \includegraphics[width=\linewidth]{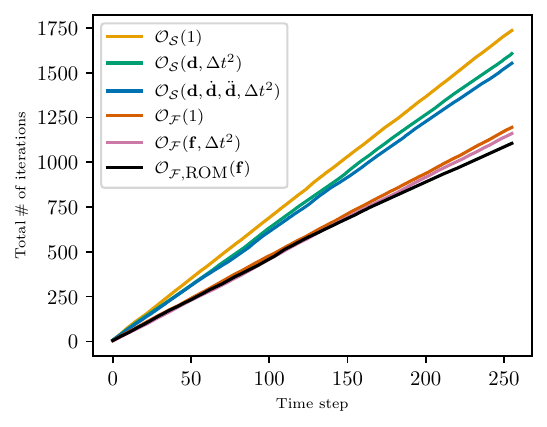}
        \caption{Total number of iterations using different schemes with $\Delta t = 0.25 s$. In this case a modified coupling convergence criterion is used, considering both the interface forces and the displacements.}
        \label{fig:lidDrivenSchemeComparison}
    \end{subfigure}
    \hfill
    \begin{subfigure}[t]{0.53\columnwidth}
        \centering
        \includegraphics[width=\linewidth]{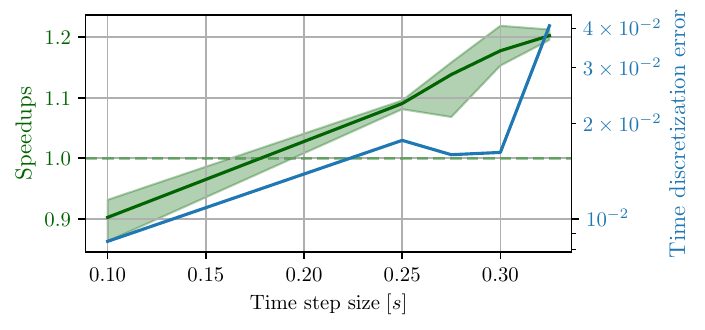}
        \caption{Speedups depending on the time step size. Shaded area indicates the 10th--90th percentile across runs.}
        \label{fig:lidDrivenSpeedups}
    \end{subfigure}
    \caption{Speedups of the proposed ROM-assisted predictors in terms of (a) number of coupling iterations and (b) the wall time speedup depending on the time step.}
    \label{fig:newFig}
\end{figure}

In order to ensure comparisons with the strongest baselines possible, we compare the total number of FSI iterations accumulated over the simulation for the different coupling schemes\footnote{We only compare here between different "sequential" Gauss-Seidel schemes. It is known from the literature \citep{degroote_performance_2009, uekermann_partitioned_2016} that the block and the parallel Jacobi schemes give similar or worse converge behaviour respectively.} and different predictors, summarized in Table \ref{tab:differentSchemes}. The accumulated iterations are shown in Fig.~\ref{fig:lidDrivenSchemeComparison}. We note that in this comparison, we modify the convergence criterion on the fluid forces in line \ref{alg:tolerance}, Algorithm \ref{globalAlg} by adding an additional criterion (residual tolerance of $10^{-6}$) on the interface displacements as well, the goal being to have a valid comparison. When comparing between the different schemes and predictors, the proposed ROM-assisted predictor, applied on the Gauss-Seidel scheme $\mathcal{F}\circ\mathcal{S}$ yields the lowest iteration count.

\begin{table}
\centering
\small
\setlength{\tabcolsep}{4pt}
\renewcommand{\arraystretch}{1.2}

\begin{tabularx}{\linewidth}{
|>{\raggedright\arraybackslash}p{1.8cm}
|>{\centering\arraybackslash}p{0.8cm}
|>{\centering\arraybackslash}p{3.1cm}
|>{\centering\arraybackslash}X
|>{\centering\arraybackslash}p{0.8cm}
|>{\centering\arraybackslash}p{3.1cm}|}
\hline
\textbf{Coupling} &
\multicolumn{3}{c|}{$\mathcal{S}\circ\mathcal{F}$} &
\multicolumn{2}{c|}{$\mathcal{F}\circ\mathcal{S}$} \\
\hline
\textbf{Notation}
& $\mathcal{O}_{\mathcal{S}}(1)$
& $\mathcal{O}_{\mathcal{S}}(\boldsymbol{d}, \Delta t^2)$
& $\mathcal{O}_{\mathcal{S}}(\boldsymbol{d}, \dot{\boldsymbol{d}}, \Delta t^2)$
& $\mathcal{O}_{\mathcal{F}}(1)$
& $\mathcal{O}_{\mathcal{F}}(\boldsymbol{f}, \Delta t^2)$ \\
\hline
\textbf{Prediction}
& $\boldsymbol{d}^{n-1}$
& $3\boldsymbol{d}^{n-1}-3\boldsymbol{d}^{n-2}+\boldsymbol{d}^{n-3}$
& $\boldsymbol{d}^{n-1}
   +\Delta t\,\dot{\boldsymbol{d}}^{n-1}
   +\frac{\Delta t^2}{2}\ddot{\boldsymbol{d}}^{n-1}$ 
& $\boldsymbol{f}^{n-1}$
& $3\boldsymbol{f}^{n-1}-3\boldsymbol{f}^{n-2}+\boldsymbol{f}^{n-3}$ \\
\hline
\end{tabularx}

\caption{Prediction strategies used in the different coupling schemes. The main practical difference between the two coupling schemes is the order of calls to the fluid and solid solver. $\mathcal{F}\circ\mathcal{S}$ calls the solid solver first, as opposed to $\mathcal{S}\circ\mathcal{F}$ that calls the fluid solver first.}
\label{tab:differentSchemes}
\end{table}

The computational speedups of the proposed ROM predictor depend on the selected time step, with larger time steps leading to greater computational speedups. It is important to note that the proposed ROM strategy works with no accuracy loss, it does not introduce any additional error beyond that of the time integration itself. To assess the viability of our approach, a reference solution is first computed using a sufficiently small time step; $\Delta t = 0.01~s$.

In Table \ref{tab:benchmark_comparison}, the reference solution is further validated against the benchmark results in \citep{dissertation-2117-94177}. In Fig. \ref{fig:comparingReferenceToOurs}, we compare the time evolution of the vertical displacement of the flexible structure's bottom midpoint, obtained with the coarsest time grid considered here ($\Delta t = 0.325~s$) and the reference solution. 

The accuracy of a target time step ($\Delta t$) is then evaluated by comparing three characteristic quantities of the structural response with those of the reference solution: (i) the mean displacement, (ii) the maximum oscillation, and (iii) the oscillation period. The relative error associated with each quantity is computed, and a composite error indicator is defined as the arithmetic mean of these three relative errors,
\begin{equation}
    E_{\Delta t}=\frac{E_{\mathrm{mean}}+E_{\mathrm{max}}+E_{\mathrm{period}}}{3}.
\end{equation}
In Fig. \ref{fig:lidDrivenSpeedups}, this error indicator is then considered together with the computational speedup obtained using the proposed coupling scheme. The objective here to identify a range of time steps for which the temporal discretization error remains within an acceptable level; below $4\%$, while the proposed scheme yields reasonable computational savings (a speedup of $20\%$). This analysis highlights the practical operating region in which the proposed approach offers the most favorable compromise between solution accuracy and computational efficiency.

\subsubsection{Wall-time analysis}

\begin{figure}
    \centering
    \includegraphics[width=\textwidth]{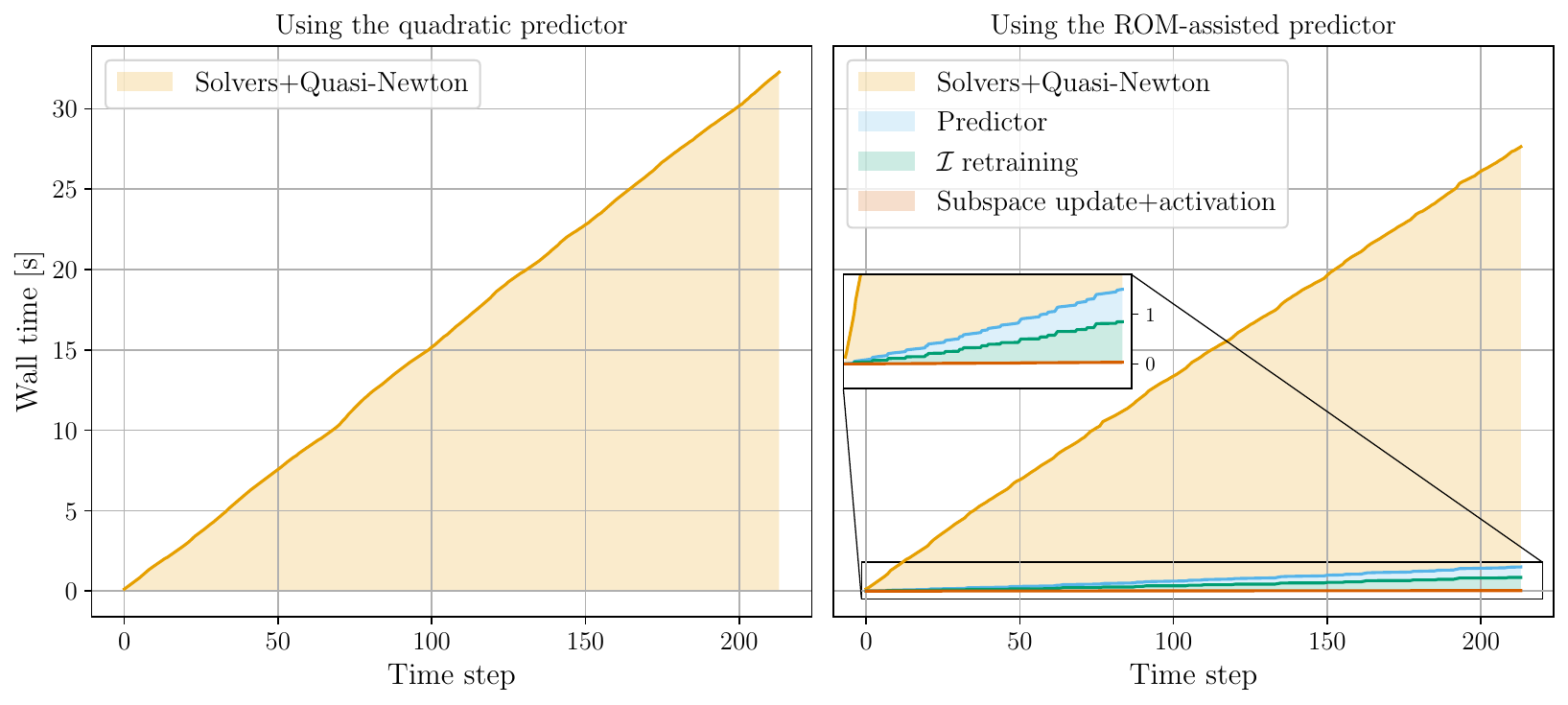}
    \caption{[Lid Driven Case] - Wall time breakdown. The wall times are computed as the median across 10 runs.}
    \label{fig:lidDrivenTimeProfiling}
\end{figure}

Fig.~\ref{fig:lidDrivenTimeProfiling} presents a breakdown of the wall-clock time into its main components: HF fluid and solid solver calls, ROM prediction (including the reduced coupling iterations), regression retraining, and subspace update and activation operations. The ROM-related overhead (prediction, retraining, and GROUSE updates) represents a negligible fraction of the total simulation time compared to the HF solver calls. The net effect is a wall-time speedup that is directly proportional to the iteration savings. In this specific case with $\Delta t = 0.3~s$, that speedup has a median of $14.36~\%$ across 10 runs. For comparison, the training wall time in this case is around $4.8~s$.

\subsection{Turek Hron Benchmark}

The simulations of the Turek--Hron benchmark were executed on a single Linux node equipped with four Intel Xeon E7-8867 v4 processors (18 cores per socket, 2.40 GHz base frequency) and 378 GB of RAM. All results were obtained using 10 shared-memory threads for the matrix assembly and linear solver.

\subsubsection{Problem setup}

\begin{figure}
    \centering
    \includegraphics[width=0.6\textwidth]{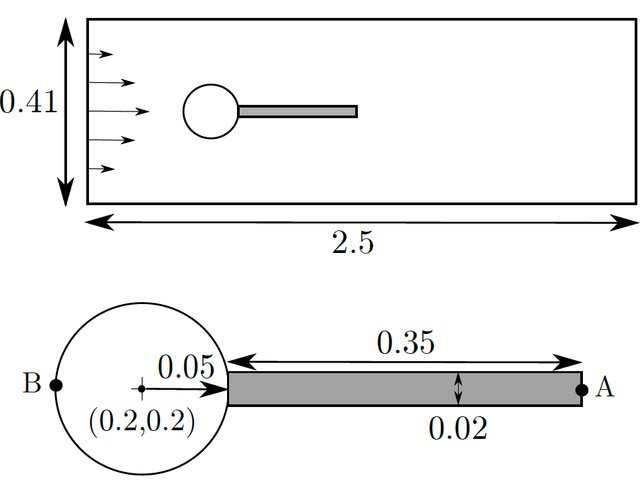}
    \caption{Geometry of the Turek-Hron benchmark problem (dimensions in $[m]$). From \citep{margetts}.}
    \label{fig:turekHronGeomDims}
\end{figure}

The test case is the well-known Turek--Hron FSI benchmark \citep{turekBench}, consisting of a thin elastic flap attached behind a rigid cylinder in a channel flow; see Fig.~\ref{fig:turekHronGeomDims} for the geometry, and Fig.~\ref{fig:turekHronSnap} for a snapshot solution at $t=12.536~s$. The fluid domain follows the same ALE, VMS-stabilized, equal-order finite element formulation and BDF2 time integration used for the lid driven cavity case, with mesh deformation similarly handled through fictitious linear elastic solid elements and a matching Bossak scheme \citep{bossak} for the mesh velocity. The structural domain is here modelled under a plane-strain with a Saint-Venant--Kirchhoff constitutive law and large-displacement kinematics; the structural parameters are density $\rho_s = 10\,000~\mathrm{kg/m^3}$, elastic modulus $E = 1.4\times10^{6}~\mathrm{Pa}$, and Poisson ratio $\nu = 0.4$. The fluid density is fixed at $\rho_\mathrm{f} = 1000~\mathrm{kg/m^3}$. A parabolic velocity profile is imposed at the left channel face, ramping up to a maximum of $1.5~\mathrm{m/s}$. The fluid and solid meshes are non-matching at the fluid--structure interface, and the mapping used for the communication of the fields is a nearest-element algorithm following \citep{dissertationWang}.

The two solvers are again coupled in a partitioned Gauss--Seidel scheme accelerated by the IQNILS quasi-Newton method \citep{degroote_performance_2009}. The FSI convergence criterion (line~\ref{alg:tolerance}, Algorithm~\ref{globalAlg}) is a tolerance on the fluid forces of $10^{-6}$. The time step is fixed at $\Delta t = 0.01~\mathrm{s}$ throughout this test case.

The dimension of the state-space is the number of degrees of freedom of the fluid forces $N_x = 898$. The parameter space is again two-dimensional, $\boldsymbol{\theta} = [c_v, \mu]$, where $c_v$ is a coefficient multiplying the amplitude of the imposed inlet velocity and $\mu$ is the fluid dynamic viscosity in $\mathrm{Kg/ms}$. The reference benchmark configuration corresponds to $\boldsymbol{\theta}_* = (1, 1)$. We use $p = 4$ training configurations, taken at the corners of the square $[0.9, 1.1]\times[0.9,1.1]$ surrounding $\boldsymbol{\theta}_*$ (detailed in Table~\ref{tab:turekParameters}, Appendix~\ref{appendix:turek}), each obtained from the reference simulation run up to $t=20~\mathrm{s}$, generating snapshots of sizes $m_k \in [ 3816, ~4686 ]\; \forall k \in \{1, \cdots p \}$. The mean training wall time across the four configurations is $33.44~\mathrm{s}$. We predict for $5$ test configurations: the benchmark parameter $\boldsymbol{\theta}_*$ itself, together with the four midpoints of the square's edges, namely $(0.9, 1)$, $(1.1, 1)$, $(1, 0.9)$, and $(1, 1.1)$. Unless indicated otherwise, the reported results of this ROM are based on the benchmark test parameter $\boldsymbol{\theta}_*$.

As indicated in (\ref{eq:equivStateFSI}), the state-space is again the discretized fluid forces at the interface. The fluid basis rank is set through the energy criterion to $r=18$, corresponding to an energy threshold of $99.99\%$, and the solid encoder latent dimension is set to $N_u = r_S = 60$. 
Fig.~\ref{fig:TurekcompareReducedSnaps3D} (Appendix~\ref{appendix:turek}) shows the parametric trajectories in the reduced coordinate spaces after Procrustes alignment.
The remaining hyperparameters are set to $z = 2$, $\xi = 0.5$, $d = 11000$, $\tau = 60$, and $K = 200$.

\subsubsection{Subspace evolution and projection error}

The subspace evolution and projection error metrics (Figures~\ref{fig:combinedAngleShiftOnlineTurek} and~\ref{fig:TurekOrthNorms}, Appendix~\ref{appendix:turek}) follow trends similar to those observed for the previous two test cases. The maximum accumulated subspace angle reaches $20^{\circ}$ here, higher than in the lid driven cavity case, reflecting a larger geometric adaptation of the fluid subspace over the course of the online simulation.

\subsubsection{Latent space predictions}

Fig.~\ref{fig:compareTurekReducedPrediction} (Appendix~\ref{appendix:turek}) compares the individual aligned predictions of each locally trained regression with the combined prediction and the reference full-order reduced trajectory. The figure exhibits the same qualitative behaviour already discussed for the previous test cases.

\subsubsection{Full-field prediction accuracy}

Fig.~\ref{fig:turekAccurac} (Appendix~\ref{appendix:turek}) reports the relative prediction error for the compared methods. Nearly the same trends as for the previous test cases are observed; the main difference here is that the proposed method's error only reaches parity with the {Global static (nonlinear)} baseline towards the end of the online simulation, remaining slightly higher for most of it. This is attributed to the more limited amount of online data available to the proposed method, given the relatively simple, near-periodic oscillatory nature of the flap dynamics in this test case.

\subsubsection{FSI convergence acceleration}

\begin{figure}
    \centering
    \includegraphics[width=.9\textwidth]{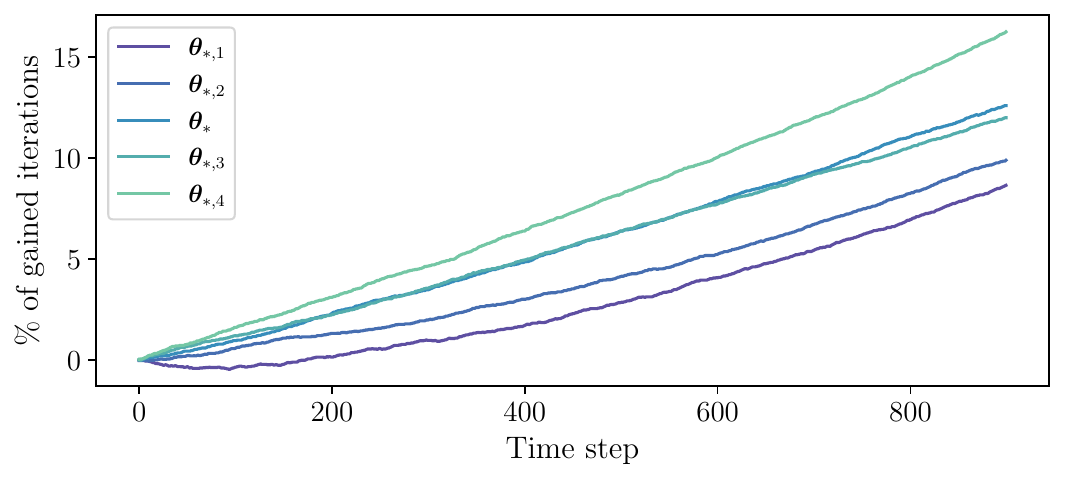}
\caption{[Turek Hron case] - $\%$ of gained iterations for the different unseen parameters $\boldsymbol{\theta}_{*}$. $\boldsymbol{\theta}_{*, 1} = (0.9, 1)$, $\boldsymbol{\theta}_{*, 2} = (1, 0.9)$, $\boldsymbol{\theta}_{*, 3} = (1, 1.1)$,
$\boldsymbol{\theta}_{*, 4} = (1.1, 1)$ and $\boldsymbol{\theta}_{*} = (1, 1)$.}
    \label{fig:turekGain}
\end{figure}

Fig.~\ref{fig:turekGain} shows the evolution of the percentage of FSI iterations gained, compared to the quadratic extrapolation predictor, for each of the five test parameters. All five runs show a net gain by the end of the simulation, ranging between a minimum of about $8\%$ and a maximum of about $16\%$.

\begin{figure}
    \centering
    \includegraphics[width=.6\textwidth]{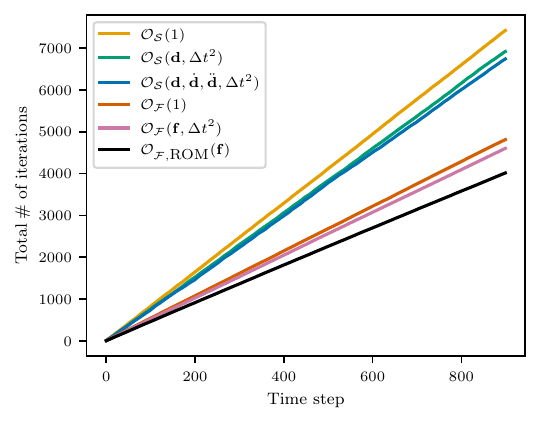}
    \caption{[Turek Hron case] - Total number of iterations using different schemes with $\Delta t = 0.01 s$.}
    \label{fig:turekSchemeComparison}
\end{figure}

We further compare the accumulated number of FSI iterations across the different coupling schemes and predictors summarized in Table~\ref{tab:differentSchemes}. For this comparison, the convergence criterion is additionally augmented with a tolerance on the interface displacements, using an absolute residual norm of $10^{-7}$. Fig.~\ref{fig:turekSchemeComparison} shows that, as in the lid driven cavity case, the proposed ROM-assisted predictor applied on the Gauss--Seidel scheme $\mathcal{F}\circ\mathcal{S}$ yields the lowest iteration count among the five compared schemes and predictors. A comparison of our reference simulation with that of the benchmark in terms of the vertical displacement is reported in Fig \ref{fig:comparingReferenceToOursTurek} .

\subsubsection{Wall-time analysis}

\begin{figure}
    \centering
    \includegraphics[width=\textwidth]{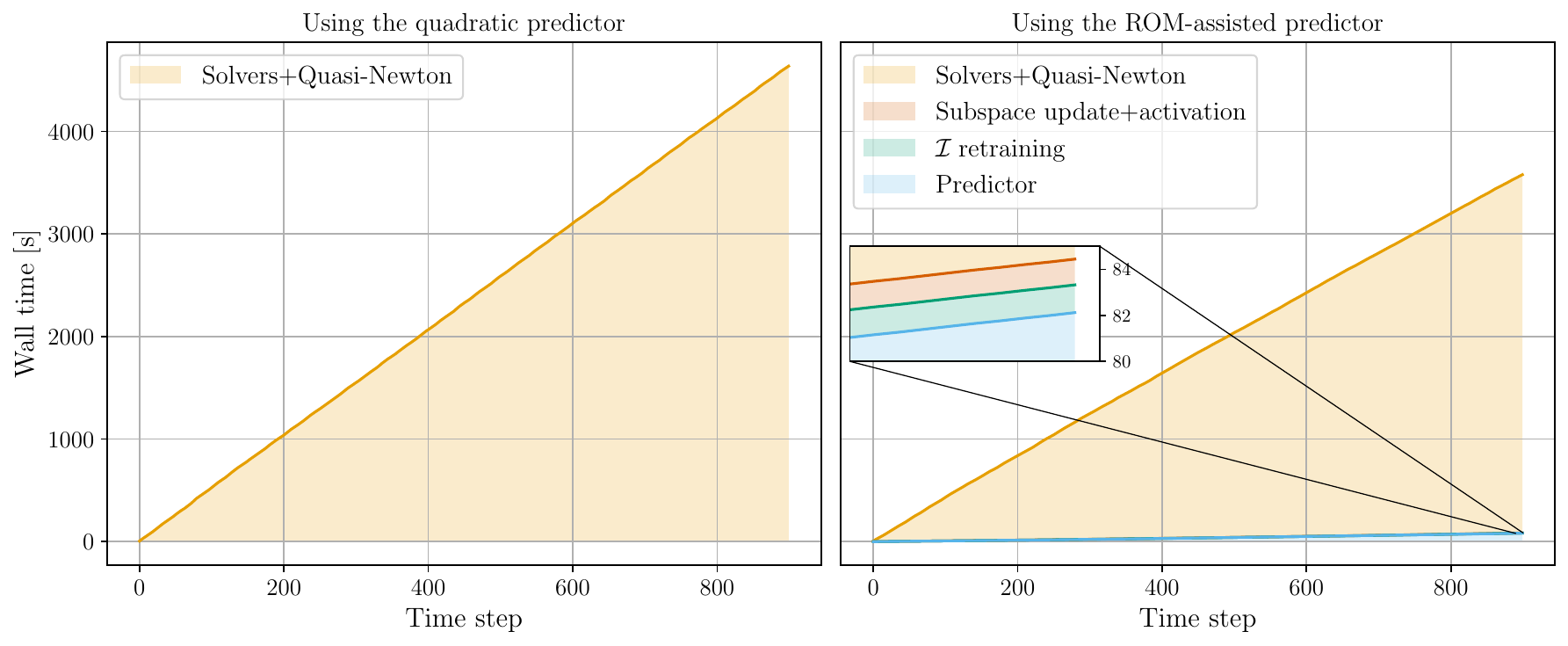}
    \caption{[Turek-Hron Benchmark] - Wall time breakdown. The wall times are computed as the median across 10 runs.}
    \label{fig:turekTimeProfiling}
\end{figure}

Fig.~\ref{fig:turekTimeProfiling} presents the wall-clock time breakdown for the benchmark test parameter $\boldsymbol{\theta}_*$. The ROM-related overhead is negligible compared to the full-order solver calls, and the resulting wall-time speedup has a median of $22.83\%$ across 10 runs. We recall that the training wall time is around $33.44~\mathrm{s}$ for this case, which is practically negligible compared to the simulation time.

\section{Conclusion}\label{section5}
Non-intrusive reduced order models offer the advantages of modularity and flexibility when used with black-box solvers. However, being constrained to purely data-driven latent dynamics penalizes them with poor generalization and reduced accuracy for strongly nonlinear problems. The current work addresses this challenge from two complementary angles. The first is the adoption of online adaptive subspaces, a strategy that has recently gained traction but has been developed almost exclusively for projection-based ROMs. This paper extends it to the non-intrusive setting: through a strategy combining Grassmann interpolation, Grassmann-weighted aggregate predictions in the reduced space, and geodesic subspace updates, the proposed approach adapts both the ROM subspace and the latent dynamics independently, and, notably, does so without requiring the storage of streaming high-dimensional data.
The second angle exploits the partitioned FSI coupling context, where these ROMs are continuously and naturally fed full-order data, both as regression input and as the source of online updates, at no additional data-acquisition cost. In that setting, the ROM predicts accurate initial guesses for the coupling iterations, achieving computational speedups with no loss of accuracy.

Evaluation on three FSI test cases showed enhanced predictive capability compared with global and local static-basis ROMs. In terms of computational gain, the ROM-assisted predictor achieved wall-time speedups up to $23~\%$ (depending on how large the time step is), while the offline training and online adaptation overhead remained negligible relative to the cost of the full-order solver calls.

This work makes a step towards hybrid, online-adaptive, non-intrusive ROMs with solid accuracy and robust generalization, and motivates two directions for further work: evaluating the approach on more challenging transport-dominated problems, and devising strategies that automatically decide when to trigger an update by reacting to the FOM evolution, rather than relying on user-defined frequencies.

\section*{Acknowledgements}

The authors would like to thank Thibault Dairay and Mohamed Elsherbiny for valuable discussions and insightful comments related to this work.

\newpage

\appendix
\section{GROUSE}

In this appendix, we recall the GROUSE rank-1 update in Algorithm \ref{alg:rank1-update}.
\begin{algorithm}
\SetKwInput{KwData}{Input}
\caption{Rank-one GROUSE update of the subspace~\citep{balzano2010online}}\label{alg:rank1-update}
\KwData{
\begin{itemize}[noitemsep, topsep=0pt, parsep=0pt, partopsep=0pt]
    \item New HF snapshot $\boldsymbol{x}^n$.
    \item Previous $\overline{\boldsymbol{\Phi}_{*}}^n$
\end{itemize}}

\KwResult{
Updated subspace $\overline{\boldsymbol{\Phi}}_{*}^n$
}

\hrulefill

    $\boldsymbol{w} = \overline{\boldsymbol{\Phi}}_{*}^{n\,T}  \boldsymbol{x}^n$
    
    $\boldsymbol{p} = \overline{\boldsymbol{\Phi}}_{*}^n  \boldsymbol{w}$
    
    $\boldsymbol{r} = \boldsymbol{x}^n - \boldsymbol{p}$
    
    
    $\rho_n = \text{asin}(||\boldsymbol{r}||_2/||\boldsymbol{x}||_2)$~\citep{balzano2015local}
    
    $\overline{\boldsymbol{\Phi}}_{*}^n \leftarrow \overline{\boldsymbol{\Phi}}_{*}^n + \left( (cos( \rho_n ) - 1 ) 
    \frac{\boldsymbol{p}}{||\boldsymbol{p}||_2} + \frac{\boldsymbol{r}}{||\boldsymbol{x}||_2} \right)   \frac{\boldsymbol{w}^T}{||\boldsymbol{w}||_2}  $

\end{algorithm}

\section{ROM-ROM coupling and use in FSI coupling}\label{appendix:FSIalgorithms}

\begin{figure}
    \centering
    \includegraphics[width=\textwidth]{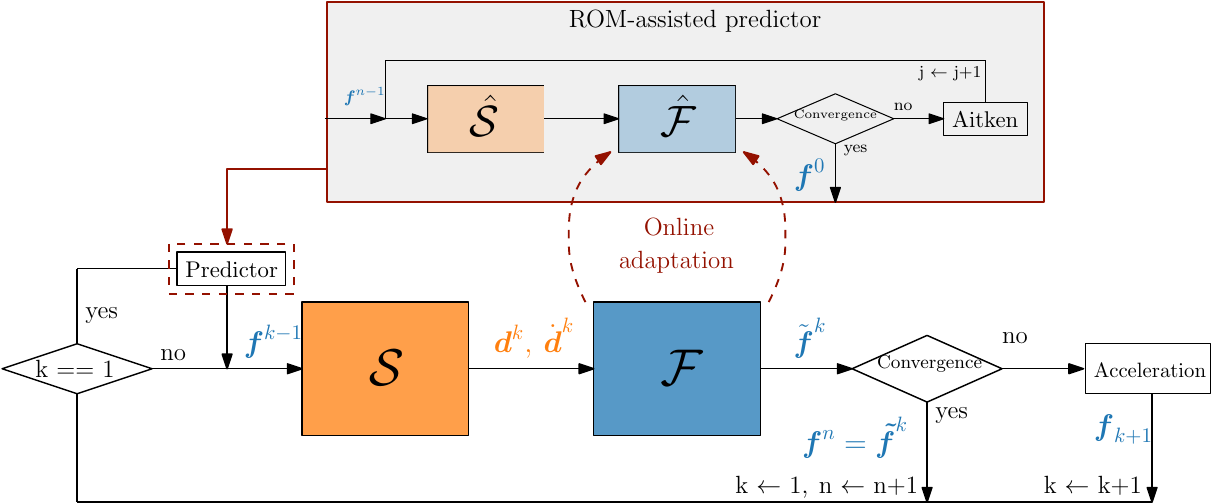}
    \caption{The proposed approach in \citep{TIBA2025109522}. In this work the adaptation is extended from the regression operator to the ROM subspace as well.}
    \label{fig:illust-pred-ROM}
\end{figure}

We recall here the approach of ROM-assisted predictors, introduced in \citep{TIBA2025109522}. A summarized illustration can be seen in Fig.\ref{fig:illust-pred-ROM}. In Algorithm \ref{local}, we recall the reduced coupling algorithm, and in Algorithm \ref{globalAlg}, we present the global FSI coupling approach, using the new proposed predictor.

\begin{algorithm}
\caption{Reduced coupling - $\operatorname{Predictor}(\cdot)$}\label{local}
\SetKwInput{KwData}{Input}
\KwData{
Fluid ROM $\widehat{F}^n_{\boldsymbol{\theta}}$

Solid ROM $\widehat{S}^n$

Initial relaxation $\beta^0$

Convergence tolerance $\delta_r$

Maximum number of iterations $M$

Previous time step interface forces, displacements and velocities $\boldsymbol{f}^{n-1}$,  $\boldsymbol{f}^{n-2}$, $\boldsymbol{f}^{n-3}$, $\boldsymbol{d}^{n-1}$,
$\dot{\boldsymbol{d}}^{n-1}$
}
\KwResult{Next initial guess $\boldsymbol{f}^{0, n}$}
\vline

\nl $\boldsymbol{f}^{0, n} = 3\boldsymbol{f}^{n-1} - 3\boldsymbol{f}^{n-2} + \boldsymbol{f}^{n-3}$\label{firstPred_local}
    
\nl $j = 1$

\nl $e_r = 1$

  \While{$e > \delta_r$ and $j<M$}{
    $\boldsymbol d^{r, j} = \widehat{S}^n\!\left( \left[ \boldsymbol{d}^{n-1}, \dot{\boldsymbol{d}}^{n-1} \right] ,\, 
    \boldsymbol{f}^{j-1, r}\right)$

    \nl $\Tilde{\boldsymbol f}^j = \widehat{F}^n_{\boldsymbol{\theta}} \left(  \boldsymbol f^{n-1}, \boldsymbol{d}^{r, j} \right)$, Algorithm \ref{alg:rom-pred}

    \nl $\boldsymbol r^j = \Tilde{\boldsymbol f}^j - \boldsymbol f^{j-1}$

    \nl $e_r = ||\boldsymbol r^j||_2/||\Tilde{\boldsymbol{f}}^j||_2$

    \uIf{$e_r > \delta_r$}{
    \nl $\beta^j = - \beta^{j-1} \frac{ \left(\boldsymbol{r}^{j-1} \right)^T (\boldsymbol{r}^{j}-\boldsymbol{r}^{j-1})}{||\boldsymbol{r}^{j}-\boldsymbol{r}^{j-1}||_2^2}$

    \nl ${\boldsymbol f^j} = \beta^j \Tilde{\boldsymbol f}^j + (1-\beta^j) \boldsymbol f^{j-1}$\label{relaxed_local}
  }\Else{
  \nl $\boldsymbol{f}^{0, n} = \Tilde{\boldsymbol{f}}^j$

  \nl \textbf{End algorithm}
  }
  \nl $j = j+1$
    }
\nl $\boldsymbol{f}^{0, n} = \boldsymbol f^{n-1}$ \textcolor{mynicegreen}{// \textit{Failure of convergence}}
\end{algorithm}

\begin{algorithm}
\caption{FSI iterative scheme}\label{globalAlg}

\nl Pre-prediction step of $F^n_{\boldsymbol\theta}$, Algorithm \ref{alg:pre-pred}

\nl $\boldsymbol{X}_*^0 = [ \,]$, $\boldsymbol{Y}_*^0 = [\, ]$

\nl $n = 1$

\nl $\kappa = 1$

\While{$n < N_t$}{

\nl $k = 1$

\nl $e = 1$

\nl Predictor : $\boldsymbol{f}^{0, n} = \operatorname{Predictor}(\boldsymbol f^{n-1}, \boldsymbol f^{n-2}, \boldsymbol f^{n-3})$ using Algorithm \ref{local}

  \While{$e > \delta$ \textit{and} $k<N_k$}{

    \nl Solid solver : ${\boldsymbol d}^k, {\dot{\boldsymbol{d}}}^k = \mathcal{S}(\boldsymbol{f}^{k-1}) $\label{callingSolid}

    \nl Fluid solver : $\Tilde{\boldsymbol f}^k = {\mathcal{F}}(\boldsymbol{d}^k, \dot{\boldsymbol{d}}^k)$\;\label{callingFluid}

    \nl Grouse update, Algorithm \ref{alg:rank1-update}

    \nl Update matrices ${\boldsymbol{X}^r_{*}}$ and $\Tilde{\boldsymbol{Y}}^r_{*}$ with $\boldsymbol{ u}^k$, $\Tilde{\boldsymbol f}^k$ and $\boldsymbol{f}^{n-1}$

    \uIf{$\kappa \bmod \tau = 0$}{\nl Retrain $\mathcal{I}_F(\cdot)$, Algorithm \ref{alg:regress-update}}
    \uIf{$\kappa \bmod K = 0$}{\nl Activate the updated basis, Algorithm \ref{alg:basis-update}}

    \nl $\kappa = \kappa + 1$

    \nl $\boldsymbol r^k = \Tilde{\boldsymbol f}^k - \boldsymbol f^{k-1}$

    \nl $e = ||\boldsymbol r^k||_2/\sqrt{N}$\label{alg:tolerance}

    \uIf{$e > \delta$}{
    \nl $\Delta \boldsymbol{f}^k = \operatorname{Acceleration}(\Tilde{\boldsymbol f}^k, \boldsymbol r^k)$ using the Quasi-Newton acceleration from \citep{degroote_performance_2009}

    \nl ${\boldsymbol f^k} = {\boldsymbol f}^{k-1} + \Delta \boldsymbol{f}^k$
  }\Else{
    \nl $\boldsymbol{f}^{n} = \Tilde{\boldsymbol f}^k$  \textcolor{mynicegreen}{// \textit{Convergence}}
  }
  \nl $k \leftarrow k+1$
    }
    \nl $n \leftarrow n + 1$}

\end{algorithm}

\section{Lid driven cavity: supplementary data and figures}\label{appendix:liddriven}

This appendix collects the detailed diagnostic figures for the lid driven cavity test case that complement the discussion in the main text.

\begin{figure}
    \centering
    \includegraphics[width=\linewidth]{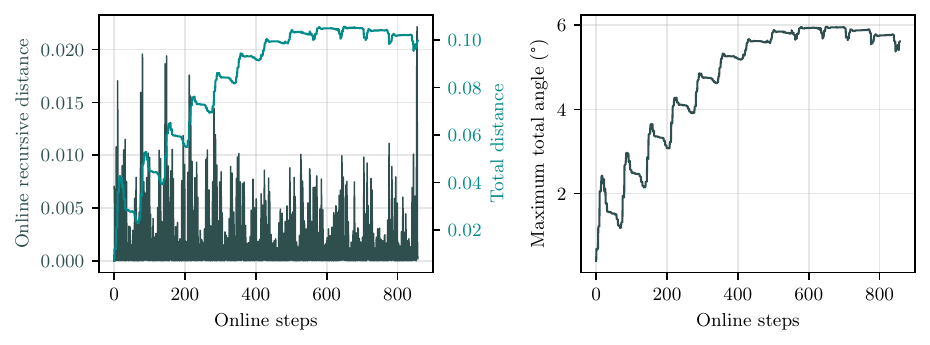}
    \caption{[Lid Driven Case] - Online recursive distance, accumulated subspace distance and maximum accumulated subspace angle through the online updates.}
    \label{fig:totalAngleRotationLidDriven}
\end{figure}

\begin{figure}
    \centering
    \includegraphics[width=.8\textwidth]{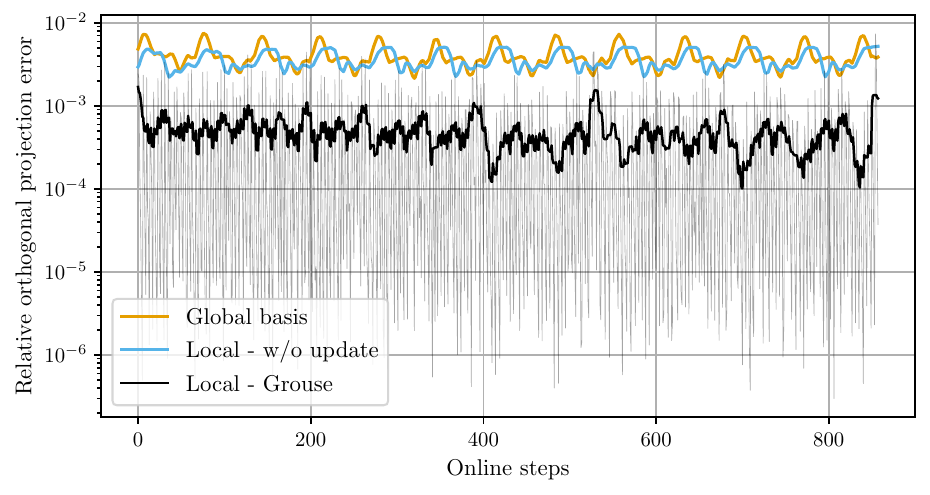}
    \caption{[Lid Driven Case] - Relative norm of the orthogonal part of the forces snapshots through the online steps, and comparing a global static basis, a local static basis, and GROUSE-updates basis. The GROUSE curve is shown in gray, and a smoothed overlay is shown in black.}
    \label{fig:LidDrivenOrthNorms}
\end{figure}

\begin{figure}
    \centering
    \includegraphics[width=\textwidth]{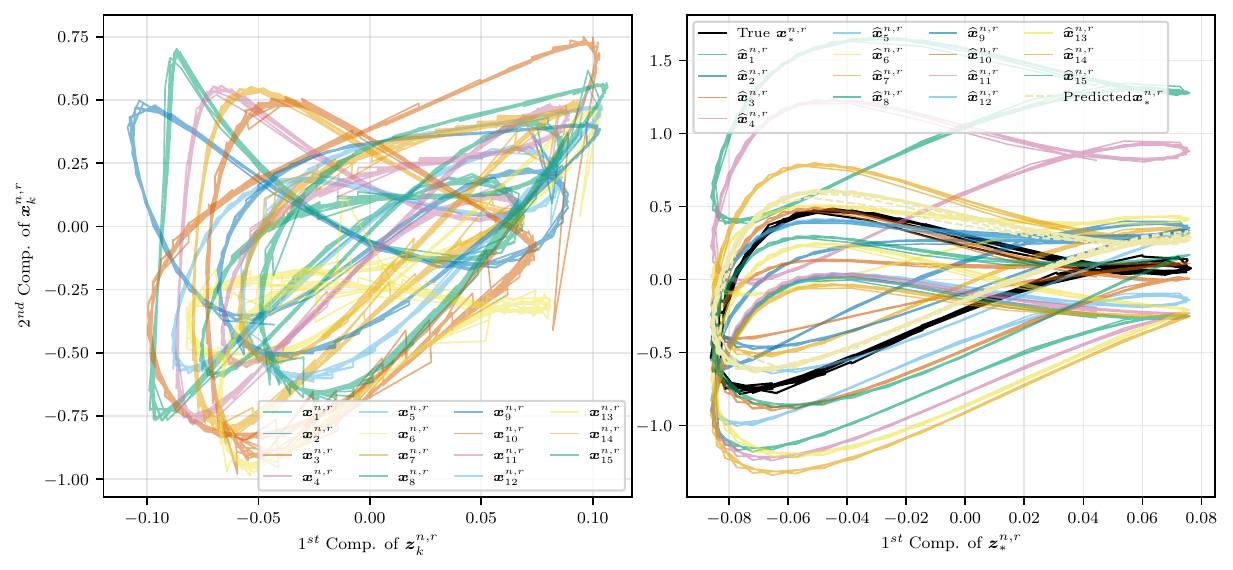}
    \caption{Lid Driven Case - Comparison of the different predicted reduced forces and their linear combination leading to the final ROM prediction in the reduced space. }
    \label{fig:compareLidDrivenReducedPrediction}
\end{figure}

Fig.~\ref{fig:LidDrivenOrthNorms} reports the relative orthogonal projection error, Fig.~\ref{fig:totalAngleRotationLidDriven} shows the accumulated subspace rotation angle, and Fig.~\ref{fig:compareLidDrivenReducedPrediction} illustrates the individual and combined predictions in the reduced space. Their interpretation follows the same patterns discussed for the VIV case in Section~\ref{section4}.

\begin{table}
\centering
\caption{Training parameter sets $(\rho,\, \mu)$.}
\label{tab:parameters}
\begin{tabular}{ccc}
\hline
\textbf{Set} & \textbf{Density} $\rho$ $[Kg/m^3]$ & \textbf{Dynamic Viscosity} $\mu$ $\times 10^{-3}$ $[Kg/ms]$ \\
\hline
 1 & $5.000 $ & $8.000$ \\
 2 & $5.000 $ & $1.250 \times 10^{1}$ \\
 3 & $5.000 $ & $1.700 \times 10^{1}$ \\
 4 & $8.750 $ & $8.000$ \\
 5 & $8.750 $ & $1.250 \times 10^{1}$ \\
 6 & $8.750 $ & $1.700 \times 10^{1}$ \\
 7 & $1.250 \times 10^{1}$  & $8.000 $ \\
 8 & $1.250 \times 10^{1}$  & $1.250 \times 10^{1}$ \\
 9 & $1.250 \times 10^{1}$  & $1.700 \times 10^{1}$ \\
10 & $1.625 \times 10^{1}$  & $8.000 $ \\
11 & $1.625 \times 10^{1}$  & $1.250 \times 10^{1}$ \\
12 & $1.625 \times 10^{1}$  & $1.700 \times 10^{1}$ \\
13 & $2.000 \times 10^{1}$  & $8.000$ \\
14 & $2.000 \times 10^{1}$  & $1.250 \times 10^{1}$ \\
15 & $2.000 \times 10^{1}$  & $1.700 \times 10^{1}$ \\
\hline
\end{tabular}
\end{table}

\begin{figure}
    \centering
    \begin{subfigure}[t]{0.48\columnwidth}
        \centering
        \includegraphics[width=\linewidth]{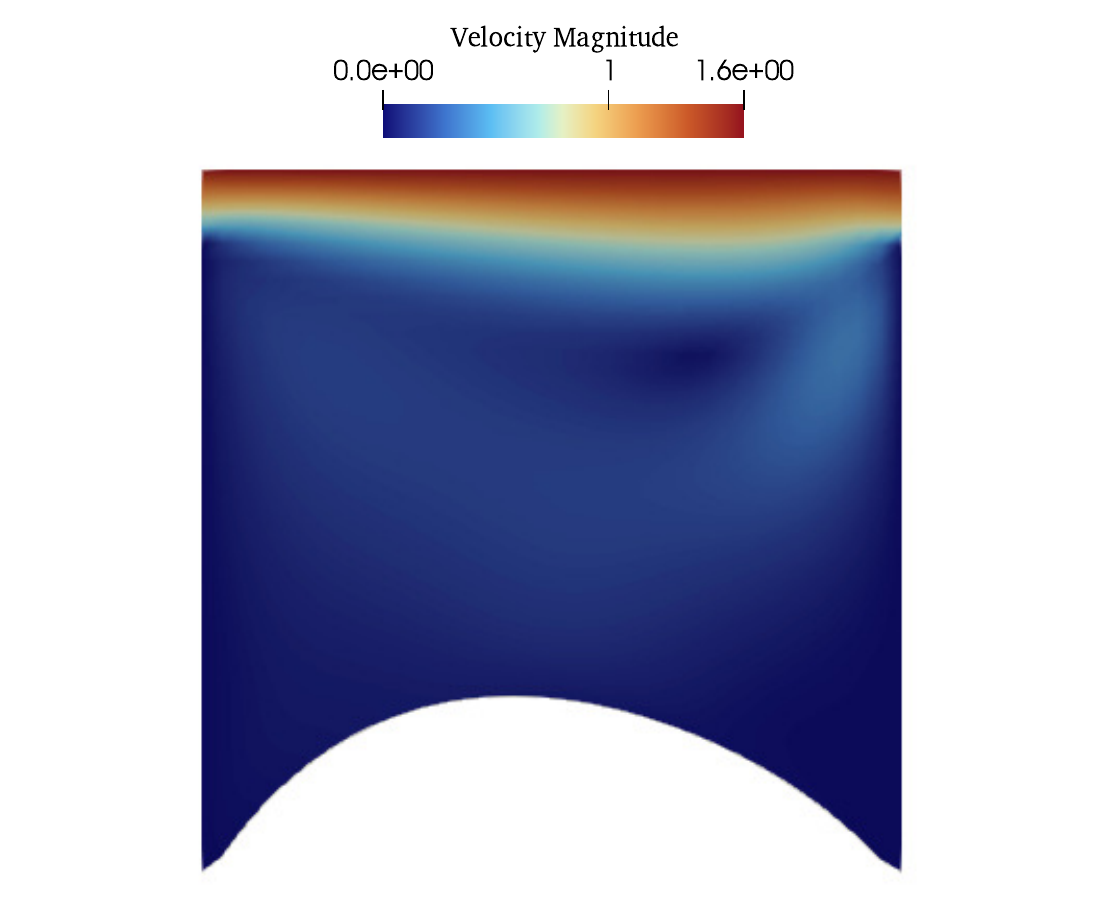}
        \caption{Velocity magnitude field in the fluid domain together with the deformed solid, at $t = 28\,\mathrm{s}$.}
        \label{fig:velocity_t28}
    \end{subfigure}
    \hfill
    \begin{subfigure}[t]{0.48\columnwidth}
        \centering
        \includegraphics[width=\linewidth]{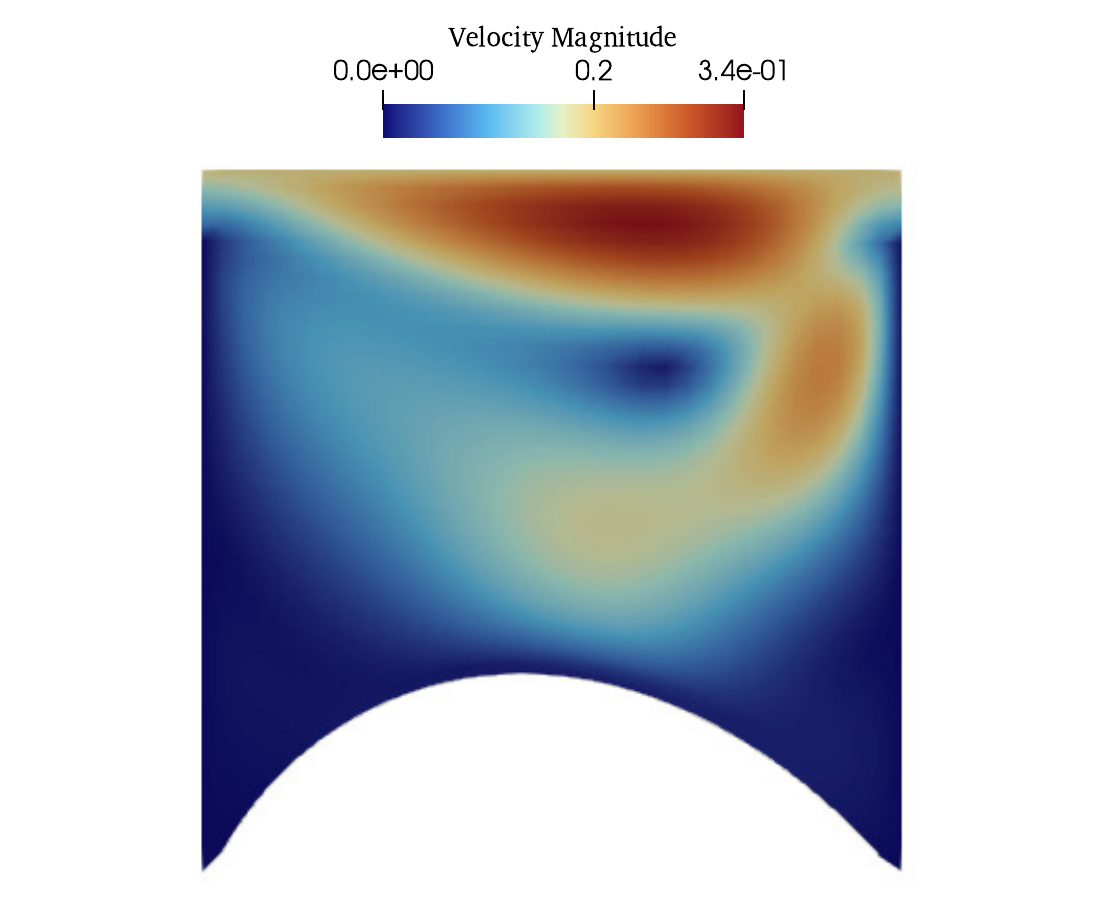}
        \caption{Velocity magnitude field in the fluid domain together with the deformed solid, at $t = 29.25\,\mathrm{s}$.}
        \label{fig:velocity_t2925}
    \end{subfigure}
    \caption{Velocity magnitude field in the fluid domain and the corresponding deformed solid configuration at two representative time instants: (a) $t = 28\,\mathrm{s}$ and (b) $t = 29.25\,\mathrm{s}$.}
    \label{fig:velocity_deformation}
\end{figure}

\begin{figure}
    \centering
    \includegraphics[width=.5\textwidth]{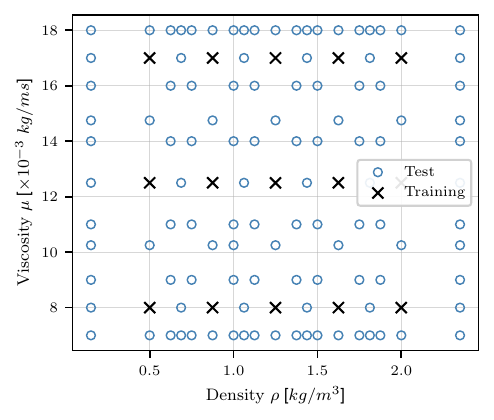}
    \caption{[Lid Driven Case] - Parameter grid.}
    \label{fig:lidDrivenGrid}
\end{figure}

\begin{figure}
    \centering
    \includegraphics[width=.8\textwidth]{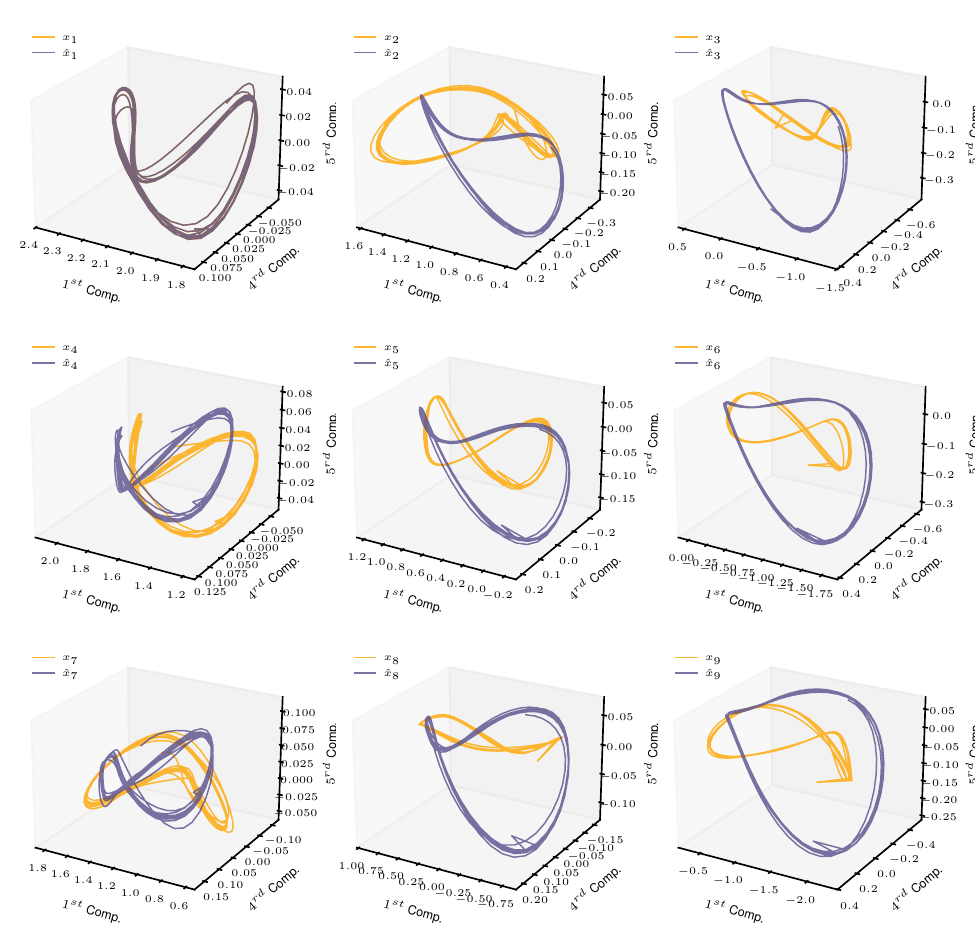}
    \caption{[Lid Driven case] - The trajectories on a chosen 3D coordinate system in the latent space}
    \label{fig:LidDRivencompareReducedSnaps3D}
\end{figure}


\begin{figure}
    \centering
    \includegraphics[width=\linewidth]{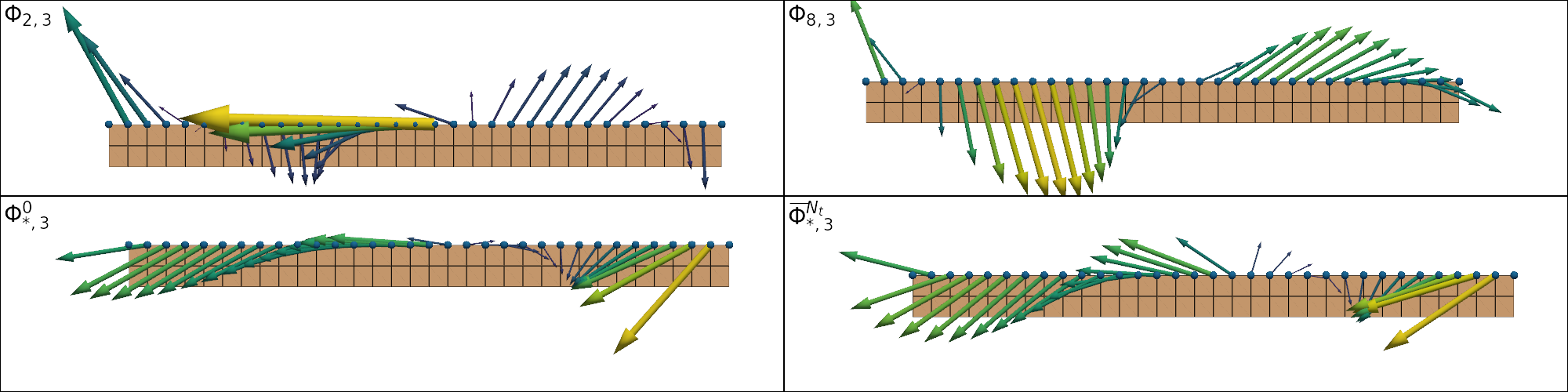}
    \caption{[Lid Driven Case] - Comparison of the POD modes of the forces' field. Here the third mode is shown. Top row: The modes associated to $\boldsymbol{\mu}_2$ and $\boldsymbol{\mu}_8$. Bottom left: The predicted local mode associated to $\boldsymbol{\mu}_{*}$. Bottom right: The updated basis at the end of the online simulation.}
    \label{fig:modesComparisonLidDriven}
\end{figure}

\begin{figure}
    \centering
    \includegraphics[width=\textwidth]{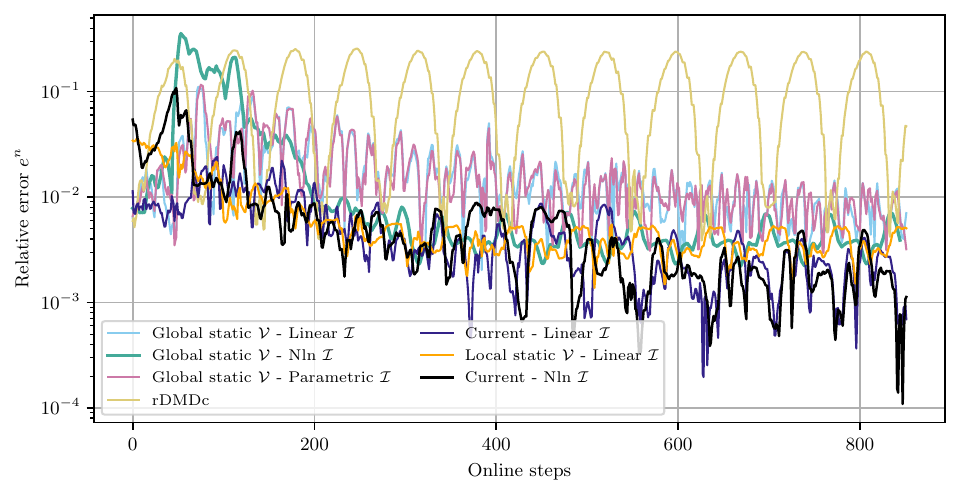}
    \caption{[Lid Driven Case] - Relative error of the forces through the online steps.}
    \label{fig:lidDrivenAccuracy}
\end{figure}

\begin{figure}
    \centering
    \includegraphics[width=\textwidth]{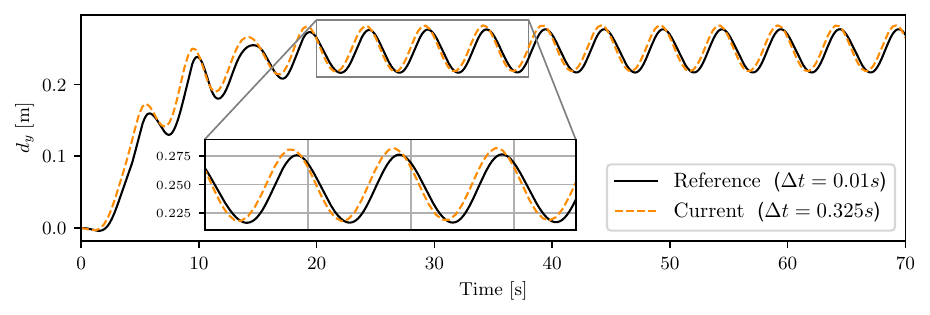}
    \caption{[Lid Driven Case] - Comparison between the reference case and the case corresponding to the biggest time step considered here.}
    \label{fig:comparingReferenceToOurs}
\end{figure}

\begin{table}
\centering
\caption{[Lid driven case] -- Comparison of the tip displacement oscillation characteristics with the benchmark solution reported in \cite{dissertation-2117-94177}.}
\label{tab:benchmark_comparison}
\begin{tabular}{lcc}
\toprule
Quantity & \citet{dissertation-2117-94177} ($\Delta t = 0.01~s$) & Current reference ($\Delta t = 0.01~s$) \\
\midrule
Period, $T$ [s] &
$T_{\mathrm{ref}} = 5.12$ &
$T_{\mathrm{sim}} = 4.75$\\

Maximum displacement, $u_{\max}$ [cm] &
$u_{\max,\mathrm{ref}} = 27.2$ &
$u_{\max,\mathrm{sim}} = 27.69$\\

Mean displacement, $\bar{u}$ [cm] &
$\bar{u}_{\mathrm{ref}} = 23.5$ &
$\bar{u}_{\mathrm{sim}} = 24.66$\\
\bottomrule
\end{tabular}
\end{table}

\section{Turek Hron benchmark: supplementary data and figures}\label{appendix:turek}

This appendix collects the detailed diagnostic figures for the Turek--Hron benchmark that complement the discussion in the main text.

\begin{figure}
    \centering
    \includegraphics[width=\textwidth]{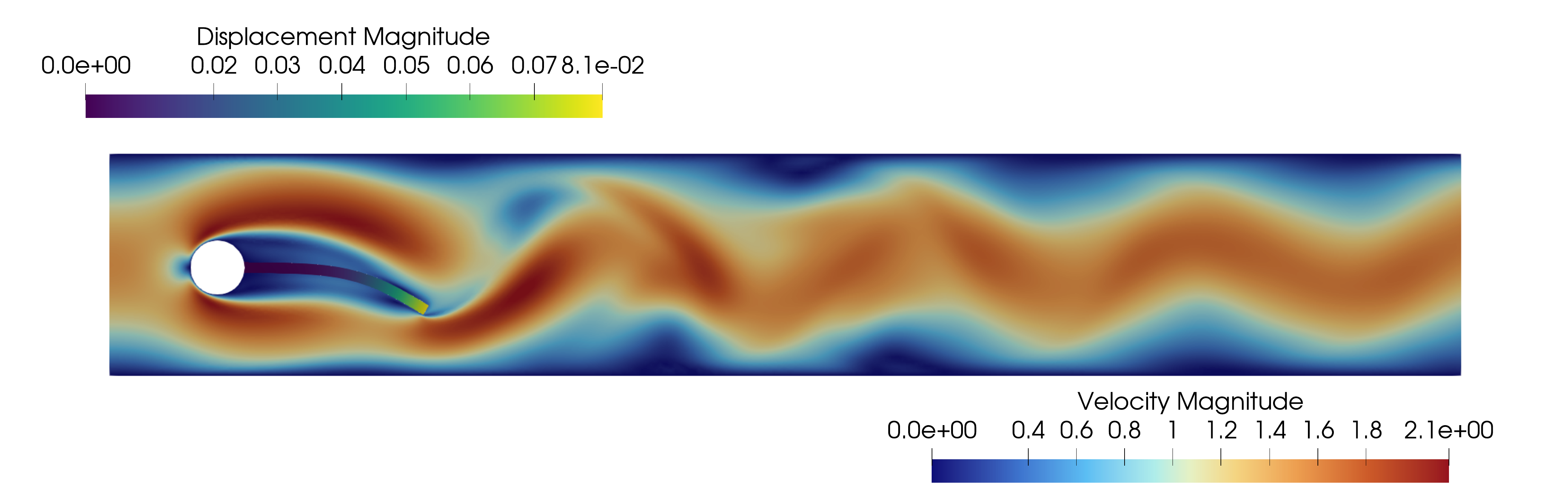}
    \caption{[Turek Hron case] - Velocity magnitude field in the fluid domain and the displacement field at the deformed solid, at $t = 12.536\,\mathrm{s}$.}
    \label{fig:turekHronSnap}
\end{figure}

\begin{table}
\centering
\caption{Training parameter sets $(c_v,\, \mu)$ for the Turek--Hron benchmark, taken at the corners of the square $[0.9,1.1]\times[0.9,1.1]$ surrounding the benchmark parameter $\boldsymbol{\theta}_* = (1,1)$.}
\label{tab:turekParameters}
\begin{tabular}{ccc}
\hline
\textbf{Set} & \textbf{Velocity coefficient} $c_v$ & \textbf{Dynamic viscosity} $\mu$ $[Kg/ms]$ \\
\hline
1 & $0.9$ & $0.9$ \\
2 & $0.9$ & $1.1$ \\
3 & $1.1$ & $0.9$ \\
4 & $1.1$ & $1.1$ \\
\hline
\end{tabular}
\end{table}

\begin{figure}
    \centering
    \includegraphics[width=.75\textwidth]{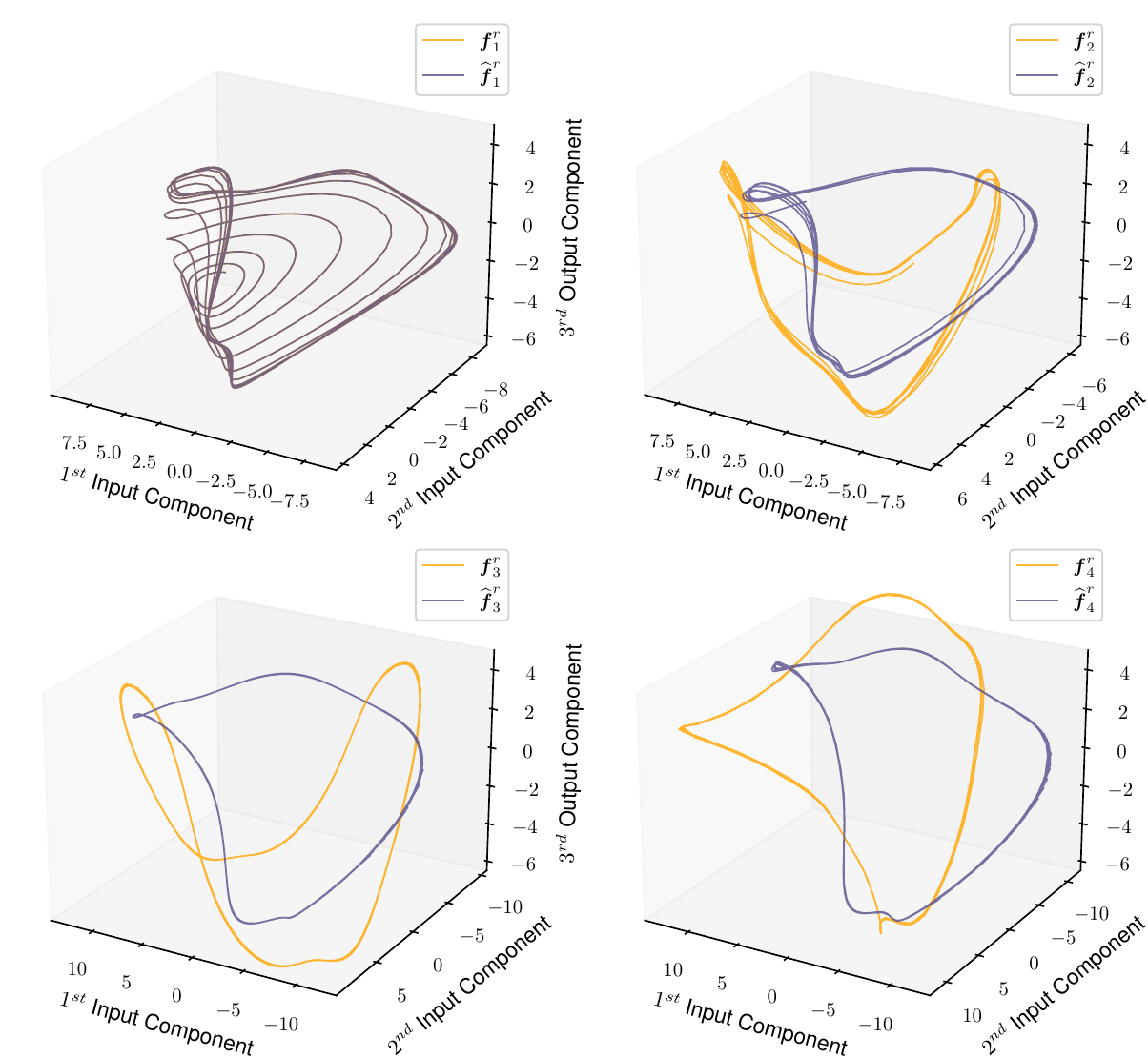}
    \caption{[Turek case] - The trajectories on a chosen 3D coordinate system in the latent space}
    \label{fig:TurekcompareReducedSnaps3D}
\end{figure}

\begin{figure}
    \centering
    \includegraphics[width=\linewidth]{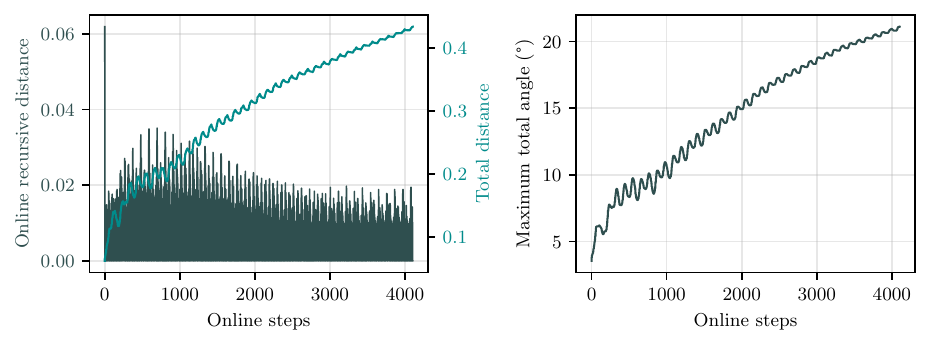}
    \caption{[Turek] -- Left: Online distance of the updated subspace from that used in the previous iteration, shown together with the the total distance of the updated subspace to the initial subspace. \, Right: The maximum subspace angle between the current $\mathcal{V}^n_{*}$ and the initial subspsace $\mathcal{V}^0_{*}$.}
    \label{fig:combinedAngleShiftOnlineTurek}
\end{figure}

\begin{figure}
    \centering
    \includegraphics[width=.8\textwidth]{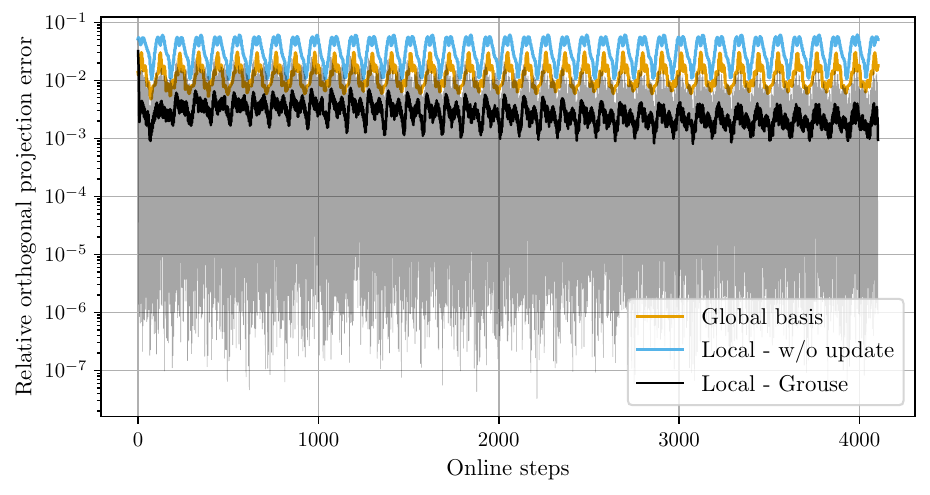}
    \caption{[Turek Case] - Relative norm of the orthogonal part of the fluid field snapshot through the online steps.}
    \label{fig:TurekOrthNorms}
\end{figure}

\begin{figure}
    \centering
    \includegraphics[width=.9\textwidth]{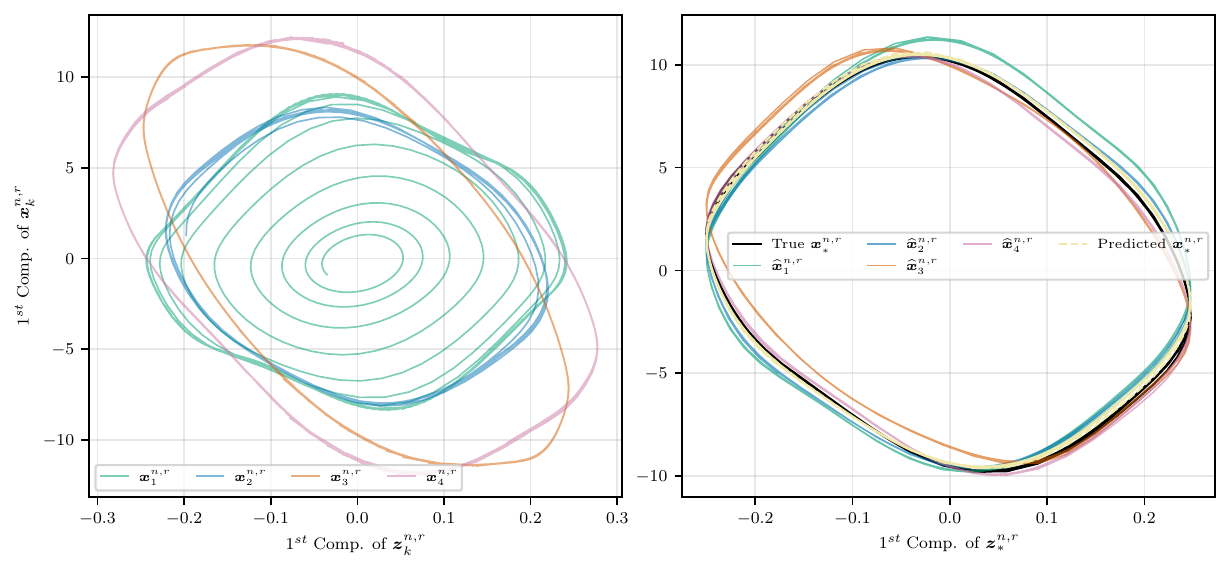}
    \caption{Turek Case - Comparison of the different predicted reduced forces and their linear combination leading to the final ROM prediction in the reduced space. }
    \label{fig:compareTurekReducedPrediction}
\end{figure}

\begin{figure}
    \centering
    \includegraphics[width=\textwidth]{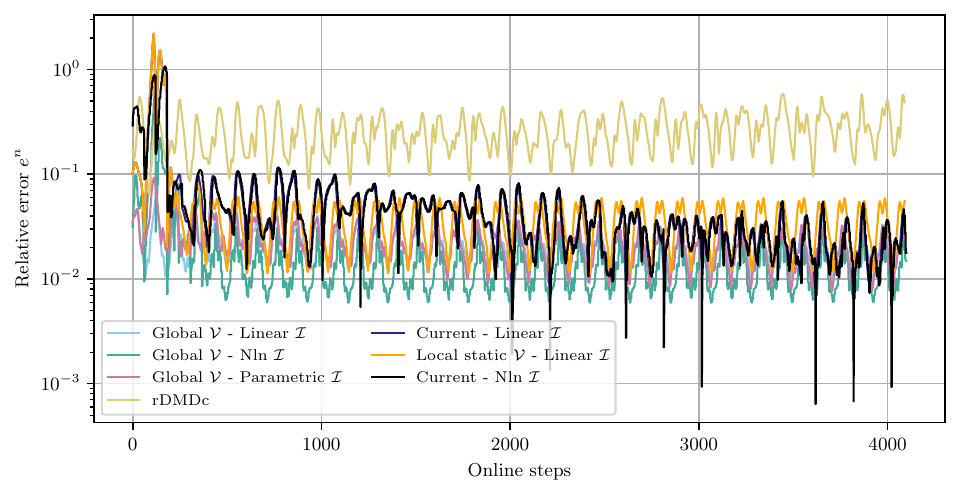}
    \caption{[Turek Case] - Relative error of the forces through the online steps.}
    \label{fig:turekAccurac}
\end{figure}

\begin{figure}
    \centering
    \includegraphics[width=\textwidth]{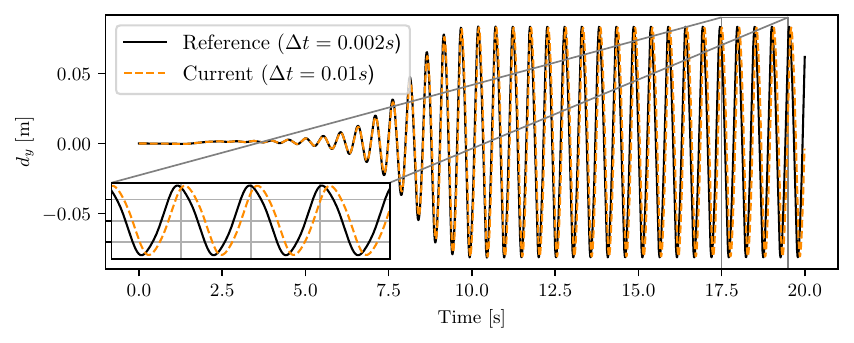}
    \caption{[Turek Case] - Comparison between the reference case and the studied case with $\Delta t = 0.01~s$.}
    \label{fig:comparingReferenceToOursTurek}
\end{figure}

In Table \ref{tab:benchmark_comparison_turek}, the reference solution is validated against the benchmark results in \citep{turekBench}. In Fig. \ref{fig:comparingReferenceToOursTurek}, we compare the time evolution of the vertical displacement of the beam's tip, obtained with $\Delta t = 0.01~s$ and the reference solution.

\begin{table}
\centering
\caption{[Turek case] -- Comparison of the vertical tip displacement oscillation characteristics with the benchmark solution reported in \cite{turekBench}.}
\label{tab:benchmark_comparison_turek}
\begin{tabular}{lcc}
\toprule
Quantity & \citet{turekBench} ($\Delta t = 0.002~s$) & Current reference ($\Delta t = 0.002~s$) \\
\midrule
Period, $T$ [s] &
$T_{\mathrm{ref}} = 0.5$ &
$T_{\mathrm{current}} = 0.5$ \\

Maximum displacement, $u_{\max}$ [cm] &
$u_{\max,\mathrm{ref}} = 8.195$ &
$u_{\max,\mathrm{current}} = 8.328$\\

Mean displacement, $\bar{u}$ [mm] &
$\bar{u}_{\mathrm{ref}} = 1.25$ &
$\bar{u}_{\mathrm{current}} = 1.02$\\
\bottomrule
\end{tabular}
\end{table}

\clearpage

\printcredits

\bibliographystyle{elsarticle-num-names}
\bibliography{cas-refs}

\end{document}